\documentclass[a4paper,11pt]{article}

\usepackage{jheppub} 

\usepackage[T1]{fontenc} 

\newcommand\re {\mathrm{Re}}
\newcommand\im {\mathrm{Im}}
\newcommand\Tr {\mathrm{Tr}}

\newcommand\arccosh {\mathrm{arccosh}}
\newcommand\Schrodinger {Schr\"{o}dinger }
\newcommand\Renyi {R\'enyi\ }

\title{\boldmath Entanglement of Harmonic Systems in Squeezed States}

\author[a]{D. Katsinis,}
\author[b,c]{G. Pastras}
\author[d]{and N. Tetradis}

\affiliation[a]{Instituto de F\'isica, Universidade de S\~ao Paulo,\\ Rua do Mat\~ao Travessa 1371, 05508-090 S\~ao Paulo, SP, Brazil}
\affiliation[b]{Institute of Nuclear and Particle Physics, NCSR `Demokritos',\\ Aghia Paraskevi 15310, Greece}
\affiliation[c]{Laboratory for Manufacturing Systems and Automation, Department of Mechanical Engineering and Aeronautics, University of Patras,\\Patra 26110, Greece}
\affiliation[d]{Department of Physics, University of Athens,\\Zographou 157 84, Greece}

\emailAdd{dkatsinis@phys.uoa.gr}
\emailAdd{pastras@lms.mech.upatras.gr}
\emailAdd{ntetrad@phys.uoa.gr}

\abstract{The entanglement entropy of a free scalar field in its ground state is dominated by an area law term. It is noteworthy, however, that the study of entanglement in scalar field theory has not advanced far beyond the ground state. In this paper, we extend the study of entanglement of harmonic systems, which include free scalar field theory as a continuum limit, to the case of the most general Gaussian states, namely the squeezed states. We find the eigenstates and the spectrum of the reduced density matrix and we calculate the entanglement entropy. We show that our method is equivalent to the correlation matrix method. Finally, we apply our method to free scalar field theory in 1+1 dimensions and show that, for very squeezed states, the entanglement entropy is dominated by a volume term, unlike the ground-state case. Even though the state of the system is time-dependent in a non-trivial manner, this volume term is time-independent. We expect this behaviour to hold in higher dimensions as well, as it emerges in a large-squeezing expansion of the entanglement entropy for a general harmonic system.}

\begin{document} 
\maketitle
\flushbottom

\section{Introduction}
\label{sec:intro}

It is well-known that entanglement entropy in scalar field theory in its ground state is dominated by an area law. This peculiar feature resembles the famous property of black hole entropy, giving rise to simple, but very fundamental, questions: Can black hole entropy be attributed to quantum entanglement? Can gravity be described as a statistical entropic force, attributed to quantum statistics due to entanglement \cite{Jacobson:1995ab,VanRaamsdonk:2010pw,Jacobson:2015hqa}?
Such a description of gravity is at least not contradictory to holographic duality. It has been shown that Einstein's equations in the bulk, or at least the linearized Einstein's equations around the pure AdS geometry, emerge as a holographic realization of the first law of entanglement thermodynamics \cite{Lashkari:2013koa,Faulkner:2013ica}, i.e. of the relation
\begin{equation}
\delta S_{\mathrm{EE}} = \delta \left< H_{\mathrm{mod}} \right> ,
\end{equation}
which is a trivial identity for any quantum system, and, thus, in the context of the holographic duality, trivially holds for the boundary theory.

The study of entanglement in scalar field theory was initiated long ago. In 1986, Bombelli et al. argued that entanglement entropy in scalar field theory should obey an area law \cite{Bombelli:1986rw}. The actual numerical calculation was performed a few years later by Srednicki \cite{srednicki}, who showed that entanglement entropy in free massless scalar field theory in its ground state does obey an area law. For a review of entanglement entropy calculations in field theory, the reader may consult \cite{Calabrese:2004eu,Casini:2009sr,Calabrese:2009qy}. A basic element of Srednicki's calculation is the discretization of the degrees of freedom of the scalar field theory on a lattice of spherical shells. This discretization gives rise to a Hamiltonian of an infinite, but countable number of degrees of freedom, i.e. a textbook quantum mechanics harmonic system. Srednicki's calculation is further based on two facts that apply to the specific ground state scenario:
\begin{enumerate}
\item The reduced density matrix can be calculated explicitly in coordinate representation via the performance of Gaussian integrals, due to the fact that the ground state of a harmonic system is a Gaussian state.
\item The reduced density matrix is also a Gaussian kernel in coordinate representation and its eigenstates and eigenvalues can be found explicitly. They resemble the eigenstates of an effective harmonic system, with as many degrees of freedom as those that have not been traced out. However, this effective harmonic system does not lie in its ground state, but in a mixed state. More specifically, each normal mode of the effective system appears to lie in a thermal state. The reduced system is not strictly thermal, as each mode is characterized by a distinct temperature. 
\end{enumerate}

Since the above hold specifically for the ground state of a harmonic system, the generalization of Srednicki's techniques to the study of entanglement in more general states presents a high level of difficulty. However, both facts listed above still hold for a massive field theory. In this case, the inverse of the mass can be used as a perturbative parameter to bypass the numerical part of Srednicki's calculation \cite{Katsinis:2017qzh}. Interestingly, both facts are still true in the case that the overall quantum harmonic system lies in a thermal state \cite{Katsinis:2019vhk,Katsinis:2019lis}. In such a case, the quantity that is proportional to the area of the entangling surface is not the entanglement entropy, but rather the mutual information of the system. Finally, both facts are also true when the harmonic system lies in a coherent state. It turns out that the entanglement entropy in such a case is identical to that when the system lies in its ground state \cite{Benedict:1995yp}. Interestingly, the time evolution of the reduced density matrix is unitary and is described by an effective quadratic Hamiltonian with explicit time-dependence \cite{Katsinis:2022fxu}.

An alternative method for the calculation of the entanglement entropy is based on the correlation function matrix. In this approach the reduced density matrix is constructed so as to reproduce the same correlation functions in subsystem under consideration as the overall density matrix. This method is particularly powerful in free field theory, where all correlation functions are determined by the two-point correlators. In the following, we shall show that the method that we develop is equivalent to the method based on the correlation function matrix.

In a seminal paper \cite{Page:1993df}, Page showed that, in an arbitrary quantum state, the entanglement entropy is close to maximal. The possible maximum is proportional to the number of degrees of freedom of the smaller of the two subsystems. The proportionality constant depends on the dimensionality of the Hilbert space of a single degree of freedom. This bound has been connected to the Page curve obeyed by the entropy of black hole radiation \cite{Page:1993wv}. In scalar field theory, the dimensionality of the Hilbert space of a local degree of freedom is infinite, rendering the above bound also infinite. However, it is still proportional to the number of degrees of freedom, implying that we should expect the entanglement entropy to be proportional to the volume of the smaller subsystem, and not the area (see also \cite{Eisert:2008ur}). In this sense, the states of scalar field theory that have been investigated, namely the ground state and coherent states, cannot be considered as arbitrary quantum states, but as states with special entanglement characteristics. Therefore, it is worth investigating how Page's argument applies to more general quantum states of scalar field theory. Srednicki's method appears to be more easily generalizable to Gaussian states, indicating as a more promising direction the study of the squeezed states. In a similar manner, random Gaussian states have been studied in fermionic systems \cite{Bianchi:2021lnp}, and it was shown that indeed the mean entanglement entropy approaches that of the Page curve.

In the case of the most general Gaussian state of the overall harmonic system, i.e. a squeezed state, the first fact that facilitates Srednicki's calculation still holds, namely the reduced density matrix can be found explicitly via Gaussian integrals. However, as we will discuss in what follows, the second fact does not apply, making the calculation of entanglement entropy a more complicated task. This problem has been studied at a more abstract level \cite{Bianchi:2015fra,Adesso:2014}. In this work, we develop a method for the calculation of the spectrum of the reduced density matrix, which is the direct generalization of Srednicki's method \cite{srednicki}. Our method is shown to be equivalent to the correlation matrix method. Then, we apply this method to the system of free massless scalar field theory in 1+1 dimensions. Furthermore, we study the dependence of entanglement entropy on squeezing.

The structure of the paper is as follows. In section \ref{sec:squeezed_review} we review basic facts about the coherent and squeezed states of the quantum harmonic oscillator. In section \ref{sec:two_oscillators} we study the special case of two coupled harmonic oscillators, which can be solved directly with Srednicki's method, and study the dependence of entanglement on squeezing. In section \ref{sec:many_oscillators} we generalize Srednicki's method to harmonic systems with an arbitrary number of degrees of freedom lying in a squeezed state. In section \ref{sec:expansions} we develop an expansion for the entanglement entropy for very squeezed states. In section \ref{sec:1dft} we apply our method to free massless scalar field theory in 1+1 dimensions. In section \ref{sec:discussion} we discuss our results. In order not to distract from the main part of the analysis, several secondary aspects, as well as technical details of the calculations, have been relegated to a series of appendices, so that they do not disturb the flow of the main text. Appendix \ref{sec:app_squeezing_entanglement} contains the quantitative analysis, with all technical details, of how squeezing affects entanglement in the simple case of the two oscillators. Appendix \ref{subsubsec:spectrum_creation_annihilation} presents an algebraic construction of the eigenstates of the reduced density matrix. Appendix \ref{sec:app_solvable} discusses the special solvable case of a harmonic system in a very specific squeezed state, which demonstrates technicalities of our method and provides a consistency check of the expansion of section \ref{sec:expansions}. In Appendix \ref{sec:app_renyi} the validity of our method is checked in a special example that can be solved in a different way, via the extension of the entanglement entropy to the family of \Renyi entanglement entropies. In appendix \ref{subsec:correlation} we demonstrate the equivalence of our method and the correlation matrix method. In appendix \ref{subsec:small_expansion} we develop an expansion for small squeezing parameters.

\section{Gaussian Solutions of the Simple Quantum Harmonic Oscillator}
\label{sec:squeezed_review}

In this section we define the basic terminology in relation to the Gaussian solutions of the simple quantum harmonic oscillator and remind the reader of some basic properties that they possess. The general Gaussian solutions are characterized as ``squeezed'' states. A special subclass of those with enhanced properties are the so-called ``coherent'' states.

Let us consider the simple quantum harmonic oscillator with Hamiltonian
\begin{equation}
\hat{H} = \frac{\hat{p}^2}{2 m} + \frac{m \omega^2 \hat{x}^2}{2} .
\label{eq:sho_Hamiltonian}
\end{equation}
The time-dependent \Schrodinger equation
\begin{equation}
\hat{H} \psi \left( t , x \right) = i \hbar \frac{\partial \psi \left( t , x \right)}{\partial t}
\end{equation}
possesses several Gaussian solutions. The most well-known class of such solutions consists of the so-called coherent states, whose wavefunction reads
\begin{equation}
\Psi \left( t , x \right) = \left( \frac{m \omega}{\pi \hbar} \right)^{\frac{1}{4}} \exp \left[ - \frac{m \omega \left( x - x_0 \left( t \right) \right)^2}{2 \hbar} + i \frac{p_0 \left( t \right) \left( x - x_0 \left( t \right) \right)}{\hbar} - i \varphi_{\mathrm{c}} \left( t \right) \right] ,
\label{eq:coherent}
\end{equation}
where 
\begin{align}
x_0 \left( t \right) &= X_0 \cos \left[ \omega \left( t - t_0 \right) \right] , \label{eq:coherent_mean_x} \\
p_0 \left( t \right) &= - P_0 \sin \left[ \omega \left( t - t_0 \right) \right] , \quad P_0 = m \omega X_0 , \label{eq:coherent_mean_p} \\
\varphi_{\mathrm{c}} \left( t \right) &= \frac{1}{2} \omega \left( t - t_0 \right) + \frac{X_0 P_0}{4 \hbar} \sin \left[ 2 \omega \left( t - t_0 \right) \right] + \varphi_0 . \label{eq:coherent_phase}
\end{align}

The coherent states possess several very interesting properties:
\begin{enumerate}
\item The mean values of position and momentum follow a classical orbit. Namely
\begin{equation}
\left< \hat{x} \right> = x_0 \left( t \right) = X_0 \cos \left[ \omega \left( t - t_0 \right) \right] , \quad \left< \hat{p} \right> = p_0 \left( t \right) = - m \omega X_0 \sin \left[ \omega \left( t - t_0 \right) \right] .
\end{equation}
\item They are states of \emph{minimal} and \emph{balanced} uncertainty. Namely
\begin{equation}
\Delta x = \sqrt{\frac{\hbar}{2 m \omega}} , \quad \Delta p = \sqrt{\frac{\hbar m \omega}{2}} , \quad \Delta x \Delta p = \frac{\hbar}{2} .
\end{equation}
\item The quadratic part of the exponent of the Gaussian wavefunction is \emph{real}.
\item The ground state is the special case of coherent state with $X_0 = 0$.
\end{enumerate}

If the \Schrodinger equation is solved with an initial condition identified with a Gaussian state of minimal and balanced uncertainties for position and momentum, then its solution is necessarily a coherent state. On the contrary, if it is solved with an initial condition which is a Gaussian state that does not obey both these conditions, i.e. either it is not a minimal uncertainty state, or the uncertainties of position and momentum are not balanced, then its solution is a squeezed state. A squeezed state is still Gaussian at all times, like a coherent state, but it is more general:
\begin{equation}
\Psi \left( t , x \right) = \left( \frac{m \re \left( w \left( t \right) \right)}{\pi \hbar} \right)^{\frac{1}{4}} \exp \left[ - \frac{m w \left( t \right) \left( x - x_0 \left( t \right) \right)^2}{2 \hbar} + i \frac{p_0 \left( t \right) \left( x - x_0 \left( t \right) \right)}{\hbar} - i \varphi_{\mathrm{s}} \left( t \right) \right] ,
\label{eq:squeezed}
\end{equation}
where
\begin{align}
w \left( t \right) &= \omega \frac{1 - i \sinh z \cos \left[ 2 \omega \left( t - t_0 \right) \right]}{\cosh z + \sinh z \sin \left[ 2 \omega \left( t - t_0 \right) \right]} , \label{eq:omega_squeezed}\\
\varphi_{\mathrm{s}} \left( t \right) &= \frac{1}{2} \arctan \frac{\tanh \frac{z}{2} + \tan \left[ \omega \left( t - t_0 \right) \right]}{1 + \tanh \frac{z}{2} \tan \left[ \omega \left( t - t_0 \right) \right]} + \frac{X_0 P_0}{4 \hbar} \sin \left[ 2 \omega \left( t - t_0 \right) \right] + \varphi_0.
\label{eq:phi_squeezed}
\end{align}
The functions $x_0 \left( t \right)$ and $p_0 \left( t \right)$ are given by equations \eqref{eq:coherent_mean_x} and \eqref{eq:coherent_mean_p}, as in the case of the coherent states. The parameter $z$ is called the squeezing parameter.

The squeezed states retain some, but not all, properties of the coherent states. Namely:
\begin{enumerate}
\item The mean position and momentum follow a classical orbit, exactly like in the case of coherent states.
\item They are not minimal uncertainty states at all times. More specifically:
\begin{equation}
\Delta x \Delta p = \frac{\hbar}{2} \sqrt{1 + \sinh^2 z \cos^2 \left[ 2 \omega \left( t - t_0 \right) \right]} .
\label{eq:squeezed_uncertainty_product}
\end{equation}
It follows that they are minimal uncertainty states exactly four times during each period of the corresponding classical harmonic oscillator, namely at times
\begin{equation}
t_{\min} = t_0 + \frac{T}{8} + n \frac{T}{4} , \quad n \in \mathbb{Z},
\end{equation}
where $T$ is the period of the oscillator, i.e. $T = 2 \pi / \omega$. At these instants $w$,
defined in (\ref{eq:omega_squeezed}), is real.
\item Even at the instants that the squeezed states are minimal uncertainty states, the uncertainties of position and momentum are \emph{not balanced}. In general
\begin{equation}
\Delta x = \sqrt{\frac{\hbar}{2 m \re \left(w\right)}} , \quad \Delta p = \sqrt{\frac{\hbar m \re \left(w\right)}{2}} \sqrt{1+\left(\frac{\im \left(w\right)}{\re \left(w\right)}\right)^2},
\label{eq:squeezed_uncertainties_minimal}
\end{equation}
implying that, at the instants that the squeezed state is a minimal uncertainty state, we have
\begin{equation}
\Delta x_{\min} = \sqrt{\frac{\hbar}{2 m \omega e^{\pm z}}} , \quad \Delta p_{\min} = \sqrt{\frac{\hbar m \omega e^{\pm z}}{2}} .
\end{equation}
\item The quadratic part of the exponent of the Gaussian wavefunction is not real.
\item The coherent states are special cases of squeezed states with $z = 0$.
\end{enumerate}

Because of the properties of the coherent and squeezed states, one can conceive the coherent states as the closest to classical states of the quantum harmonic oscillator, whereas the squeezed states as the next to closest. For this reason, the study of entanglement in systems lying in squeezed states presents a certain interest, since it reveals the behaviour of the system in states which are not very close to classicality. For example, it is known that the spectrum of the reduced density matrix for an arbitrary coherent state is identical to that for the ground state. This does not hold for the squeezed states.

As we will discuss in what follows, the study of entanglement in squeezed states of the overall system is much more difficult than the study of entanglement in coherent states, because the quadratic part of the exponent of the Gaussian state is not real. Although this fact does not complicate the explicit calculation of the reduced density matrix via the use of Gaussian integrals, the calculation of its spectrum is much more involved.

\section{The Special Case of Two Oscillators}
\label{sec:two_oscillators}

Let us consider the case of two coupled oscillators. From now on for simplicity we use units where $\hbar = 1$. Without loss of generality we consider that the mass of each oscillator is equal to one. Furthermore, for simplicity of the presentation of this toy case, we consider identical self couplings of the two oscillators. The Hamiltonian of the system is
\begin{equation}\label{eq:hamiltonian_2_osc}
H = \frac{1}{2} \left[ p_1^2 + p_2^2 + k_0 \left( x_1^2 + x_2^2 \right) + k_1 \left( x_1 - x_2 \right)^2 \right] .
\end{equation}
In terms of the canonical coordinates
\begin{equation}\label{eq:normal_coords}
x_\pm = \frac{1}{\sqrt{2}} \left( x_1 \pm x_2 \right)
\end{equation}
the Hamiltonian assumes the form
\begin{equation}
H = \frac{1}{2} \left( p_+^2 + p_-^2 + \omega_+^2 x_+^2 + \omega_-^2 x_-^2 \right) ,
\end{equation}
where $\omega_+ = \sqrt{k_0}$ and $\omega_- = \sqrt{k_0 + 2 k_1}$. As expected, the normal modes are decoupled.

\subsection{The Reduced Density Matrix}
\label{subsec:two_oscillators_density}

We consider the overall system lying in a squeezed state; by that we mean that \emph{each normal mode} is described by a wavefunction of the form of equation \eqref{eq:squeezed}. It follows that, at any given time, the state of the two-oscillator system can be written as
\begin{equation}
\begin{split}
\Psi \left( x_+, x_- \right) = \left( \frac{\re \left(w_+\right) \re \left(w_-\right) }{\pi^2} \right)^{\frac{1}{4}} \exp \bigg[ - \frac{1}{2} \left(w_+ \left(x_+-x_{0+}\right)^2 + w_- \left(x_-- x_{0-}\right)^2 \right) \\
+ i p_{0+} \left( x_+ - x_{0+} \right)+i p_{0-} \left( x_- - x_{0-} \right)- i \varphi_{\mathrm{s}+} - i \varphi_{\mathrm{s}-} \bigg] .
\end{split}
\end{equation}
Obviously, we demand that $\re \left(w_+\right) > 0$ and $\re \left(w_-\right) > 0$, so that the wavefunction is normalizable. The system's density matrix is given by
\begin{multline}
\rho \left( x_+, x_- ; x_+^\prime , x_-^\prime \right) = \left( \frac{\re\left(w_+\right) \re\left( w_-\right)}{\pi^2} \right)^{\frac{1}{2}} \\
\times \exp \bigg[ - \frac{1}{2} \left( w_+\left(x_+ -x_{0+}\right)^2 + w_-\left(x_- -x_{0-}\right)^2 + w_+^* \left(x_+^\prime -x_{0+}\right)^2 + w_-^* \left(x_-^\prime -x_{0-}\right)^2 \right) \\
+i p_{0+} \left( x_+ - x_+^\prime \right)+i p_{0-} \left( x_- - x_-^\prime \right)\bigg] .
\end{multline}

We would like to trace out the second oscillator and find the reduced density matrix of the first one. In order to do so, we need to express the density matrix in terms of the original coordinates $x_1$ and $x_2$. It is convenient to define
\begin{equation}
x_{0\pm} = \frac{1}{\sqrt{2}} \left( x_{01} \pm x_{02} \right),\qquad p_{0\pm} = \frac{1}{\sqrt{2}} \left( p_{01} \pm p_{02} \right)
\end{equation}
in a similar manner to \eqref{eq:normal_coords}.

The reduced density matrix assumes the form
\begin{multline}
\rho \left( x_1, x_2 ; x_1^\prime , x_2^\prime \right) = \left( \frac{\re\left( w_+\right) \re\left( w_-\right)}{\pi^2} \right)^{\frac{1}{2}} \\
\times \exp \bigg[ - \frac{1}{4} \big( \left( w_+ + w_- \right) \left( y_1^2 + y_2^2 \right) + \left( w_+ + w_- \right)^* \left( y_1^{\prime 2} + y_2^{\prime 2} \right) \\
+ 2 \left( w_+ - w_- \right) y_1 y_2 + 2 \left( w_+ - w_- \right)^* y_1^\prime y_2^\prime \big)+i p_{01}\left(y_1-y_1^\prime\right)+i p_{02}\left(y_2-y_2^\prime\right) \bigg],
\end{multline}
where we defined
\begin{equation}
y_i \equiv x_i - x_{0i} , \quad y^\prime_i \equiv x^\prime_i - x_{0i} , \quad i = 1 , 2 .
\end{equation}

The reduced density matrix for the first oscillator, $\rho_1 \left( x_1 ; x_1^\prime \right) = \int dx_2 \rho \left( x_1, x_2 ; x_1^\prime , x_2 \right)$, can be easily calculated via Gaussian integrals. We only need to complete the square for the coordinate that is integrated. After some simple algebra we find
\begin{equation}
\rho_1 \left( x_1 ; x_1^\prime \right) = \left( \frac{\re \left(\gamma\right) - \beta}{\pi} \right)^{\frac{1}{2}} \exp \left[ - \frac{1}{2} \left( \gamma y_1^2 + \gamma^* y_1^{\prime 2} \right) + \beta y_1 y_1^\prime +i p_{01}\left(y_1-y_1^\prime\right)\right] ,
\label{eq:two_osc_reduced}
\end{equation}
where
\begin{align}
\gamma &= \frac{4 w_+ w_- + \left| w_+ + w_-\right|^2}{4 \re \left( w_+ + w_- \right)} , \label{eq:two_osc_gamma} \\
\beta &= \frac{\left| w_+ - w_-\right|^2}{4 \re \left( w_+ + w_- \right)} . \label{eq:two_osc_beta}
\end{align}
The reduced density matrix is appropriately normalized, i.e. $\Tr \rho_1 = 1$. Notice that it does not depend on the parameter $p_{02}$ at all.
 
It is easy to show that
\begin{align}
\re \left(\gamma\right) - \beta &= \frac{ 2\re \left( w_+\right) \re \left(w_- \right)}{\re \left( w_+ + w_- \right)} ,
\label{eq:gamma-beta} \\
\re \left(\gamma\right) + \beta &= \frac{\left\vert w_+ + w_-^* \right\vert^2}{2\re \left( w_+ + w_- \right)}.
\label{eq:gamma+beta}
\end{align}
As a result, for any $w_+$ and $w_-$, as long as their real parts are positive, which is required for the normalizability of the wavefunction of the overall system, $\re \left(\gamma\right)$ is always positive and larger than $\vert \beta\vert$.

\subsection{The Spectrum of the Reduced Density Matrix}
\label{subsec:two_oscillators_spectrum}

The coefficient $\beta$ is real, similarly to the ground or coherent state case. This is enforced by the fact that the reduced density matrix is Hermitian, i.e. $\rho_1 \left( x_1 ; x_1^\prime \right) = \rho_1^*\left( x_1^\prime ; x_1 \right)$. Even though this property of the reduced density matrix is sufficient in order to set the parameter $\beta$ real in the case of a reduced system with a single degree of freedom, it is not sufficient in the general case. In other words, this is a special property of systems where all oscillators \emph{but one} are traced out.

However, the simple example of the two oscillators exposes an interesting difference to the ground state case or the more general coherent state case: The coefficient $\gamma$ is complex, namely
\begin{equation}
\gamma = \frac{4 \re \left( w_+ w_- \right) + \left| w_+ + w_-\right|^2}{4 \re \left( w_+ + w_- \right)} + i \frac{4 \im \left( w_+ w_- \right)}{4 \re \left( w_+ + w_- \right)} .
\end{equation}
This is a direct consequence of the fact that the coefficient of the quadratic term $w$ in the exponent of a squeezed state is complex. 

Nevertheless, the eigenvalues of the reduced density matrix \emph{do not depend} on the imaginary part of the coefficient $\gamma$. Moreover, they do not depend on the parameter $p_{01}$. We may write the reduced density matrix $\rho_1$ as
\begin{equation}
\rho_1 \left( x_1 ; x_1^\prime \right) = \tilde{\rho}_1 \left( x_1 ; x_1^\prime \right) \exp \left[ - \frac{i}{2} \im \left(\gamma\right) \left( y_1^2 - y_1^{\prime 2} \right) +i p_{01}\left(y_1-y_1^\prime\right) \right] ,
\end{equation}
where
\begin{equation}
\tilde{\rho}_1 \left( x_1 ; x_1^\prime \right) = \left( \frac{\re \left(\gamma\right) - \beta}{\pi} \right)^{\frac{1}{2}} \exp \left[ - \frac{1}{2} \re \left(\gamma\right) \left( y_1^2 + y_1^{\prime 2} \right) + \beta y_1 y_1^\prime \right] ,
\label{eq:reduced_density_matrix}
\end{equation}
i.e. $\tilde{\rho}_1$ is the same as $\rho_1$, where we have set the imaginary part of the coefficient $\gamma$ and the coefficient $p_{01}$ equal to zero by hand. Let $\tilde{f}\left( x_1 \right)$ be an eigenstate of $\tilde{\rho}_1$ with eigenvalue $\lambda$, namely
\begin{equation}
\int dx_1^\prime \tilde{\rho}_1 \left( x_1 ; x_1^\prime \right) \tilde{f} \left( x_1^\prime \right) = \lambda \tilde{f} \left( x_1 \right) .
\end{equation}
Then, the function $f \left( x_1 \right) = \exp \left( - i \im \left(\gamma\right) y_1^2 / 2 +i p_{01}x_1\right) \tilde{f} \left( x_1 \right)$ is an eigenfunction of the reduced density matrix $\rho_1$ with the \emph{same} eigenvalue. Therefore, the spectrum of the reduced density matrix does not depend on the imaginary part of $\gamma$ and the parameter $p_{01}$. In order to specify its spectrum, it suffices to specify the spectrum of the matrix $\tilde{\rho}_1$.

The matrix $\tilde{\rho}_1 \left( x_1 ; x_1^\prime \right)$ is of the same form as the reduced density matrix in the case of the ground \cite{srednicki} or coherent states \cite{Katsinis:2022fxu}. It is well known that its normalized eigenstates are
\begin{equation} 
\tilde{f}_n \left( x \right) = \frac{1}{\sqrt{2^n n!}} \left( \frac{\alpha}{\pi} \right)^{1/4} H_n \left( \sqrt{\alpha} \left(x - x_{01} \right) \right) e^{- \frac{1}{2} \alpha \left(x - x_{01} \right)^2} ,
\label{eq:two_eigenfunctions_tilde}
\end{equation}
where 
\begin{equation}
\alpha := \sqrt{\re \left( \gamma \right)^2 - \beta ^2}
\end{equation}
and $H_n$ is the Hermite polynomial of order $n$. Notice that the parameter $\alpha$ is always real, since $\re \left( \gamma \right) > \left| \beta \right|$. The corresponding eigenvalues are
\begin{equation}
p_n = \left( 1 - \xi \right) \xi^n ,
\label{eq:two_eigenvalues}
\end{equation}
where
\begin{equation}
\xi := \frac{\beta}{\re \left(\gamma\right) + \alpha} .
\end{equation}
The eigenvalues are properly normalized, since obviously $\sum\limits_{n = 0}^\infty  {{p_n}}  = 1$. It directly follows that the normalized eigenstates of the reduced density matrix $\rho_1 \left( x_1 ; x_1^\prime \right)$ are
\begin{equation}
f_n \left( x \right) = \frac{1}{\sqrt{2^n n!}} \left( \frac{\alpha}{\pi} \right)^{1/4} H_n \left( \sqrt{\alpha} \left(x - x_{01} \right) \right) e^{- \frac{1}{2} \left( \alpha + i \im \left(\gamma\right) \right) \left(x-x_{01} \right)^2+i p_{01}\left(x - x_{01} \right)} ,
\label{eq:two_eigenfunctions}
\end{equation}
with corresponding eigenvalues given by equation \eqref{eq:two_eigenvalues}. The above imply that the reduced density matrix assumes the form
\begin{equation}
\rho_1 \left( x_1 ; x_1^\prime \right) = \left( 1 - \xi \right) \sum_{n=0}^\infty \xi^n f_n(x_1) f_n(x_1^\prime)^* .
\label{eq:density_as_f_eigenstates}
\end{equation}

As in the case of the ground state, the entanglement entropy is given by
\begin{equation}\label{eq:entanglment_formula}
S = - \ln \left( {1 - \xi } \right) - \frac{\xi }{{1 - \xi }}\ln \xi .
\end{equation}

The parameter $\xi$ and the spectrum of the reduced density matrix depend on time, unlike the case of the ground state or a coherent state. This implies that the time evolution of the reduced density matrix is non-unitary.

\subsection{Squeezing and Entanglement}
\label{subsec:squeezing_entanglement}
We have obtained the spectrum of the reduced density matrix, as well as the entanglement entropy, for a system of two degrees of freedom in a squeezed state. This state is in some sense ``less classical'' than a coherent state (see discussion in section \ref{sec:squeezed_review}), so it is natural to ask how the entanglement entropy is altered by squeezing. For the case of two simple coupled harmonic oscillators, we study some 
representative examples. The details are presented in appendix \ref{sec:app_squeezing_entanglement}, while here we only summarize the conclusions.

\begin{figure}[ht]
\centering
\begin{picture}(52,34)
\put(4.5,0){\includegraphics[angle=0,width=0.45\textwidth]{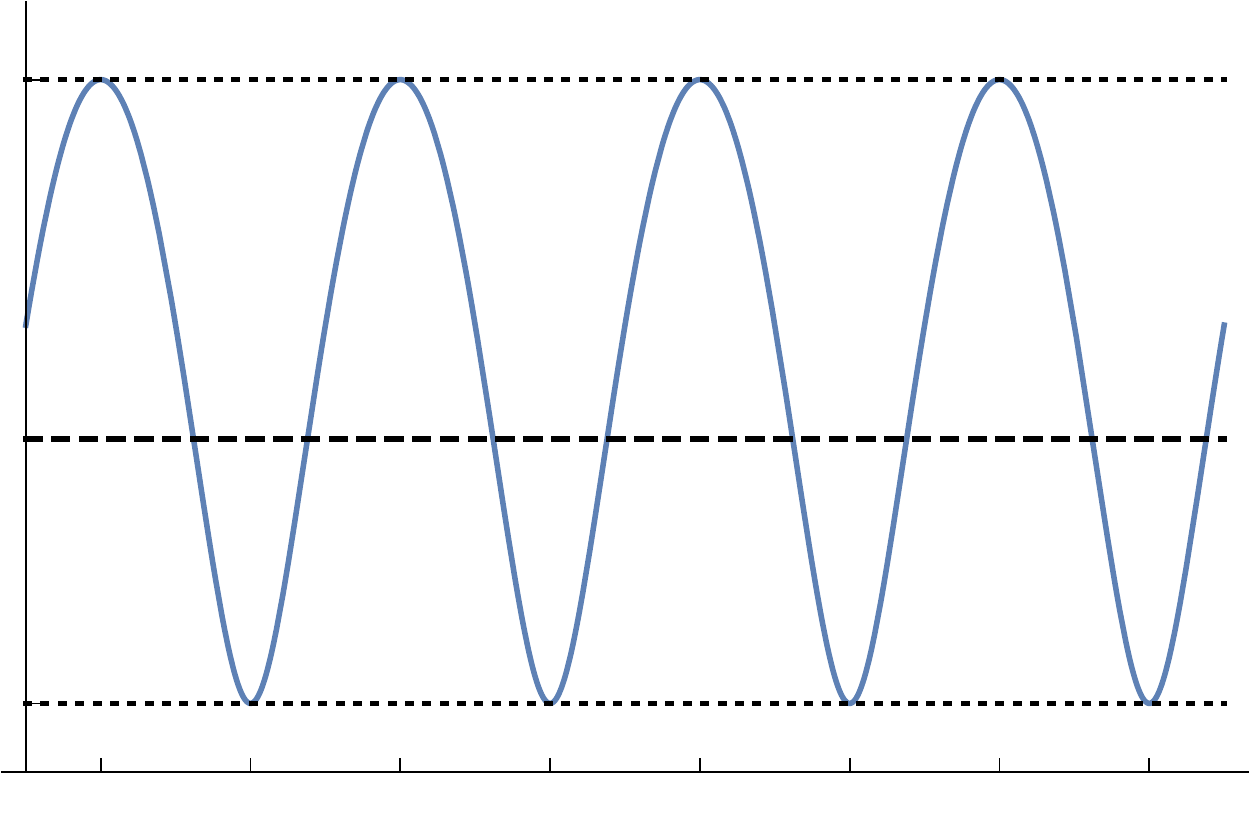}}
\put(50.25,1.25){$t$}
\put(0.5,3.75){$S_{\min}$}
\put(2.5,13.5){$S_0$}
\put(0.25,26.25){$S_{\max}$}
\put(4,30.5){$S_{\mathrm{EE}}$}
\put(5,-0.25){$T_+ / 8$}
\put(15,-0.25){$5 T_+ / 8$}
\put(26,-0.25){$9 T_+ / 8$}
\put(36,-0.25){$13 T_+ / 8$}
\end{picture}
\caption{The entanglement entropy as a function of time when only the symmetric mode is squeezed. $S_0$ is the entanglement entropy at the ground state of the system.}
\label{fig:two_osc_S_t}
\end{figure}
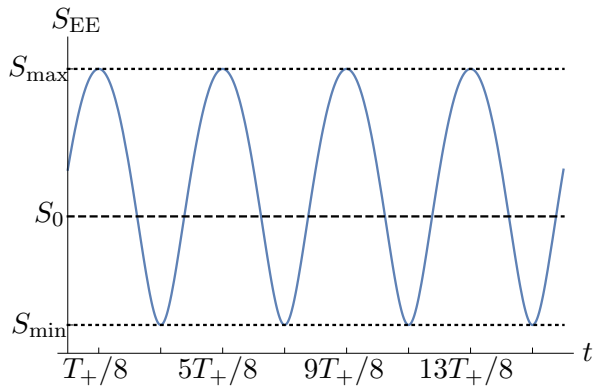

When the overall system is described by a squeezed state, the entanglement entropy depends on time. When only one mode is squeezed, while the other lies in the ground or a coherent state, this dependence results in a periodic oscillation between a maximal and a minimal value, as shown in figure \ref{fig:two_osc_S_t}.

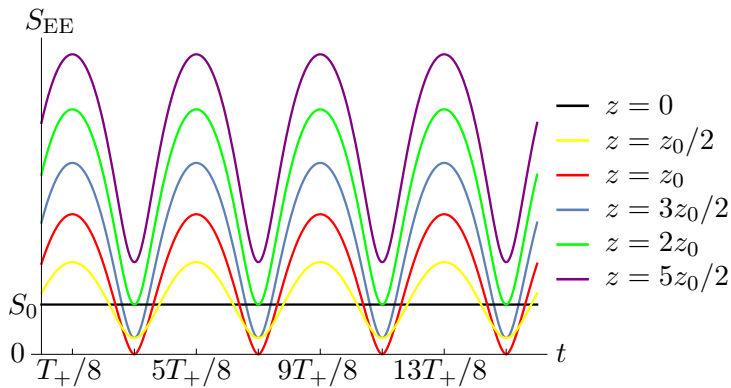
\begin{figure}[ht]
\centering
\begin{picture}(70,34)
\put(2.5,0){\includegraphics[angle=0,width=0.45\textwidth]{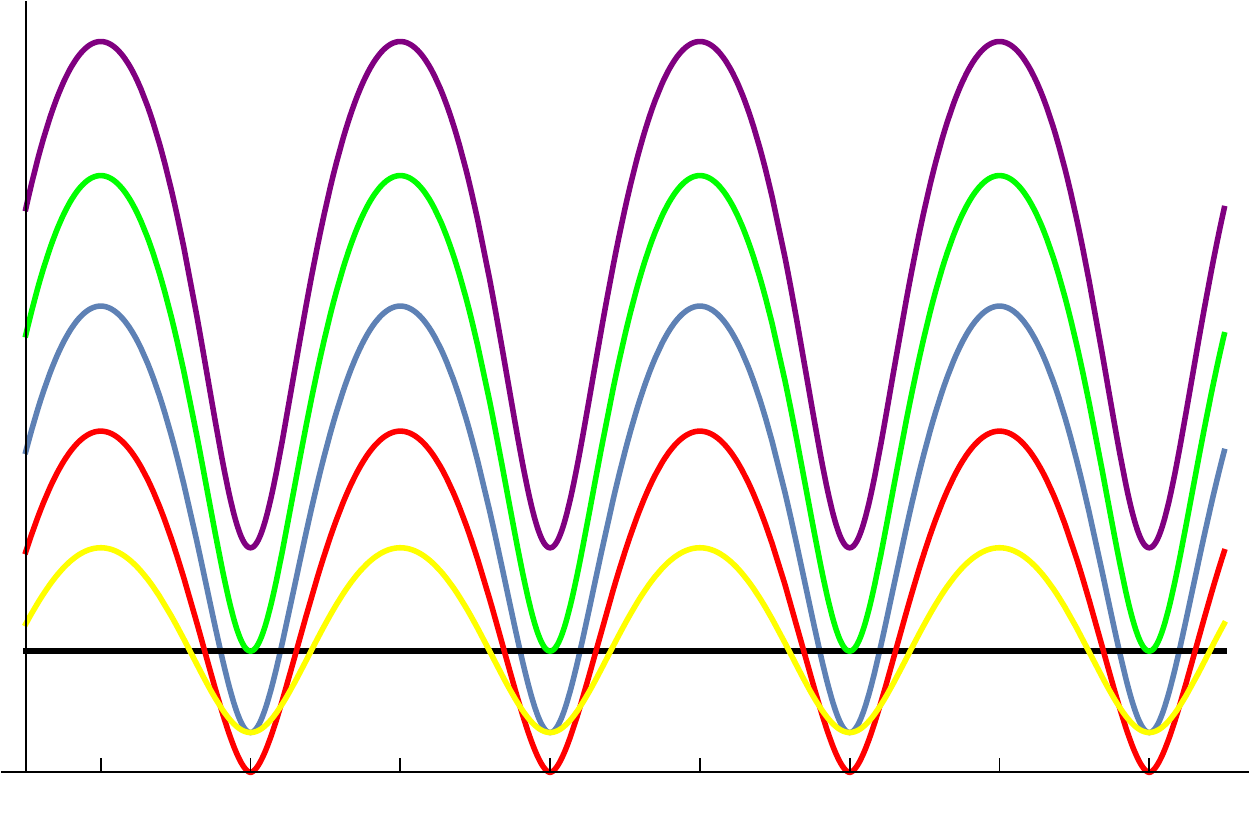}}
\put(47.5,6.5){\includegraphics[angle=0,width=0.055\textwidth]{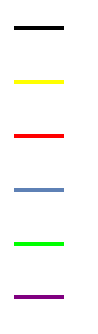}}
\put(48.25,1.25){$t$}
\put(2,30.5){$S_{\mathrm{EE}}$}
\put(0.5,5.5){$S_0$}
\put(0.75,1.25){$0$}
\put(3,-0.25){$T_+ / 8$}
\put(13,-0.25){$5 T_+ / 8$}
\put(24,-0.25){$9 T_+ / 8$}
\put(34,-0.25){$13 T_+ / 8$}
\put(52.5,8){$z = 5 z_0 / 2$}
\put(52.5,11){$z = 2 z_0$}
\put(52.5,14){$z = 3 z_0 / 2$}
\put(52.5,17){$z = z_0$}
\put(52.5,20){$z = z_0 / 2$}
\put(52.5,23){$z = 0$}
\end{picture}
\caption{The entanglement entropy as a function of time for various values of the squeezing parameter $z$. $S_0$ is the entanglement entropy at the ground state of the system.}
\label{fig:two_osc_S_t_z}
\end{figure}

The contribution of the squeezed mode to entanglement grows as the squeezing parameter increases. However, this contribution may act constructively or destructively on 
the contribution of the mode that lies in its ground state. As a result, the maximal value of entanglement entropy always increases as the squeezing parameter increases, whereas the minimal value in not monotonous with the squeezing parameter. As the latter increases, the minimal value decreases, up to a critical value of the squeezing parameter $z_0 = \ln\frac{\omega_-}{\omega_+}$, where it vanishes. Further increase of the squeezing parameter results in an increase of the minimal value of the entanglement entropy. These features are depicted in figure \ref{fig:two_osc_S_t_z}.
The contribution of the squeezed mode is constructive at the instants when the squeezed state is a minimal uncertainty state with maximal position uncertainty, and destructive at the instants when it is a minimal uncertainty state with minimal position uncertainty.

\begin{figure}[ht]
\centering
\begin{picture}(50,34)
\put(2.5,0){\includegraphics[angle=0,width=0.45\textwidth]{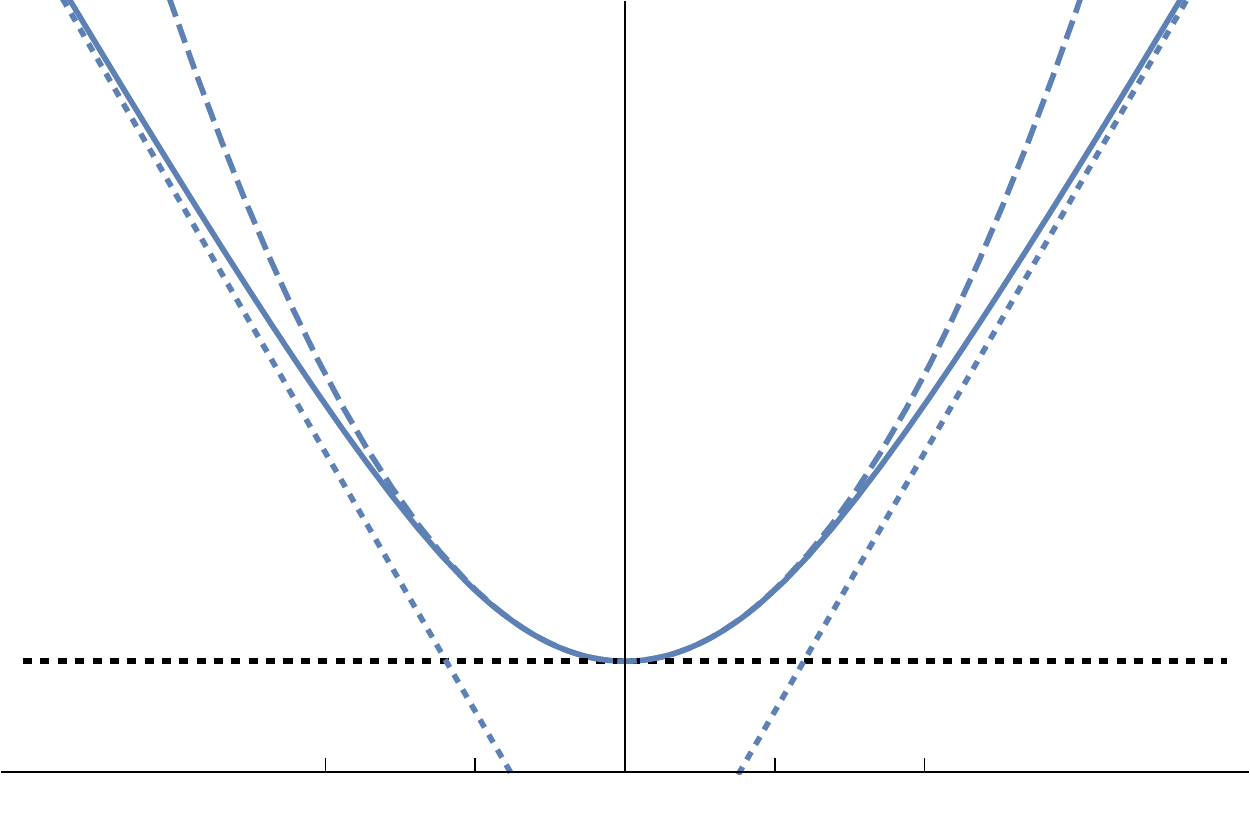}}
\put(48.25,1.25){$z$}
\put(0.5,5.5){$S_0$}
\put(24,30.5){$\bar{S}$}
\put(10.5,-0.25){$-z_{\mathrm{vac}}$}
\put(17,-0.25){$-z_0$}
\put(29.5,-0.25){$z_0$}
\put(34,-0.25){$z_{\mathrm{vac}}$}
\end{picture}
\caption{The mean entanglement entropy as a function of the squeezing parameter $z$. The dashed line shows the approximation of the mean entanglement entropy for small squeezing, whereas the dotted line the approximation for large squeezing, both provided by equation \eqref{eq:two_mean_SEE_asymtotics}.}
\label{fig:two_osc_mean_S}
\end{figure}
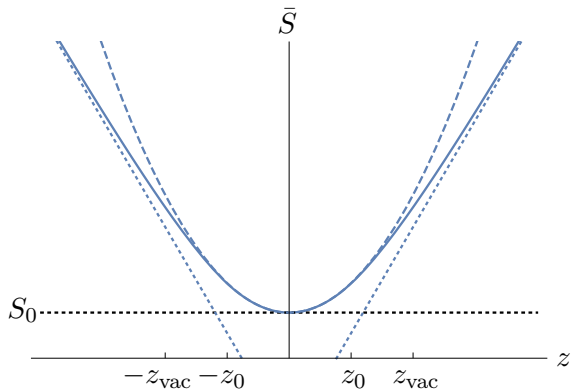

The dependence of the minimal and maximal entanglement entropy on the squeezing parameter is such that the mean entanglement entropy always increases with the squeezing parameter. After some tedious algebra that is included in appendix \ref{subsubsec:squeeze_one}, it turns out that the mean entanglement entropy is given by equation \eqref{eq:average_S}. For small squeezing the mean entanglement entropy is quadratic in the squeezing parameter, whereas for large squeezing it becomes a linear function: 
\begin{equation}
\bar{S}=\begin{cases}
S_0 - \frac{z^2}{16} \left( 1 + \frac{1 + 4 \xi_0 + \xi_0^2}{1 - \xi_0^2} \ln \xi_0 \right) , \quad & z \ll 1 , \\
\frac{z}{2} + \ln \left( \sqrt{\frac{\omega_+}{\omega_-}} + \sqrt{\frac{\omega_-}{\omega_+}} \right) + 1 - 3 \ln2 , \quad & z \gg 1 ,
\end{cases}
\label{eq:two_mean_SEE_asymtotics}
\end{equation}
where $\xi_0$ and $S_0$ are the parameter $\xi$ and the entanglement entropy when the system lies in its ground state, respectively. The mean entanglement entropy as a function of the squeezing parameter is depicted in figure \ref{fig:two_osc_mean_S}.

\begin{figure}[ht]
\centering
\begin{picture}(50,47)
\put(6,2.5){\includegraphics[angle=0,width=0.425\textwidth]{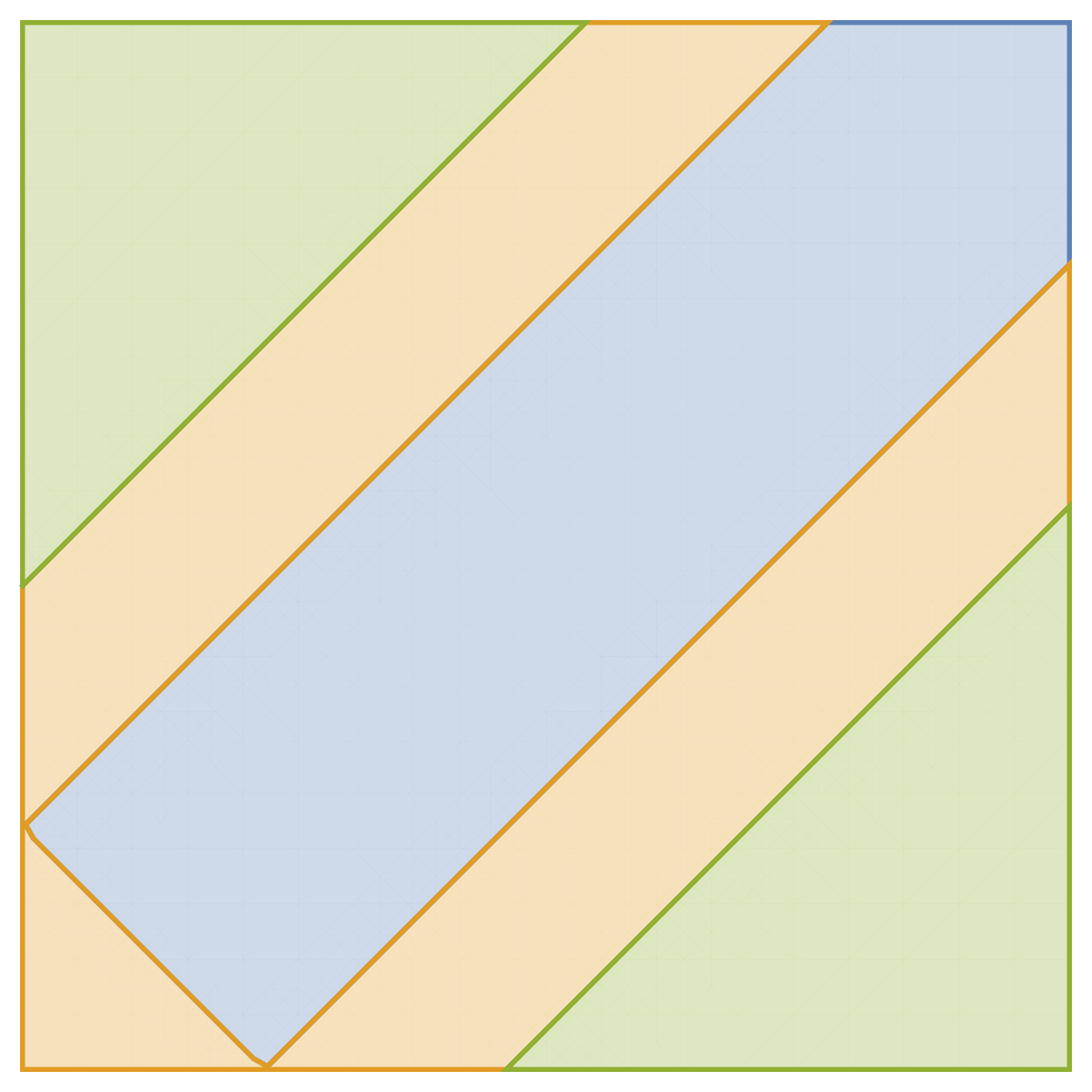}}
\put(48,3){$z_+$}
\put(1,12.25){$\ln \frac{\omega_-}{\omega_+}$}
\put(-0.75,21.5){$2 \ln \frac{\omega_-}{\omega_+}$}
\put(6,44.75){$z_-$}
\put(13.5,0.75){$\ln \frac{\omega_-}{\omega_+}$}
\put(22,0.75){$2 \ln \frac{\omega_-}{\omega_+}$}
\put(5.25,2.75){$0$}
\put(6.25,1.25){$0$}
\put(25,22){\rotatebox{45}{$S_{\min} = 0$}}
\put(28.5,11){\rotatebox{45}{$0 < S_{\min} < S_0$}}
\put(36.5,6.5){\rotatebox{45}{$S_{\min} > S_0$}}
\end{picture}
\caption{The minimal entanglement entropy relatively to $S_0$ as a function of the squeezing parameters.}
\label{fig:two_osc_regions}
\end{figure}
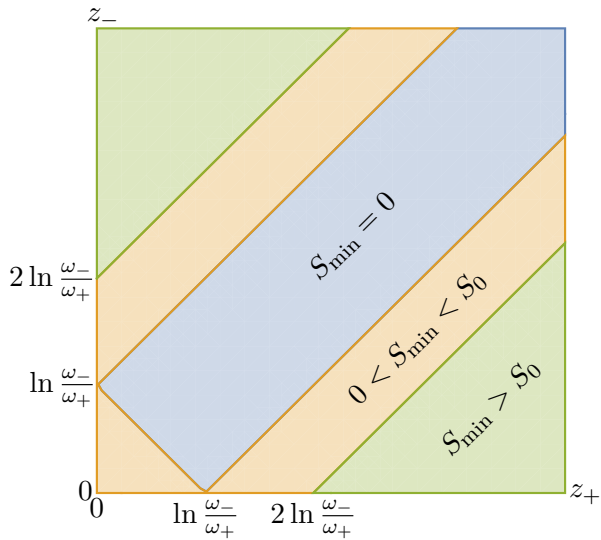

This wave-like addition of the contributions of the two modes to the entanglement entropy persists when both modes are squeezed. As a result, even when both squeezing parameters are large, the minimal entanglement entropy may be small or even vanishing, when the difference of the two squeezing parameters is small. This is depicted in figure \ref{fig:two_osc_regions}.

On general grounds, the contribution to entanglement by the two modes is the maximal possible at instants when both modes are minimal uncertainty states, but one of those has maximal position uncertainty and the other has minimal position uncertainty. When they are both minimal uncertainty states and have both either maximal or minimal position uncertainty, squeezing acts competitively and their contribution to entanglement is the minimal possible. When both modes are squeezed, the dependence of the entanglement entropy on time is in general not periodic, as the ratio of the eigenfrequencies of the two modes may be irrational.
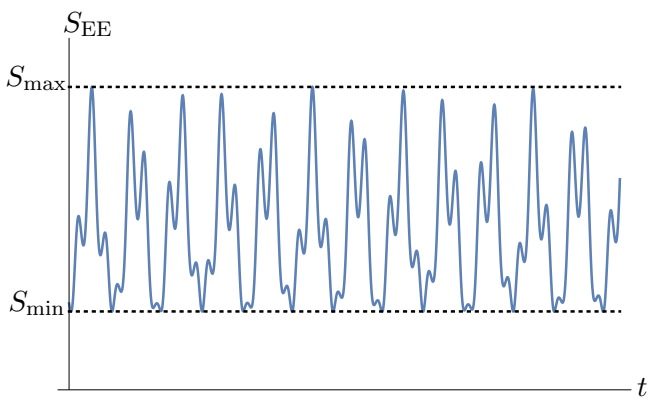
\begin{figure}[ht]
\centering
\begin{picture}(67,36)
\put(4,2.5){\includegraphics[angle=0,width=0.5\textwidth]{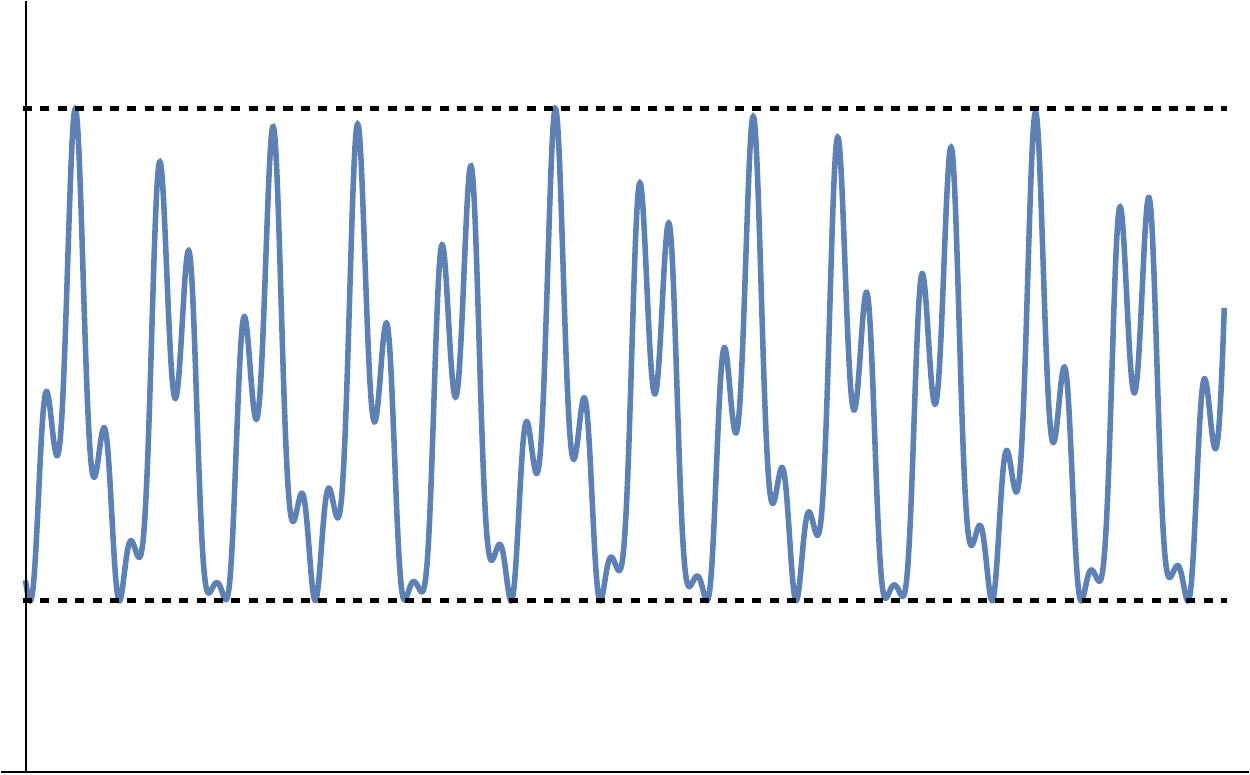}}
\put(54.5,2){$t$}
\put(-0.25,9.5){$S_{\min}$}
\put(-0.5,29.25){$S_{\max}$}
\put(4.5,34){$S_{\mathrm{EE}}$}
\end{picture}
\caption{The entanglement entropy as a function of time for  $\omega_+ = 1$, $\omega_- = \sqrt{2}$, $z_+ = 1$ and $z_- = 5 / 4$.}
\label{fig:two_osc_time}
\end{figure}
However, the constructive and destructive addition of the contributions of the two modes sets an upper and a lower bound for the entanglement entropy, as shown in figure \ref{fig:two_osc_time}.

\begin{figure}[ht]
\centering
\begin{picture}(67,30)
\put(4,2.5){\includegraphics[angle=0,width=0.6\textwidth]{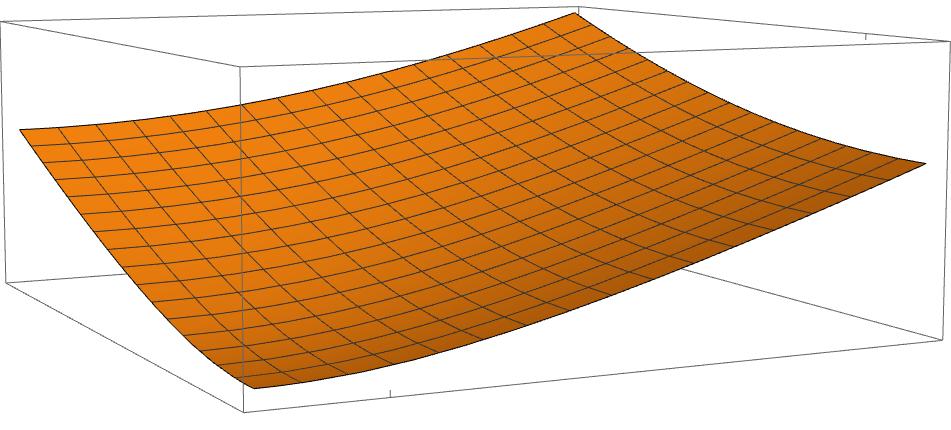}}
\put(63.75,7.5){$z_+$}
\put(1.75,12){$z_-$}
\put(18.25,25.75){$\bar{S}$}
\put(25.5,1.75){$\ln \frac{\omega_-}{\omega_+}$}
\end{picture}
\caption{The mean entanglement entropy as a function of the squeezing parameters for $\omega_+ = 1$, $\omega_- = \sqrt{2}$.}
\label{fig:2_modes_S_mean}
\end{figure}
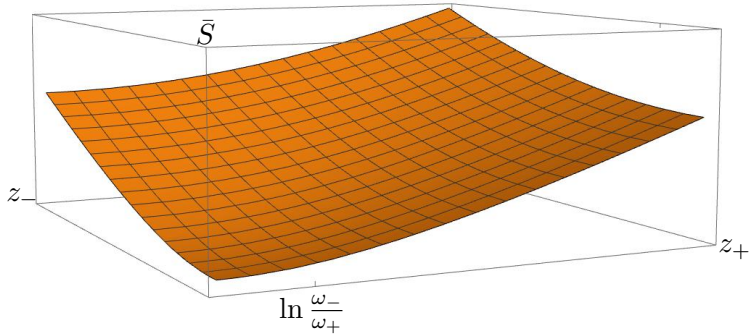

Unlike the case where a single mode is squeezed, it is not possible to obtain an analytic formula for the mean value of entanglement entropy. However, it appears that the latter is an increasing function of both $z_\pm$, as shown in the numerical calculation depicted in figure \ref{fig:2_modes_S_mean}.
Indicatively, the small and large squeezing parameter expansions of the mean entanglement entropy are
\begin{equation}
\bar{S}=\begin{cases}
S_0 - \frac{z_+^2 + z_-^2}{16} \left( 1 + \frac{1 + 4 \xi_0 + \xi_0^2}{1 - \xi_0^2} \ln \xi_0 \right) +\mathcal{O} \left( z^3 \right), \quad & z_\pm \ll 1 , \\
\frac{z_+ + z_-}{2} + \mathcal{O} \left( z^0 \right) , \quad & z_\pm \gg 1 ,
\end{cases}
\label{eq:2_mode_mean_SEE_asymtotics}
\end{equation}
in line with the above statement.

\section{General Harmonic System at a Squeezed State}
\label{sec:many_oscillators}

Having studied the system of two coupled harmonic oscillators in the previous section, we can now proceed to study a more general harmonic system with an arbitrary number of degrees of freedom. We consider a system of $N$ coupled quantum harmonic oscillators described by the Hamiltonian
\begin{equation}
H = \frac{1}{2} \sum\limits_{i = 1}^N {p_i^2} + \frac{1}{2} \sum\limits_{i,j = 1}^N {x_i K_{ij} x_j} = \frac{1}{2} \mathbf{p}^T \mathbf{p} + \frac{1}{2} \mathbf{x}^T K \mathbf{x} ,
\label{eq:many_Hamiltonian}
\end{equation}
where we use the vector notation
\begin{equation}
\mathbf{x} = \left( \begin{array}{c} x_1 \\ x_2 \\ \vdots \\ x_N \end{array} \right) , \quad \mathbf{p} = \left( \begin{array}{c} p_1 \\ p_2 \\ \vdots \\ p_N \end{array} \right) .
\end{equation}
Similarly to section \ref{sec:two_oscillators} and without loss of generality, we have assumed that all oscillators have unit mass. The matrix $K$ is symmetric and positive definite, so that it describes an oscillatory system with $N$ degrees of freedom around a stable equilibrium position.

There is an orthogonal transformation $O$, relating the coordinates $x_i$ to the normal coordinates $\tilde{x}_i$, which diagonalizes the matrix $K$, reducing the system to a set of decoupled harmonic oscillators, one for each normal mode\footnote{Throughout this work, the tilded symbols refer to quantities related to the normal coordinates.}. In other words,
\begin{equation}
H = \frac{1}{2} \sum\limits_{j = 1}^N {\tilde{p}_j^2} + \frac{1}{2} \sum\limits_{j = 1}^N {\omega_j^2 \tilde{x}_j^2} = \frac{1}{2} \mathbf{\tilde{p}}^T \mathbf{\tilde{p}} + \frac{1}{2} \mathbf{\tilde{x}}^T \tilde{K} \mathbf{\tilde{x}} ,
\end{equation}
where
\begin{equation}
\mathbf{\tilde{x}} = O \mathbf{x} , \quad \mathbf{\tilde{p}} = O \mathbf{p} .
\end{equation}
The diagonal matrix $\tilde{K}$ contains the squares of the eigenfrequencies of the normal modes
\begin{equation}
\tilde{K}_{ij} = \omega_i^2 \delta _{ij}.
\end{equation}
It is obviously related to the initial matrix $K$ as
\begin{equation}
K = O^T \tilde{K} O .
\end{equation}

We consider states of the system where all normal modes lie in a squeezed state
\begin{equation}
\begin{split}
\Psi \left( \mathbf{\tilde{x}} \right) &= \left[ \prod_{j = 1}^N \left( \frac{\re\left( w_j\right)}{\pi} \right)^{\frac{1}{4}} \right] \exp \left[ \sum_{j = 1}^N \left( - \frac{1}{2} w_j \left(\tilde{x}_j - \tilde{x}_{0j} \right)^2 + i \tilde{p}_{0j} \left(\tilde{x}_j - \tilde{x}_{0j} \right)- i \varphi_{\mathrm{s}j} \right) \right] \\
&= \left( \frac{\det \re \left( \tilde{W} \right)}{\pi^N} \right)^{\frac{1}{4}} \exp \left[ - \frac{1}{2} \left( \mathbf{\tilde{x}} - \mathbf{\tilde{x}}_0 \right)^T \tilde{W} \left(\mathbf{\tilde{x}} - \mathbf{\tilde{x}}_0 \right) + i \mathbf{\tilde{p}}_{0}^T \left( \mathbf{\tilde{x}} - \mathbf{\tilde{x}}_0 \right) - i \sum_{j = 1}^N \varphi_{\mathrm{s}j} \right] ,
\end{split}
\label{eq:many_overall_state_modes}
\end{equation}
where $\tilde{W}$ is the diagonal matrix whose diagonal elements are the complex values $w_i$, i.e.
\begin{equation}
\tilde{W}_{ij} = w_i \delta_{ij} ,
\end{equation}
with $w_i$ given by the application of formula \eqref{eq:omega_squeezed} for each normal mode.

We define as subsystem $1$ the set of $n$ oscillators described by the coordinates $x_j$, where $j \leq n$. The $N - n$ oscillators described by coordinates $x_j$, where $j > n$, comprise the complementary subsystem, which we call subsystem 2. We would like to trace out subsystem 1 in order to find the reduced density matrix for subsystem 2 and the corresponding entanglement entropy.

The bulk of this section is quite technical. For this reason we would like to provide first the reader with the summary of the results and point out the basic differences with respect to the case of the ground state, which has been extensively studied in the literature.

\subsection{Summary}
The reduced density matrix, which describes a subsystem of the overall harmonic system, turns out to be of the form
\begin{multline}
\rho_2 \left( \mathbf{x}_2 ; \mathbf{x}_2^\prime \right) = \left( \frac{\det \re \left( \gamma - \beta \right)}{\pi^{N - n}} \right)^{\frac{1}{2}} \exp \bigg[ - \frac{1}{2} \left( \mathbf{y}_2^T \gamma \mathbf{y}_2 + \mathbf{y}_2^{\prime T} \gamma^* \mathbf{y}_2^\prime \right) \\
+ \mathbf{y}_2^{\prime T} \beta \mathbf{y}_2 + i \mathbf{p}_{02}^T \left( \mathbf{y}_2 - \mathbf{y}_2^\prime \right) \bigg] ,
\end{multline}
where $\gamma$ is a complex symmetric matrix, $\beta$ is a Hermitian matrix and $\mathbf{y}_2=\mathbf{x}_2-\mathbf{x}_{02}$. The vector $\mathbf{x}_2$ contains the coordinates of the degrees of freedom that have not been traced out, while the vectors $\mathbf{x}_{02}$ and $\mathbf{p}_{02}$ contain the corresponding parameters $x_{0i}$ and $p_{0i}$.

The imaginary part of the matrix $\gamma$ does not affect the eigenvalues of the reduced density matrix and can be set to zero by hand. The same holds for the parameters in the vectors $\mathbf{x}_{02}$ and $\mathbf{p}_{02}$. Therefore, we are left with the problem of the specification of the eigenvalues of a matrix of the form
\begin{equation}
\tilde{\rho}_2 \left( \mathbf{x}_2 ; \mathbf{x}_2^\prime \right) = \left( \frac{\det \re \left( \gamma - \beta \right)}{\pi^{N - n}} \right)^{\frac{1}{2}} \exp \left[ - \frac{1}{2} \left( \mathbf{x}_2^T \gamma \mathbf{x}_2 + \mathbf{x}_2^{\prime T} \gamma \mathbf{x}_2^\prime \right) + \mathbf{x}_2^{\prime T} \beta \mathbf{x}_2 \right] ,
\end{equation}
where $\gamma$ is a real symmetric matrix and $\beta$ is a Hermitian matrix. This is very similar to the reduced density matrix in the case that the overall harmonic system lies in its ground state. However, there is one important difference: for the ground state, the matrix $\beta$ is \emph{not just Hermitian, but real and symmetric}.

In the case of the ground state, the fact that both matrices $\gamma$ and $\beta$ are real and symmetric allows the specification of the spectrum of the reduced density matrix in a trivial manner. One has to perform three coordinate transformations. The first one is an orthogonal transformation that diagonalizes the matrix $\gamma$. The second is a coordinate rescaling that sets the matrix $\gamma$ equal to the identity matrix. The last one is an orthogonal transformation that diagonalizes the matrix $\beta$. After these transformations, 
the density matrix is written as the tensor product of $N - n$ matrices of the form
\begin{equation}
\rho \left( \hat{x}_i ; \hat{x}_i^\prime \right) = \left( \frac{1 - \hat{\beta}_i}{\pi} \right)^{\frac{1}{2}} \exp \left[ - \frac{1}{2} \left( \hat{x}_i^2 + \hat{x}_i^{\prime 2} \right) + \hat{\beta}_i \hat{x}_i \hat{x}_i^\prime \right] ,
\end{equation}
describing \emph{one} degree of freedom each. The coordinates $\hat{x}_i$ are the coordinates describing the reduced system after the three linear transformations that we performed above, i.e. real linear combinations of the original coordinates. Namely, they equal $\hat{x}_i = \mathbf{v}_i^T \mathbf{x}_2$, where $\mathbf{v}_i$ are the normalized eigenvectors of the real symmetric matrix
\begin{equation}
\hat{\beta} = \gamma^{-\frac{1}{2}} \beta \gamma^{-\frac{1}{2}} .
\label{eq:beta_hat_def}
\end{equation}
The parameters $\hat{\beta}_i$ are the corresponding eigenvalues of the matrix $\hat{\beta}$.

The above density matrix is of the form of equation \eqref{eq:reduced_density_matrix}, i.e. of the form that we met in the simple case of two coupled harmonic oscillators. We know that its eigenfunctions are given by equation \eqref{eq:two_eigenfunctions} and its eigenvalues by equation \eqref{eq:two_eigenvalues}. The eigenfunctions are identical to the eigenstates of an effective simple harmonic oscillator with eigenfrequency equal to $\alpha_i = \sqrt{1 - \hat{\beta}_i^2}$. This fact, combined with the form of the eigenvalues, implies that the above reduced density matrix is identical to a thermal density matrix describing the effective harmonic oscillator at a temperature 
\begin{equation}
e^{-\frac{\alpha_i}{T_i}} = \frac{\hat{\beta}_i}{1 + \alpha_i} \equiv \xi_i .
\label{eq:effective_temperatures}
\end{equation}

Returning to the reduced system, the fact that the reduced density matrix can be factored to matrices of the above form, describing one degree of freedom each, implies that it is identical to the density matrix describing an effective harmonic system in a quasi-thermal state. The coordinates $\hat{x}_i$ are the ``canonical'' coordinates of this harmonic system, whereas the values $\alpha_i$ are the corresponding eigenfrequencies. The eigenstates of the system are trivially given by
\begin{equation}
\tilde{f}_{\{m_1 , m_2 , \ldots , m_{N - n} \}}\left( \mathbf{x}_2 \right) = \prod_{i = 1}^{N - n} \frac{1}{\sqrt{2^{m_i} m_i!}} \left( \frac{\alpha_i}{\pi} \right)^{1/4} H_{m_i} \left( \sqrt{\alpha_i} \left( \hat{x}_i \right) \right) e^{- \frac{1}{2} \alpha_i \left( \hat{x}_i \right)^2} .
\end{equation}
The state is quasi-thermal is the sense that each normal mode is in a thermal state, but has its own temperature. Such a state is not unexpected, considering that the normal modes do not interact. In an obvious manner, the spectrum of the reduced density matrix is of the form
\begin{equation}
p_{\{m_1 , m_2 , \ldots , m_{N - n} \}} = \prod_{i=1}^{N-n} \left( 1 - \xi_i \right) \xi_i^{m_i} .
\label{eq:reduced_ground_eigenvalues}
\end{equation}

In our case, the overall system does not lie in its ground state, but rather in a squeezed state. The matrix $\beta$ is not real; it is Hermitian. We may diagonalize the matrix $\gamma$ via an orthogonal transformation and even rescale the coordinates in order to set $\gamma$ equal to the identity matrix. Nevertheless it is not possible to diagonalize the matrix $\beta$ through another orthogonal transformation.
 
It follows that the reduced density matrix cannot be factored to the tensor product of matrices, each describing a single degree of freedom. However, we can still perform the first two coordinate transformations and express the reduced density matrix in the form\footnote{Throughout this work, the hatted symbols refer to quantities defined in the coordinates of the reduced system where the matrix $\re \left( \gamma \right)$ has been set to the identity matrix.}
\begin{equation}
\tilde{\rho}_2 \left( \mathbf{x}_2 ; \mathbf{x}_2^\prime \right) = \sqrt{\frac{\det \left( I - {\rm Re} \left(\hat{\beta}\right) \right)}{\pi^{N - n}}} \exp \left[ - \frac{1}{2} \left( \mathbf{\hat{x}}_2^T \mathbf{\hat{x}}_2 + \mathbf{\hat{x}}_2^{\prime T} \mathbf{\hat{x}}_2^\prime \right) + \mathbf{\hat{x}}_2^{\prime T} \hat{\beta} \mathbf{\hat{x}}_2 \right] ,
\label{eq:reduced_many_summary}
\end{equation}
where $\hat{\beta} = \re\left( \gamma \right)^{-\frac{1}{2}} \beta\, \re\left( \gamma \right)^{-\frac{1}{2}}$. Rather surprisingly, it turns out that the general properties of the eigenstates and eigenvalues of the reduced density matrix remain the same as when $\beta$ is real and symmetric.

Even though we cannot factor the system, we may search for eigenfunctions of the reduced density matrix that are of similar form to those in the case of the ground state. First, there is a Gaussian ``ground'' eigenstate
\begin{equation}
\Psi_0 \left( \mathbf{x} \right) \sim \exp \left( - \frac{1}{2} \mathbf{x}^T {\cal A} \mathbf{x} \right) ,
\label{eq:eigenstate_ground}
\end{equation}
where the matrix ${\cal A}$ satisfies the quadratic equation
\begin{equation}
{\cal A} = I - \hat{\beta}^T \left( I + {\cal A} \right)^{- 1} \hat{\beta} .
\label{eq:omega_eq}
\end{equation}
This equation has many solutions, but only one gives rise to a normalizable Gaussian eigenstate.

Second, there are $N - n$ ``first excited'' eigenstates
\begin{equation}
\Psi_{1i} \left( \mathbf{x} \right) \sim \mathbf{v}_i^T \mathbf{x} \exp \left( - \frac{1}{2} \mathbf{x}^T {\cal A} \mathbf{x} \right) ,
\label{eq:spectrum_first excited_state_summary}
\end{equation}
where ${\cal A}$ satisfies \eqref{eq:omega_eq} and the vectors $\mathbf{v}_i$ are the right eigenvectors of the matrix
\begin{equation}
\Xi = \hat{\beta}^T \left( I + {\cal A} \right)^{- 1} .
\label{eq:spectrum_Xi_def_summary}
\end{equation}
Let us call $\xi_i$ the eigenvalue of the matrix $\Xi$ that corresponds to the eigenvector $\mathbf{v}_i$. Then, it turns out that the eigenvalue of the eigenstate $\Psi_{1i}$ of the reduced density matrix is $\lambda_0 \xi_i$, where $\lambda_0$ is the eigenvalue of the ``ground'' eigenstate \eqref{eq:eigenstate_ground}. The vectors $\mathbf{v}_i$ are in general complex. The matrix $\Xi$ has no specific symmetry property, nevertheless, its eigenvalues are real.

If the overall system lay in its ground state, we would upgrade the linear combinations of the coordinates $\mathbf{v}_i^T \mathbf{x}$ to Hermite polynomials of those and construct the whole tower of states. In our case, this is not possible. However, testing the function
\begin{equation}
\psi_{\left\{ m_1 , m_2 , \ldots , m_n \right\}} \left( \mathbf{x} \right) \sim \left( \mathbf{v}_1^T \mathbf{x} \right)^{m_1} \left( \mathbf{v}_2^T \mathbf{x} \right)^{m_2} \ldots \left( \mathbf{v}_{N-n}^T \mathbf{x} \right)^{m_{N-n}} \exp \left( - \frac{1}{2} \mathbf{x}^T {\cal A} \mathbf{x} \right) ,
\label{eq:spectrum_general_excited_state_summary}
\end{equation}
it can be shown that, although it is not an eigenstate, there is always a unique way to add terms of \emph{lower} order to the polynomial $\left( \mathbf{v}_1^T \mathbf{x} \right)^{m_1} \left( \mathbf{v}_2^T \mathbf{x} \right)^{m_2} \ldots \left( \mathbf{v}_{N-n}^T \mathbf{x} \right)^{m_{N-n}}$, so that we obtain an eigenstate of the reduced density matrix. The corresponding eigenvalue is
\begin{equation}
\lambda_{\left\{ m_1 , m_2 , \ldots , m_n \right\}} = \lambda_0 \xi_1^{m_1} \xi_2^{m_2} \ldots \xi_{N-n}^{m_{N-n}} .
\label{eq:spectrum_reduced_eigenvalues_summary}
\end{equation}

It follows that the spectrum of the reduced density matrix in the case of squeezed states has the same form as in the case of the ground state, namely \eqref{eq:reduced_ground_eigenvalues}. The difference is the following: in the case of the ground state, the values of the parameters $\xi_i$ are determined by the eigevalues of the matrix $\hat{\beta}$, defined in equation \eqref{eq:beta_hat_def}, via the formula \eqref{eq:effective_temperatures}. In the case of the squeezed state, the $\xi_i$ are the eigenvalues of the matrix $\Xi$, defined in equation \eqref{eq:spectrum_Xi_def_summary}. In the coherent limit of the squeezed states the two definitions become equivalent.

The matrix $\Xi$ is defined via the matrix ${\cal A}$. This introduces an extra difficulty: in order to calculate the matrix ${\cal A}$ one needs to solve the \emph{non-linear matrix equation} \eqref{eq:omega_eq} and \emph{specify which solution gives rise to normalizable eigenstates of the reduced density matrix}. It turns out that there is only a single admissible solution for ${\cal A}$ and thus for the matrix $\Xi$. The eigenvalues of the admissible matrix $\Xi$ can be specified as the solutions of the equation
\begin{equation}
\det \left(2 I - \lambda \hat{\beta} - \frac{1}{\lambda} \hat{\beta}^T \right) = 0
\end{equation}
that are smaller than 1. The above equation has in general $2 \left( N - n \right)$ solutions that come in pairs of the form $\left( \lambda , 1 / \lambda \right)$. Therefore $N - n$ of its solutions are smaller than 1 and the other $N - n$ are larger than 1.

The next four subsections contain all the technical details of the calculation of the eigenfunctions and eigenvalues of the reduced density matrix. The reader who is not interested in these details may skip them and move to the next section. In appendix \ref{subsubsec:spectrum_creation_annihilation} we present a systematic iterative method to construct the eigenfunctions of the reduced density matrix based on creation and annihilation operators. In appendix \ref{sec:app_renyi} we present an explicit realization of this calculational process in a toy example. We confirm that the result is in agreement with that obtained through an alternative method using the \Renyi entropies, which is not applicable to the general case, but is feasible in this simple example.

\subsection{The Reduced Density Matrix}
\label{subsec:many_oscillators_density}
In order to trace out the degrees of freedom of subsystem 1, we need to express the state \eqref{eq:many_overall_state_modes} in terms of the original coordinates $\mathbf{x}$,
\begin{equation}
\Psi \left( \mathbf{x} \right) = \left( \frac{\det \re \left( W \right)}{\pi^N} \right)^{\frac{1}{4}} \exp \left[ - \frac{1}{2} \left(\mathbf{x} - \mathbf{x}_0 \right)^T W \left(\mathbf{x} - \mathbf{x}_0 \right) + i \mathbf{p}_{0}^T \left( \mathbf{x} - \mathbf{x}_0 \right) - i \sum_{j = 1}^N \varphi_{\mathrm{s}j} \right] ,
\label{eq:many_overall_state}
\end{equation}
where obviously $W = O^T \tilde{W} O$, $\mathbf{\tilde{x}}_0 = O \mathbf{x}_0 $ and $\tilde{\mathbf{p}}_{0}=O\mathbf{p}_{0}$. The matrix $W$ is a \emph{complex symmetric} matrix. The density matrix describing the overall system assumes the form
\begin{multline}
\rho \left( \mathbf{x} ; \mathbf{x}^\prime \right) = \left( \frac{\det \re \left( W \right)}{\pi^N} \right)^{\frac{1}{2}} \exp \bigg[ - \frac{1}{2} \Big( \left( \mathbf{x} - \mathbf{x}_0 \right)^T W \left( \mathbf{x} - \mathbf{x}_0 \right) \\
+ \left( \mathbf{x}^\prime - \mathbf{x}_0 \right)^T W^* \left( \mathbf{x}^\prime - \mathbf{x}_0 \right) \Big) \bigg] \exp \left[ i \mathbf{p}_{0}^T \left( \mathbf{x} - \mathbf{x}^\prime \right) \right] .
\end{multline}
We use the block form notation
\begin{equation}
W = \left( \begin{array}{cc} A & B \\ B^T & C \end{array} \right) , \quad \mathbf{x} = \left( \begin{array}{c} \mathbf{x}_1 \\ \mathbf{x}_2 \end{array} \right) , \quad \mathbf{x}_0 = \left( \begin{array}{c} \mathbf{x}_{01} \\ \mathbf{x}_{02} \end{array} \right) , \quad \mathbf{p}_0 = \left( \begin{array}{c} \mathbf{p}_{01} \\ \mathbf{p}_{02} \end{array} \right) ,
\label{eq:many_blocks_def}
\end{equation}
where the matrix $A$ is an $n \times n$ matrix, the matrix $C$ is an $\left( N - n \right) \times \left( N - n \right)$ matrix and so on. Notice that the matrices $A$ and $C$ are complex symmetric matrices, whereas the matrix $B$ is not even a square matrix. 

Using this block form notation, the reduced density matrix describing subsystem 2, $\rho_2 \left( \mathbf{x}_2 ; \mathbf{x}_2^\prime \right) = \int d^n \mathbf{x}_1 \rho \left( \mathbf{x}_1 , \mathbf{x}_2 ; \mathbf{x}_1 , \mathbf{x}_2^\prime \right)$, can be easily found via the application of multidimensional Gaussian integrals. It is a matter of simple algebra to show that
\begin{multline}
\rho_2 \left( \mathbf{x}_2 ; \mathbf{x}_2^\prime \right) = \left( \frac{\det \re \left( \gamma - \beta \right)}{\pi^{N - n}} \right)^{\frac{1}{2}} \exp \bigg[ - \frac{1}{2} \left( \mathbf{y}_2^T \gamma \mathbf{y}_2 + \mathbf{y}_2^{\prime T} \gamma^* \mathbf{y}_2^\prime \right) \\
+ \mathbf{y}_2^{\prime T} \beta \mathbf{y}_2 + i \mathbf{p}_{02}^T \left( \mathbf{y}_2 - \mathbf{y}_2^\prime \right) \bigg] ,
\end{multline}
where
\begin{align}
\gamma &= C - \frac{1}{2} B^T \re \left( A \right)^{-1} B , \label{eq:many_gamma_def}\\
\beta &= \frac{1}{2} B^\dagger \re \left( A \right)^{-1} B . \label{eq:many_beta_def}
\end{align}
and $\mathbf{y}_i=\mathbf{x}_i-\mathbf{x}_{0i}$.

Notice that the matrix $\gamma$ is by definition a \emph{complex symmetric} matrix; it obeys $\gamma^T = \gamma$. On the contrary, the matrix $\beta$ is by definition a \emph{Hermitian} matrix; it obeys $\beta^\dagger = \beta$. For this reason, in the case of the two oscillators, where the matrices $\gamma$ and $\beta$ were numbers, the coefficient $\gamma$ was complex, whereas the coefficient $\beta$ was forced to be real, as we saw in section \ref{sec:two_oscillators}.

\subsection{The Eigenproblem for the Reduced Density Matrix}
\label{subsec:many_oscillators_spectrum}

Similarly to the case of the two oscillators of section \ref{sec:two_oscillators}, the imaginary part of the matrix $\gamma$ does not affect the eigenvalues of the reduced density matrix. We may write this matrix as
\begin{multline}
\rho_2 \left( \mathbf{x}_2 ; \mathbf{x}_2^\prime \right) = \left( \frac{\det \re \left( \gamma - \beta \right)}{\pi^{N - n}} \right)^{\frac{1}{2}} \exp \left[ - \frac{1}{2} \left( \mathbf{y}_2^T \re \left( \gamma \right) \mathbf{y}_2 + \mathbf{y}_2^{\prime T} \re \left( \gamma \right) \mathbf{y}_2^\prime \right) + \mathbf{y}_2^{\prime T} \beta \mathbf{y}_2 \right. \\
\left. - \frac{i}{2} \left( \mathbf{y}_2^T \im \left( \gamma \right) \mathbf{y}_2 - \mathbf{y}_2^{\prime T} \im \left( \gamma \right) \mathbf{y}_2^\prime \right) + i \mathbf{p}_{02}^T \left( \mathbf{y}_2 - \mathbf{y}_2^\prime \right) \right] .
\end{multline}
Consider the matrix $\tilde{\rho}_2$ which is identical to the reduced density matrix $\rho_2$, where we have set by hand the imaginary part of $\gamma$ and the shifts $\mathbf{x}_{02}$ and $\mathbf{p}_{02}$ to zero, i.e.
\begin{multline}
\tilde{\rho}_2 \left( \mathbf{x}_2 ; \mathbf{x}_2^\prime \right) = \left( \frac{\det \re \left( \gamma - \beta \right)}{\pi^{N - n}} \right)^{\frac{1}{2}} \\
\times \exp \left[ - \frac{1}{2} \left( \mathbf{x}_2^T \re \left( \gamma \right) \mathbf{x}_2 + \mathbf{x}_2^{\prime T} \re \left( \gamma \right) \mathbf{x}_2^\prime \right) + \mathbf{x}_2^{\prime T} \beta \mathbf{x}_2 \right] .
\end{multline}
Furthermore, consider that $\tilde{f} \left( \mathbf{x}_2 \right)$ is an eigenfunction of $\tilde{\rho}_2$ with eigenvalue $\lambda$, i.e.
\begin{equation}
\int d^{N - n} \mathbf{x}_2^\prime \tilde{\rho}_2 \left( \mathbf{x}_2 ; \mathbf{x}_2^\prime \right) \tilde{f} \left( \mathbf{x}_2^\prime \right) = \lambda \tilde{f} \left( \mathbf{x}_2 \right) .
\end{equation}
Then, the function
\begin{equation}
f \left( \mathbf{x}_2 \right) = \exp \left[ - \frac{i}{2} \left( \mathbf{x}_2 - \mathbf{x}_{02} \right)^T \im \left( \gamma \right) \left( \mathbf{x}_2 - \mathbf{x}_{02} \right) + i \mathbf{p}_{02}^T \left( \mathbf{x}_2 - \mathbf{x}_{02} \right) \right] \tilde{f} \left( \mathbf{x}_2 - \mathbf{x}_{02} \right)
\label{eq:eigenrho_eigenrhotilde}
\end{equation}
is trivially an eigenfunction of the reduced density matrix $\rho_2$ with the same eigenvalue.

Therefore, it is sufficient to find the spectrum of the simpler matrix $\tilde{\rho}_2$. We may further simplify $\tilde{\rho}_2$ via the following linear transformations of the coordinates:
\begin{enumerate}
\item A real orthogonal transformation of the coordinates $\mathbf{x}_2$, which diagonalizes the matrix $\gamma$.
\item A rescaling of the coordinates $\mathbf{x}_2$, so that the matrix $\re\left( \gamma \right)$ becomes the identity matrix.
\end{enumerate}
Let us denote the coordinates after these two transformations as $\mathbf{\hat{x}}_2$. In an obvious manner $\mathbf{\hat{x}}_2 = \re\left( \gamma \right)^{-\frac{1}{2}} \mathbf{x}_2$. Then, the matrix $\tilde{\rho}_2$ assumes the form
\begin{equation}
\tilde{\rho}_2 \left( \mathbf{\hat{x}}_2 ; \mathbf{\hat{x}}_2^\prime \right) = \sqrt{\frac{\det \left( I - \re \left(\hat{\beta}\right) \right)}{\pi^{N - n}}} \exp \left[ - \frac{1}{2} \left( \mathbf{\hat{x}}_2^T \mathbf{\hat{x}}_2 + \mathbf{\hat{x}}_2^{\prime T} \mathbf{\hat{x}}_2^\prime \right) + \mathbf{\hat{x}}_2^{\prime T} \hat{\beta} \mathbf{\hat{x}}_2 \right] ,
\end{equation}
where
\begin{equation}
\hat{\beta} = \re\left( \gamma \right)^{-\frac{1}{2}} \beta \, \re\left( \gamma \right)^{-\frac{1}{2}} .
\end{equation}
The calculation of its spectrum cannot continue along the same path as in the case of the ground or coherent state. In such a case, the matrices $\gamma$ and $\beta$ would be both real and symmetric and so would be the matrix $\hat{\beta}$. So, we would apply a final real orthogonal transformation, which would diagonalize $\hat{\beta}$ and effectively factorize the problem to problems of a single degree of freedom, rendering the calculation of the density matrix eigenstates and eigenvalues trivial. This is not possible in our case. The matrix $\hat{\beta}$ is not real and symmetric, but rather it is a Hermitian matrix, and thus \emph{it cannot be diagonalized via a real orthogonal transformation}. 

In the following, we drop the index 2 from the coordinates that describe the degrees of freedom of subsystem 2.

\subsection{The Eigenstates of the Reduced Density Matrix}
\label{subsec:many_eigenstates}
In the case that the matrix $\hat{\beta}$ is real and symmetric, we know the form of the eigenstates of the reduced density matrix. They are the Fock space states of an effective system of coupled harmonic oscillators.

This is not the case when the matrix $\hat{\beta}$ is Hermitian and not real. The structure of the eigenstates is deformed. However, there are several characteristics that remain invariant and allow the specification of the spectrum of the reduced density matrix.

\subsubsection{The ``Ground'' Eigenstate of the Reduced Density Matrix}
First, let us investigate whether there is a ``ground'' state similar to the case of real $\hat{\beta}$, i.e. a Gaussian state. Its existence is supported by the fact that the reduced density matrix is also Gaussian. Consider the normalized wavefunction
\begin{equation}
\Psi_0 \left( \mathbf{x} \right) = c_0 \exp \left( - \frac{1}{2} \mathbf{x}^T {\cal A} \mathbf{x} \right) , \quad c_0=\left(\frac{\det\left(\re\left(\mathcal{A}\right)\right)}{\pi^{N-n}}\right)^{1/4} .
\label{eq:spectrum_ground_state}
\end{equation}
The matrix ${\cal A}$ is in general a complex symmetric matrix, which needs to be specified so that $\Psi_0$ is an eigenstate of the reduced density matrix. It is a matter of algebra to show that
\begin{equation}
\begin{split}
\tilde{\rho}_2 \Psi_0 \left( \mathbf{x} \right) &= \int d^n \mathbf{x^\prime} \tilde{\rho}_2 \left( \mathbf{x} ; \mathbf{x}^\prime \right) \Psi_0 \left( \mathbf{x^\prime} \right) \\
&= c c_0 \int d^n \mathbf{x^\prime} \exp \left[ - \frac{1}{2} \left( \mathbf{x}^T \mathbf{x} + \mathbf{x}^{\prime T} \left( I + {\cal A} \right) \mathbf{x}^\prime \right) + \mathbf{x}^{\prime T} \hat{\beta} \mathbf{x} \right] ,
\end{split}
\end{equation}
where
\begin{equation}
c = \sqrt{\frac{\det \left( I - \re \left(\hat{\beta}\right) \right)}{\pi^{N - n}}} .
\end{equation}
Performing this integral yields
\begin{equation}
\tilde{\rho}_2 \Psi_0 \left( \mathbf{x} \right) = c c_0 c_{\mathrm{int}} \exp \left[ - \frac{1}{2} \mathbf{x}^T \left( I - \hat{\beta}^T \left( I + {\cal A} \right)^{- 1} \hat{\beta} \right) \mathbf{x} \right] ,
\end{equation}
where
\begin{equation}
c_{\mathrm{int}} = \sqrt{\frac{\left( 2 \pi \right)^{N - n} }{\det \left( I + {\cal A} \right)}} .
\end{equation}

It follows that the Gaussian state \eqref{eq:spectrum_ground_state} is indeed an eigenstate of the reduced density matrix, if
\begin{equation}
{\cal A} = I - \hat{\beta}^T \left( I + {\cal A} \right)^{- 1} \hat{\beta} .
\label{eq:spectrum_omega_def}
\end{equation}
Then, the corresponding eigenvalue is
\begin{equation}
\lambda_0 = c c_{\mathrm{int}} = \sqrt{\frac{2^{N - n} \det \left( I - \re \left(\hat{\beta}\right) \right)}{\det \left( I + {\cal A} \right)}} .
\label{eq:spectrum_lambda_0}
\end{equation}
Since the reduced density matrix is Hermitian, and thus, it has real eigenvalues, the above implies that the matrix $I + {\cal A}$ has a real determinant, although in general it is a complex symmetric matrix.

\subsubsection{The ``First Excited'' Eigenstates of the Reduced Density Matrix}

In the case that the matrix $\hat{\beta}$ is real, there exist eigenstates that are the first excited states of the effective harmonic system. These states are the product of a Gaussian with a linear combination of the coordinates that is the corresponding normal coordinate of the mode that is excited. So let us investigate whether there are eigenstates of the form
\begin{equation}
\psi_1 \left( \mathbf{x} \right) = c_1 \mathbf{v}^T \mathbf{x} \exp \left( - \frac{1}{2} \mathbf{x}^T {\cal A} \mathbf{x} \right) ,
\label{eq:spectrum_first excited_state}
\end{equation}
where $\mathbf{v}$ is a constant vector. The algebra is similar to the case of the Gaussian eigenstate:
\begin{equation}
\begin{split}
\tilde{\rho}_2 \psi_1 \left( \mathbf{x} \right) &= \int d^n \mathbf{x^\prime} \tilde{\rho}_2 \left( \mathbf{x} ; \mathbf{x}^\prime \right) \psi_1 \left( \mathbf{x^\prime} \right) \\
&= c c_1 \int d^n \mathbf{x}^\prime \mathbf{v}^{\prime T} \mathbf{x} \exp \left[ - \frac{1}{2} \left( \mathbf{x}^T \mathbf{x} + \mathbf{x}^{\prime T} \left( I + {\cal A} \right) \mathbf{x}^\prime \right) + \mathbf{x}^{\prime T} \hat{\beta} \mathbf{x} \right] .
\end{split}
\end{equation}
In order to perform this integral, we complete the square in the exponent, as in the case of the ``ground'' eigenstate. This yields
\begin{equation}
\tilde{\rho}_2 \psi_1 \left( \mathbf{x} \right) = c c_1 c_{\mathrm{int}} \mathbf{v}^T \left( I + {\cal A} \right)^{- 1} \beta \mathbf{x} \exp \left[ - \frac{1}{2} \mathbf{x}^T \left( I - \hat{\beta}^T \left( I + {\cal A} \right)^{- 1} \hat{\beta} \right) \mathbf{x} \right] .
\end{equation}

It follows that the state \eqref{eq:spectrum_first excited_state} is indeed an eigenstate of the reduced density matrix, as long as it has the same matrix ${\cal A}$ as the Gaussian eigenstate, i.e. the solution of equation \eqref{eq:spectrum_omega_def}, and furthermore the vector $\mathbf{v}$ is a right eigenvector of the matrix
\begin{equation}
\Xi := \hat{\beta}^T \left( I + {\cal A} \right)^{- 1} .
\label{eq:spectrum_Xi_def}
\end{equation}
Notice that the matrix $\Xi$ in general is not symmetric or real or Hermitian. Let $\mathbf{v}_i$ be the eigenvectors of the matrix $\Xi$ and $\xi_i$ the corresponding eigenvalues, i.e.
\begin{equation}
\Xi \mathbf{v}_i = \xi_i \mathbf{v}_i .
\end{equation}
Then, we have found $N - n$ ``first excited'' eigenstates of the reduced density matrix. They read
\begin{equation}
\Psi_{1i} \left( \mathbf{x} \right) = c_{1i} \mathbf{v}_i^T \mathbf{x} \exp \left( - \frac{1}{2} \mathbf{x}^T {\cal A} \mathbf{x} \right) 
\label{eq:spectrum_first excited_eigenstate}
\end{equation}
and the corresponding eigenvalues are
\begin{equation}
\lambda_{1i} = c c_{\mathrm{int}} \xi_i = \lambda_0 \xi_i .
\end{equation}

The fact that the density matrix is Hermitian implies that its eigenvalues are real; therefore the eigenvalues $\xi_i$ of the matrix $\Xi$ \emph{are real}, although this matrix has no particular symmetry property. The fact that the density matrix is Hermitian further implies that these eigenstates are orthogonal, although the eigenvectors of the matrix $\Xi$ are not necessarily orthogonal either in the real or the complex sense. It is a matter of simple algebra to show that demanding that the states $\Psi_{1i}$ are not only orthogonal, but also normalized, yields
\begin{equation}
\delta_{ij} = c_{1i} c_{1j}^* \int d^n \mathbf{x} \left( \mathbf{v}_i^T \mathbf{x} \right) \left( \mathbf{v}_j^T \mathbf{x} \right)^* \exp \left( - \mathbf{x}^T \re\left( {\cal A} \right) \mathbf{x} \right) = \frac{c_{1i} c_{1j}^*}{2 c_0^2} \mathbf{v}^\dagger_j \re \left( {\cal A} \right)^{-1} \mathbf{v}_i .
\label{eq:spectrum_first_escited_orthonormality}
\end{equation}
This implies that the eigenvectors of the matrix $\Xi$ are orthogonal in the complex sense upon the introduction of a real metric, which is equal to the inverse of the real part of the matrix ${\cal A}$. In what follows we always choose the eigenvectors of the matrix $\Xi$ to be orthonormal in this sense, i.e.
\begin{equation}
\mathbf{v}^\dagger_j \re \left( {\cal A} \right)^{-1} \mathbf{v}_i = \delta_{ij}.
\end{equation}
Obviously, this definition of the vectors $\mathbf{v}_i$, combined with the equation \eqref{eq:spectrum_first_escited_orthonormality}, implies that the appropriate choice for the normalization constant of the ``first excited'' eigenstates \eqref{eq:spectrum_first excited_eigenstate} is
\begin{equation}
c_{1i} = \sqrt{2} c_0 .
\label{eq:spectrum_c1i}
\end{equation}

The above also imply that the matrix
\begin{equation}
\Xi^\prime = \re \left( {\cal A} \right)^{- \frac{1}{2}} \Xi \, \re \left( {\cal A} \right)^{\frac{1}{2}}
\label{eq:spectrum_Xi_prime}
\end{equation}
has the same eigenvalues as $\Xi$, and its eigenvectors, which are simply $\mathbf{v}_i^\prime = \re \left( {\cal A} \right)^{- \frac{1}{2}} \mathbf{v}_i$, are orthogonal is the usual complex sense, i.e.
\begin{equation}
\mathbf{v}^{\prime \dagger}_j \mathbf{v}^\prime_i = \delta_{ij} .
\end{equation}
It follows that the matrix $\Xi^\prime$ is Hermitian.

\subsubsection{The Tower of Eigenstates of the Reduced Density Matrix}
\label{subsubsec:spectrum_tower}

In the case of real $\hat{\beta}$ the construction of the rest of the tower of eigenstates is trivial, since they constitute the Fock space of an effective harmonic system. In our case of interest, namely that of complex $\hat{\beta}$, the construction of the whole tower of states is not that simple. Let as consider the ``second excited'' state
\begin{equation}
\psi_{2ij} \left( \mathbf{x} \right) = c_{2ij} \mathbf{v}_i^T \mathbf{x} \mathbf{v}_j^T \mathbf{x} \exp \left( - \frac{1}{2} \mathbf{x}^T {\cal A} \mathbf{x} \right) .
\label{eq:spectrum_second excited_state}
\end{equation}
The indices $i$ and $j$ may coincide or not. In the case of real $\hat{\beta}$, we would expect that this is an eigenstate if $i \neq j$, whereas, if $i = j$, the square $\left( \mathbf{v}_i^T \mathbf{x} \right)^2$ should be corrected to the second order Hermite polynomial of $\mathbf{v}_i^T \mathbf{x}$. In the case of complex $\hat{\beta}$, the above do not hold. Let us study the action of the density matrix on this state:
\begin{equation}
\begin{split}
\tilde{\rho}_2 \psi_{2ij} \left( \mathbf{x} \right) &= \int d^n \mathbf{x^\prime} \tilde{\rho}_2 \left( \mathbf{x} ; \mathbf{x}^\prime \right) \psi_{2ij} \left( \mathbf{x^\prime} \right) \\
&= c c_{2ij} \int d^n \mathbf{x}^\prime \mathbf{v}_i^T \mathbf{x}^\prime \mathbf{v}_j^T \mathbf{x}^\prime \exp \left[ - \frac{1}{2} \left( \mathbf{x}^T \mathbf{x} + \mathbf{x}^{\prime T} \left( I + {\cal A} \right) \mathbf{x}^\prime \right) + \mathbf{x}^{\prime T} \hat{\beta} \mathbf{x} \right] .
\end{split}
\end{equation}
We complete the square in the exponent as in the case of the Gaussian eigenstate. This yields
\begin{equation}
\begin{split}
\tilde{\rho}_2 \psi_{2ij} \left( \mathbf{x} \right) &= \lambda_0 \xi_i \xi_j c_{2ij} \mathbf{v}_i^T \mathbf{x} \mathbf{v}_j^T \mathbf{x} \exp \left( - \frac{1}{2} \mathbf{x}^T {\cal A} \mathbf{x} \right) + c c_{2ij} c_{\mathrm{int}}^{2ij} \exp \left( - \frac{1}{2} \mathbf{x}^T {\cal A} \mathbf{x} \right) \\
&= \lambda_0 \xi_i \xi_j \psi_{2ij} + \frac{c c_{2ij} c_{\mathrm{int}}^{2ij}}{c_0} \Psi_0 ,
\end{split}
\end{equation}
where
\begin{equation}
c_{\mathrm{int}}^{2ij} = \int d^n \mathbf{x}^\prime \mathbf{v}_i^T \mathbf{x}^\prime \mathbf{v}_j^T \mathbf{x}^\prime \exp \left[ - \frac{1}{2} \mathbf{x}^{\prime T} \left( I + {\cal A} \right) \mathbf{x}^\prime \right] =\sqrt{\frac{\left(2\pi\right)^{N-n}}{\det\left(I+\mathcal{A}\right)}} v_i^T\left(I+\mathcal{A}\right)^{-1}v_j.
\end{equation}

The above integral does not vanish either when $i = j$ or when $i \neq j$, unlike the case of real $\hat{\beta}$. Therefore, the state $\psi_{2ij}$ is not an eigenstate of the reduced density matrix, as the action of the latter on $\psi_{2ij}$ gives a linear combination of $\psi_{2ij}$ and the Gaussian ``ground'' eigenstate $\Psi_0$. Nevertheless, it follows that it is trivial to construct an eigenstate of the reduced density matrix by taking an appropriate linear combination of $\psi_{2ij}$ and $\Psi_0$, namely 
\begin{equation}
\Psi_{2ij} = \psi_{2ij} + c_{0ij} \Psi_0 ,
\end{equation}
where
\begin{equation}
c_{0ij} = \frac{c_{2ij} c_{\mathrm{int}}^{2ij}}{c_0 c_{\mathrm{int}}} \frac{1}{\xi_i \xi_j - 1} .
\end{equation}
More interestingly though, we do not need to explicitly calculate the coefficient $c_{0ij}$ in order to specify the eigenvalue of $\Psi_{2ij}$. This is the coefficient of $\psi_{2ij}$ in $\tilde{\rho}_2 \psi_{2ij}$. The addition of $c_{0ij} \Psi_0$ cannot alter this term; it only corrects the subleading terms. Therefore, there is a set of ``second excited'' eigenstates, the states $\Psi_{2ij}$, with corresponding eigenvalues
\begin{equation}
\lambda_{2ij} = \lambda_0 \xi_i \xi_j .
\end{equation}

It is not difficult to show that this argument holds inductively. It is always possible to build an eigenstate $\Psi_{\left\{ m_1 , m_2 , \ldots , m_n \right\}} \left( \mathbf{x} \right)$, whose higher-order term is
\begin{multline}
\psi_{\left\{ m_1 , m_2 , \ldots , m_n \right\}} \left( \mathbf{x} \right) \\
= c_{\left\{ m_1 , m_2 , \ldots , m_n \right\}} \left( \mathbf{v}_1^T \mathbf{x} \right)^{m_1} \left( \mathbf{v}_2^T \mathbf{x} \right)^{m_2} \ldots \left( \mathbf{v}_n^T \mathbf{x} \right)^{m_n} \exp \left( - \frac{1}{2} \mathbf{x}^T {\cal A} \mathbf{x} \right) ,
\label{eq:spectrum_general_excited_state}
\end{multline}
with eigenvalue
\begin{equation}
\lambda_{\left\{ m_1 , m_2 , \ldots , m_n \right\}} = \lambda_0 \xi_1^{m_1} \xi_2^{m_2} \ldots \xi_n^{m_n} .
\label{eq:spectrum_reduced_eigenvalues}
\end{equation}

The definition of the matrix $\Xi$ \eqref{eq:spectrum_Xi_def}, combined with the defining equation of the matrix ${\cal A}$ \eqref{eq:spectrum_omega_def}, implies that
\begin{equation}
I - {\cal A} = \Xi \left( I + {\cal A} \right) \Xi^T .
\end{equation}
Using the definition of the matrix $\Xi$, the trick $I = \frac{1}{2} \left( I + {\cal A} \right) + \frac{1}{2} \left( I - {\cal A} \right)$ and the above relation, yields
\begin{equation}
I - \re \left( \hat{\beta} \right) = \frac{1}{2} \left( I - \Xi \right) \left( I + {\cal A} \right) \left( I - \Xi^T \right) .
\end{equation}
This, combined with equation \eqref{eq:spectrum_lambda_0}, implies that
\begin{equation}
\lambda_0 = \sqrt{\det \left( I - \Xi \right) \det \left( I - \Xi^T \right)}
\end{equation}
or
\begin{equation}
\lambda_0 = \left( 1 - \xi_1 \right) \left( 1 - \xi_2 \right) \ldots \left( 1 - \xi_n \right) .
\end{equation}
Thus, the eigenvalues of the eigenfunctions that we have already discovered are equal to
\begin{equation}
\lambda_{\left\{ m_1 , m_2 , \ldots , m_n \right\}} = \left( 1 - \xi_1 \right) \left( 1 - \xi_2 \right) \ldots \left( 1 - \xi_n \right) \xi_1^{m_1} \xi_2^{m_2} \ldots \xi_n^{m_n} .
\label{eq:spectrum_final_spectrum}
\end{equation}
It follows that
\begin{equation}
\sum_{\left\{ m_1 , m_2 , \ldots , m_n \right\}} \lambda_{\left\{ m_1 , m_2 , \ldots , m_n \right\}} = 1 .
\end{equation}
Therefore, we have discovered all the eigenfunctions of the reduced density matrix, or at least all the eigenfunctions with non-vanishing eigenvalues.

Finally, equation \eqref{eq:spectrum_final_spectrum} implies that the entanglement entropy is given by the same formula as in the case of the ground or coherent states, namely
\begin{equation}
S_{\mathrm{EE}} = - \sum_i \left( \ln \left( 1 - \xi_i \right) + \frac{\xi_i}{1 - \xi_i} \ln \xi_i \right) .
\label{eq:spectrum_final_SEE}
\end{equation}
As in the simple case of the two oscillators, the parameters $\xi_i$ are time-dependent, implying that the time evolution of the reduced density matrix includes a non-unitary part.

The eigenfunctions of the matrix $\rho_2$ can also be constructed algebraically with the use of appropriate creation and annihilation operators. The details of this construction are presented
in appendix \ref{subsubsec:spectrum_creation_annihilation}. The structure of the eigenvalues of the reduced density matrix implies that this matrix assumes the form
\begin{equation}
\rho_2 \sim \exp \left[ \sum_i - \mu_i N_i \right] ,
\end{equation}
where $N_i$ are the occupation numbers resulting from the creation and annihilation operators that construct algebraically the eigenfunctions, while $\mu_i = - \ln \xi_i$. This structure corresponds to a generalized Gibbs ensemble (GGE). However, there are several subtle features:
\begin{enumerate}
\item The creation and annihilation operators are linear functions of the local positions and momenta. However, the linear combination of positions is not conjugate to the linear combination of momenta appearing in the same operator.
\item The overall system does not lie in an equilibrium state. The squeezed states are time-dependent and they display an oscillatory behaviour. However, the reduced density matrix corresponds to a generalized Gibbs ensemble with time-dependent chemical potentials. It is interesting that the subsystem can be described by notions of equilibrium thermodynamics, such as the GGE.
\item If we wanted to use the form of this distribution in order to estimate scaling properties of the entanglement entropy, we would need to know the spectrum of the matrix $\Xi$ analytically. Unfortunately, this is a task that can only be performed numerically. However, in Section \ref{sec:expansions}, we show that an analytical treatment is possible in the limit that the squeezing parameter is large.
\end{enumerate}

\subsection{A Comment on the Symmetry Property of Entanglement Entropy}
\label{subsec:symmetry}

We found above that the spectrum of the reduced density matrix is given by equation \eqref{eq:spectrum_final_spectrum}, where $\xi_i$ are the eigenvalues of the square matrix $\Xi$, defined in \eqref{eq:spectrum_Xi_def}, whose dimension is equal to the number of degrees of freedom of the reduced system. It follows that the entanglement entropy is given by the formula \eqref{eq:spectrum_final_SEE}.

It is well known that, when the overall system lies in a pure state, which is the case for 
the system we study in this work, the entanglement entropy has a symmetry property: the entanglement entropy calculated by the reduced density matrix of subsystem $A$ is identical to the entanglement entropy calculated by the reduced density matrix of the complementary subsystem $A^C$,
\begin{equation}
S_{\mathrm{EE}} \left( \rho_A \right) = S_{\mathrm{EE}} \left( \rho_{A^C} \right) .
\end{equation}
Within the framework that we performed our calculation of the spectrum of the reduced density matrix, this property may appear peculiar, since the larger of the two subsystems would be characterized by a larger number of parameters $\xi_i$.
Actually, the symmetry property does not only require equality of the two entanglement entropies. The two reduced density matrices have identical spectra; the one with the larger dimension has the same eigenvalues as the one with the smaller dimension, plus vanishing eigenvalues.

Given that the spectrum of the reduced density matrix in our case is given by \eqref{eq:spectrum_final_spectrum}, these facts imply the following: The matrix $\Xi$ of the subsystem with the larger number of degrees of freedom, namely $\max \left( n , N - n \right)$, has $\min \left( n , N - n \right)$ eigenvalues, which are identical to the $\min \left( n , N - n \right)$ eigenvalues of the matrix $\Xi$ of the subsystem with the smaller number of degrees of freedom. The remaining $\max \left( n , N - n \right) - \min \left( n , N - n \right)$ eigenvalues are vanishing.

As long as we are interested in the entanglement entropy or even the spectrum of the reduced density matrix, it is simpler to consider the reduced density matrix of the smaller of the two subsystems. This simplifies numerical calculations, since these are performed with matrices of smaller dimension. More importantly, considering the reduced density matrix for the smaller subsystem eliminates the presence of vanishing eigenvalues in the spectrum of matrices such as $\beta$ or $\Xi$, which would render them non-invertible.

\subsection{The Eigenvalues of the Reduced Density Matrix}
\label{subsec:spectrum_M_method}

The problem of the specification of the spectrum of the matrix $\tilde{\rho}_2$, and, thus, of the reduced density matrix, has been reduced to the problem of the specification of the eigenvalues of the matrix $\Xi$, defined in equation \eqref{eq:spectrum_Xi_def}. The matrix $\Xi$ has no specific symmetry property, but nevertheless it has real eigenvalues; we have shown that it is similar to a Hermitian matrix (see equation \eqref{eq:spectrum_Xi_prime}). A problem that appears in this task is that the matrix $\Xi$ is defined in terms of the matrix ${\cal A}$, which is a solution of the quadratic equation \eqref{eq:spectrum_omega_def}. As such, there are many matrices ${\cal A}$, and thus matrices $\Xi$. However, not all of them correspond to normalizable eigenstates of the reduced density matrix. We need to find a systematic way to distinguish which ${\cal A}$ corresponds to normalizable eigenstates and then calculate the eigenvalues of the matrix $\Xi$ that correspond to this specific choice.

We write \eqref{eq:spectrum_omega_def} in the form of a matrix Riccati equation as
\begin{equation}
I - 2 \Xi^T \hat{\beta}^{-1} + \Xi^T \hat{\beta}^{-1} \hat{\beta}^T \Xi^T = 0 .
\end{equation}
The solutions of equations of the form
\begin{equation}\label{eq:matrix_Riccati}
M_{21} + M_{22} W - W M_{11} - W M_{12} W = 0
\end{equation}
are constructed as follows: We define the matrix
\begin{equation}
M=\begin{pmatrix}
M_{11} & M_{12} \\
M_{21} & M_{22}
\end{pmatrix}.
\end{equation}
Then the original equation is equivalent to the ``eigenvalue" problem
\begin{equation}
\begin{pmatrix}
M_{11} & M_{12} \\
M_{21} & M_{22}
\end{pmatrix}\begin{pmatrix}
I \\ W
\end{pmatrix}=\begin{pmatrix}
I \\ W
\end{pmatrix} Z,
\end{equation}
where $Z$ is a matrix. The first line of this equation implies that $Z=M_{11}+M_{12}W$ and thus the second line is equivalent to the original Riccati equation \eqref{eq:matrix_Riccati}. Given the solutions of the ordinary eigenvalue problem
\begin{equation}\label{eq:eig_M}
\begin{pmatrix}
M_{11} & M_{12} \\
M_{21} & M_{22}
\end{pmatrix}
\begin{pmatrix}
\chi_j \\ \psi_j
\end{pmatrix}=\lambda_j \begin{pmatrix}
\chi_j \\ \psi_j
\end{pmatrix},
\end{equation}
the solutions of \eqref{eq:matrix_Riccati} are
\begin{equation}\label{eq:matrix_Riccati_sol}
W=\psi\chi^{-1},
\end{equation}
where the matrices $\chi$ and $\psi$ are constructed using some of the $\chi_j$ and $\psi_j$ as columns, i.e.
\begin{equation}
\chi_{ij}=\left(\chi_j\right)_i\qquad \psi_{ij}=\left(\psi_j\right)_i.
\end{equation}
In particular when all matrices of \eqref{eq:matrix_Riccati} are square $d \times d$ matrices, there are $(2d)!/(d!)^2$ combinations. Of course not all combinations correspond to valid solutions, since the matrix $\chi$ should be invertible. 

In our case, the matrix $M$ is $2 \min \left( n , N - n \right) \times 2 \min \left( n , N - n \right)$ and reads
\begin{equation}
M=
\begin{pmatrix}
2 \hat{\beta}^{-1} & -\hat{\beta}^{-1}\hat{\beta}^T\\
I & 0
\end{pmatrix}.
\label{eq:many_M_def}
\end{equation}
Notice that 
\begin{equation}
\det M = \det \left(\hat{\beta}^{-1} \right) \det \left( \hat{\beta}^T \right) = 1 .
\end{equation}
The eigenvalues of the matrix $M$ are specified by the equation
\begin{equation}
\det \left(2 I - \lambda \hat{\beta} - \frac{1}{\lambda} \hat{\beta}^T \right) = 0 .
\label{eq:spectrum_eigenvalues_Xi_equation}
\end{equation}
Since the determinant is invariant under transposition, the eigenvalues of $M$ \emph{come in pairs} of the form $(\lambda,1/\lambda)$.

In order to identify admissible solutions for $\Xi$ we have to understand the relation between the eigenvalues of the matrix $\Xi$ and the eigenvalues of the matrix $M$. It turns out that this relation is quite simple. Recall that the spectrum of the reduced density matrix is given by \eqref{eq:spectrum_final_spectrum}. It follows that the eigenvalues of $\Xi$ (and thus the eigenvalues of $\Xi^T$) should all be not only real and positive, but also smaller than $1$.

The eigenvalue problem \eqref{eq:eig_M} implies that the matrices $\chi$ and $\psi$ obey
\begin{align}
2 \chi - \hat{\beta}^T \psi &= \hat{\beta} \chi \lambda_D , \\
\chi \lambda_D^{-1} &= \psi , \label{eq:block_eq_2}
\end{align}
where $\lambda_D$ is a diagonal matrix containing the eigenvalues of $M$ which correspond to the eigenvectors that we used in order to construct the matrices $\chi$ and $\psi$. Equation \eqref{eq:matrix_Riccati_sol} along with \eqref{eq:block_eq_2} implies that the matrix $\Xi^T$ reads
\begin{equation}
\Xi^T = \chi \lambda_D^{-1} \chi^{-1} .
\end{equation}
As a direct consequence, the eigenvalues of $\Xi$ coincide with the inverse of the eigenvalues of $M$ which correspond to the eigenvectors that we used in order to construct the matrices $\chi$ and $\psi$. This implies that $M$ has at least $\min \left( n , N - n \right)$ eigenvalues that are real and larger than 1. Since the eigenvalues of the matrix $M$ come in pairs of the form $(\lambda,1/\lambda)$, it follows that exactly $\min \left( n , N - n \right)$ eigenvalues of the matrix $M$ are real and larger than 1 and the other $\min \left( n , N - n \right)$ are real, positive and smaller than 1. This also implies that:
\begin{itemize}
\item There is a single admissible matrix $\Xi$. It is constructed using the $\min \left( n , N - n \right)$ eigenvectors of $M$ which correspond to its eigenvalues that are larger than $1$.
\item The matrix ${\cal A}$, which corresponds to this admissible $\Xi$, is the only one which gives rise to normalizable eigenstates of the reduced density matrix.
\item The eigenvalues of the admissible matrix $\Xi$ are simply the $\min \left( n , N - n \right)$ solutions of the equation \eqref{eq:spectrum_eigenvalues_Xi_equation} that are smaller than 1.
\end{itemize}

A solvable example that demonstrates the structure of the eigenvalues of the matrix $M$ is presented in appendix \ref{sec:app_solvable}. In this example, although the system lies in a squeezed state, the phases of the modes are selected in a very particular way (they are all equal to zero) and as a result the matrix $\beta$ is real. Another example, where the matrix $\beta$ is complex is presented in appendix \ref{sec:app_renyi}, where it is verified that the calculated entanglement entropy is in agreement with a calculation based on the \Renyi extension of entanglement entropy. Finally, in appendix \ref{subsec:correlation} we relate our calculation to the calculation of the spectrum of the reduced density matrix using the correlation matrix method. There, we also show that we can relate the eigenvalues $\lambda$ of the matrix $M$ to the eigenvalues $\tilde{\lambda}$, of the $N\times N$ matrix $\tilde{M}$, defined as
\begin{equation}
\tilde{M}=\re\left(W\right)^{-1}\begin{pmatrix}
-\re\left(A\right) & i\im\left(B\right)\\
-i\im\left(B\right)^T & \re\left(C\right)
\end{pmatrix}.
\end{equation}
Their relation is
\begin{equation}
\lambda=\frac{\tilde{\lambda}-1}{\tilde{\lambda}+1}.
\end{equation}
The matrix $\tilde{M}$ is well defined independently of whether we trace out the larger or the smaller subsystem. It turns out that the admissible spectrum of $\tilde{M}$, which satisfies $\tilde{\lambda}>1$, is equivalent to the spectrum deduced from the correlation matrix.

The numerical analysis in the following sections has been performed with three equivalent approaches, namely the method based on the matrix $M$, the correlation matrix method, and the method based on the matrix $\tilde{M}$, discussed in appendix \ref{subsec:correlation}. All three methods give results that are identical up to at least ten significant digits.

\section{Large-Squeezing Expansion}
\label{sec:expansions}
In section \ref{sec:two_oscillators} we studied the system of two coupled oscillators, where we traced out one of them. We showed that in general entanglement entropy increases with squeezing. For small squeezings, the entanglement entropy is a quadratic function of the squeezing parameters, whereas for large squeezings, the entanglement entropy becomes a linear function of the squeezing parameters, see e.g. equation \eqref{eq:2_mode_mean_SEE_asymtotics}. 

Although it is very difficult to find exact formulae for the case of a general harmonic system with an arbitrary number of degrees of freedom, we would like to find asymptotic expressions in order to check whether the form of dependence of entanglement entropy on squeezing in the system of two oscillators persists in the general harmonic system. In the following we discuss this expansion for large squeezing parameters, which will be relevant for the interpretation of the numerical results of section \ref{sec:1dft}. The expansion for small squeezing parameters is presented in appendix \ref{subsec:small_expansion}.

The most important parameter that defines the state of the overall system and consequently the entanglement between subsystems is the coefficient $w$ of the quadratic part of the exponent of the wavefunction of a squeezed mode. We remind the reader that, in the case of the ground state or a coherent state, this coefficient is trivially real and equal to the eigenfrequency of the mode, whereas in the case of a squeezed state, it is complex, depends on the squeezing parameter and is not constant in time, as shown in equation \eqref{eq:omega_squeezed}. Defining $\epsilon= \exp(-z)$, we may rewrite this equation as
\begin{equation}
w = \frac{\omega}{1 +\sin \left[ 2 \omega \left( t - t_0 \right) \right]} \frac{2\epsilon - i \left(1-\epsilon^2\right) \cos \left[ 2 \omega \left( t - t_0 \right) \right]}{1+\epsilon^2\frac{1-\sin \left[ 2 \omega \left( t - t_0 \right) \right]}{1 +\sin \left[ 2 \omega \left( t - t_0 \right) \right]}}.
\end{equation}
This can be written as a series in powers of $\epsilon$ as,
\begin{multline}
w = \frac{2\omega}{1 +\sin \left[ 2 \omega \left( t - t_0 \right) \right]}\sum_{k=0}^\infty\left(-1\right)^k\left(\frac{1-\sin \left[ 2 \omega \left( t - t_0 \right) \right]}{1 +\sin \left[ 2 \omega \left( t - t_0 \right) \right]}\right)^k \epsilon^{2k+1}\\
+\frac{i\omega \cos \left[ 2 \omega \left( t - t_0 \right) \right]}{1-\sin \left[ 2 \omega \left( t - t_0 \right) \right]}-\frac{2i\omega}{\cos \left[ 2 \omega \left( t - t_0 \right) \right]}\sum_{k=0}^\infty\left(-1\right)^k\left(\frac{1-\sin \left[ 2 \omega \left( t - t_0 \right) \right]}{1 +\sin \left[ 2 \omega \left( t - t_0 \right) \right]}\right)^k \epsilon^{2k} ,
\label{eq:large_w_expansion}
\end{multline}
i.e. a series forming a large-squeezing expansion for the parameter $w$.

Notice that this series is convergent only when 
\begin{equation}
\frac{1-\sin \left[ 2 \omega \left( t - t_0 \right) \right]}{1 +\sin \left[ 2 \omega \left( t - t_0 \right) \right]}\epsilon^2<1
\end{equation}
or
\begin{equation}
\sin \left[ 2 \omega \left( t - t_0 \right) \right]> -\tanh z.
\end{equation}
When taking the limit $z \to \infty$, the above inequality is satisfied for all times except a very small time period centred around the specific instant that the mode in question is a minimal uncertainty state with minimal position uncertainty. For a system containing a large number of modes, the dominant effect to the mean entanglement entropy arises from the bulk of modes, for which this inequality is satisfied.

When we study an arbitrary harmonic system with $N$ degrees of freedom, the coefficient $w$ is upgraded to the $N \times N$ matrix $W$, see equation \eqref{eq:many_overall_state}. The reduced density matrix is expressed directly in terms of the blocks of the matrix $W$ defined in \eqref{eq:many_blocks_def}. We assume that all modes are characterized by large squeezing, i.e. $z_i \gg 1$, and write the mode squeezing parameters as $z_i = z + \zeta_i$, where $z$ is the mean squeezing parameter. We focus on the case that the parameters $\zeta_i$ are subleading to the mean parameter $z$, i.e. $\zeta_i \ll z$. We may now define a single parameter $\epsilon= \exp(-z)$, so that the matrix $W$, as well as its blocks, have an expansion of the form
\begin{equation}
W= i \sum_{i=0}^\infty \epsilon^{2i} W_I^{(2i)}+\sum_{i=0}^\infty \epsilon^{2i+1} W_R^{(2i+1)} .
\end{equation}
Notice that the imaginary part of $W$ contains only even powers of $\epsilon$, whereas the real part of $W$ contains only odd powers of $\epsilon$. The leading contribution in $\epsilon$ is imaginary. In the following, we use the same notation for the expansions of the blocks of $W$ and the matrices $A$, $B$ and $C$.

The reduced density matrix is expressed in terms of the matrices $\gamma$ and $\beta$ defined in equations \eqref{eq:many_gamma_def} and \eqref{eq:many_beta_def}. The above expansion implies that $\gamma$ and $\beta$ are given by
\begin{align}
\gamma &=\frac{1}{2 \epsilon} \left( B_I^{(0)} \right)^T \left( A_R^{(1)} \right)^{-1} B_I^{(0)} \nonumber\\
&+ i\left[ C_I^{(0)} - \frac{1}{2} \left( \left( B_R^{(1)} \right)^T \left( A_R^{(1)} \right)^{-1} B_I^{(0)} + \left( B_I^{(0)} \right)^T \left( A_R^{(1)} \right)^{-1} B_R^{(1)}\right) \right] + {\mathcal O}\left( \epsilon \right), \\
\beta &= \frac{1}{2 \epsilon} \left( B_I^{(0)} \right)^T \left( A_R^{(1)} \right)^{-1} B_I^{(0)} \nonumber\\
&+ \frac{i}{2} \left( \left( B_R^{(1)} \right)^T \left( A_R^{(1)} \right)^{-1} B_I^{(0)} - \left( B_I^{(0)} \right)^T \left( A_R^{(1)} \right)^{-1} B_R^{(1)} \right) + {\mathcal O} \left( \epsilon \right).
\end{align}
In the following, we use the notation
\begin{align}
\gamma &= \frac{\gamma^{(-1)}}{\epsilon} + i \gamma^{(0)} + \ldots, \\
\beta &= \frac{\beta^{(-1)}}{\epsilon} + i \beta^{(0)} + \ldots
\end{align}
The important property of the expansions of the matrices $\gamma$ and $\beta$ is the fact that the leading contributions are identical, i.e. $\beta^{(-1)}=\gamma^{(-1)}$.

Recall that the entanglement entropy is determined by the eigenvalues of the matrix $\Xi$ that corresponds to the normalizable eigenstates of the reduced density matrix. These are identical to the eigenvalues of the matrix $M$, defined in equation \eqref{eq:many_M_def}, which are smaller than 1. This matrix can be written as
\begin{equation}
M = \begin{pmatrix}
2 \hat{\beta}^{-1} & -\hat{\beta}^{-1}\hat{\beta}^T\\
I & 0
\end{pmatrix}
= \begin{pmatrix}
 \re\left(\gamma\right)^{\frac{1}{2}} & 0\\
 0 &  \re\left(\gamma\right)^{\frac{1}{2}}
\end{pmatrix}
M^\prime
\begin{pmatrix}
 \re\left(\gamma\right)^{-\frac{1}{2}} & 0\\
 0 &  \re\left(\gamma\right)^{-\frac{1}{2}}
\end{pmatrix} ,
\label{eq:Mprime_def}
\end{equation}
where
\begin{equation}
M^\prime=\begin{pmatrix} 2 \beta^{-1} \re\left(\gamma\right)& -\beta^{-1}\beta^T\\
I & 0
\end{pmatrix} .
\label{eq:large_M_prime_def}
\end{equation}
The matrix $M^\prime$ is similar to $M$ and thus, it has the same eigenvalues. It has the expansion
\begin{equation}
M^\prime = \begin{pmatrix} 2 I - 2 i \epsilon \left( \beta^{(-1)} \right)^{-1} \beta^{(0)} + \epsilon^2 M_{11}^{(2)} & - I + 2 i \epsilon \left( \beta^{(-1)} \right)^{-1} \beta^{(0)} + \epsilon^2 M_{12}^{(2)} \\
I & 0
\end{pmatrix} + \mathcal{O} \left( \epsilon^3 \right) .
\end{equation}
It is a matter of algebra to show that
\begin{multline}
\det \left( M^\prime - \lambda I \right) = \det \bigg[ \left( 1 - \lambda \right)^2 I - 2 i \epsilon \left( 1 - \lambda \right) \left( \beta^{(-1)} \right)^{-1} \beta^{(0)} \\
- \epsilon^2 \left( \lambda M_{11}^{(2)} + M_{12}^{(2)} \right) + \mathcal{O} \left( \epsilon^3 \right) \bigg] .
\end{multline}
It follows that the eigenvalues of the matrix $M^\prime$, and thus of $M$, are of the form
\begin{equation}
\lambda = 1 - \epsilon \lambda^{(1)} + \mathcal{O} \left( \epsilon^2 \right) ,
\label{eq:large_eigen_expansion}
\end{equation}
where $\lambda^{(1)}$ solves the equation
\begin{equation}
\det \left[ \left( \lambda^{(1)} \right)^2 I - 2 i \lambda^{(1)} \left( \beta^{(-1)} \right)^{-1} \beta^{(0)} - \left( M_{11}^{(2)} + M_{12}^{(2)} \right) \right] = 0 .
\end{equation}
Bear in mind that the non-vanishing eigenvalues of the matrix $\Xi$ are as many as the number of degrees of freedom of the smaller subsystem, namely $\min\left(n,N-n\right)$. The above equation is of order $2 \min\left(n,N-n\right)$. We know that the eigenvalues of $M$ come in pairs of the form $\left( \lambda , 1 / \lambda \right)$. It can be shown that the solutions of the above equation also come in pairs of the form $\left( \lambda^{(1)} , - \lambda^{(1)} \right)$. It follows that exactly $\min\left(n,N-n\right)$ of these solutions are positive. Therefore, the eigenvalues of the matrix $\Xi$ that corresponds to the normalizable eigenstates of the reduced density matrix are $\xi_i = 1 - \epsilon\lambda^{(1)}_i$, where $i = 1 , 2 , \ldots , \min\left(n,N-n\right)$ and $\lambda^{(1)}_i > 0$.

In appendix \ref{sec:app_solvable}, the example of a squeezed state with all modes having the same squeezing parameter and vanishing phases is presented. This case is exactly solvable as this specific selection leads to a real matrix $\beta$. In this solvable example, the eigenvalues of the matrix $\Xi$ for large squeezing parameters have indeed the form of equation \eqref{eq:large_eigen_expansion}. This provides a consistency check for our large-squeezing expansion.

To leading order in $\epsilon$, the entanglement entropy reads
\begin{equation}
\begin{split}
S &= \min\left(n,N-n\right)\left(-\ln\epsilon+1\right)-\sum_{i=1}^{\min\left(n,N-n\right)}\ln \lambda^{(1)}_i+\mathcal{O}\left(\epsilon\right)\\
&= \min\left(n,N-n\right)\left(z+1\right)-\sum_{i=1}^{\min\left(n,N-n\right)}\ln\lambda^{(1)}_i+\mathcal{O}\left(\epsilon\right) .
\end{split}
\label{eq:large_squeezing_leading}
\end{equation}
This formula shows that, for large squeezing, the entanglement entropy has a linear dependence on the squeezing parameter. It also shows that the leading term, which depends linearly on the squeezing parameter, is \emph{time-independent}. The formula is in agreement with
\eqref{eq:2_mode_mean_SEE_asymtotics}. In that case, $z = \frac{z_+ + z_-}{2}$ and $N - n = 1$.

Formula \eqref{eq:large_squeezing_leading} suggests something very interesting. The leading contribution to the entanglement entropy is proportional to $\min\left(n,N-n\right)$, i.e. the number of degrees of freedom of the smaller subsystem. It follows that in a continuous harmonic system, like free scalar field theory, in the large squeezing limit, the leading contribution to the entanglement entropy is proportional to the \emph{volume} of the smaller subsystem. In other words, squeezing generates a \emph{violation} of the famous area-law property of entanglement entropy. This property apparently holds only when the system lies in a coherent state, which is a closest-to-classical state.

\section{A Field Theory Example}
\label{sec:1dft}

We are particularly interested in the application of the method that we developed in section \ref{sec:many_oscillators} to the harmonic system of scalar quantum field theory. Our interest is enhanced by the fact that the large squeezing expansion, which we developed in section \ref{sec:expansions}, suggests that the area-law property of entanglement entropy may not persist when the theory lies in a squeezed state. 

The calculation of entanglement entropy in scalar field theory in 3+1 dimensions presents several technical difficulties. The usual discretization of the degrees of freedom, which is also employed in the original calculation at the ground state \cite{srednicki}, relies on the expansion of the field in spherical harmonic moments. This is obviously a suitable choice when we desire to introduce a spherical entangling surface. However, such a choice makes it difficult to preserve a uniform density of the degrees of freedom. Therefore, an elegant regularization scheme is required, so that both area and volume terms are detectable. On the other hand, a uniform square lattice would solve this problem, but then the entangling surface would not be smooth, giving rise to additional universal terms in the entanglement entropy. For these reasons, we restrict here our attention to scalar field theory in 1+1 dimensions, where these problems do not appear, and leave the study of the 3+1 dimensional system for future work.

The Hamiltonian of a free scalar field in 1+1 dimensions reads
\begin{equation}
H=\frac{1}{2}\int dx\left[\pi^2(x)+\left(\frac{\partial}{\partial x}\varphi(x)\right)^2+\mu^2\varphi^2(x)\right].
\end{equation} 
We discretize the degrees of freedom, introducing a uniform lattice in space as
\begin{equation}
\begin{split}
x&\rightarrow j a,\\
\varphi(x)&\rightarrow \varphi_j,\\
\left.\frac{\partial\varphi(x)}{\partial x}\right|_{r=ja}&\rightarrow \frac{\varphi_{j+1}-\varphi_{j}}{a},\\
\pi(x)&\rightarrow \frac{\pi_j}{a},\\
\int_{0}^{(N+1)a}dx&\rightarrow a\sum_{j=0}^N .
\end{split}
\end{equation}
The discretized Hamiltonian that we obtain reads
\begin{equation}
H=\frac{1}{2a}\sum_{j=0}^N\left[\pi^2_j+\left(\varphi_{j+1}-\varphi_j\right)^2+\mu^2 a^2\varphi^2_j\right],
\end{equation}
where we set the boundary conditions $\varphi_0 = \varphi_{N + 1} = 0$. We introduce this kind of boundary conditions in order to avoid the existence of a zero-frequency mode. This Hamiltonian describes $N$ coupled harmonic oscillators, exactly as studied in section \ref{sec:many_oscillators}. Their Hamiltonian is of the form \eqref{eq:many_Hamiltonian}, where the couplings matrix $K$ is given by
\begin{equation}
K_{ij} = \frac{1}{a} \left[ \left(2+\mu^2 a^2\right) \delta_{i,j}-\delta_{i+1,j}-\delta_{i,j+1} \right] .
\end{equation}

In the following we use a lattice with $N = 60$. We furthermore consider the case of free massless scalar field theory in 1+1 dimensions, i.e. we assume that $\mu = 0$. In all cases we divide the system in two complementary subsystems; the first one contains the degrees of freedom $\varphi_j$ with $1 \leq j \leq n$, and the second one those with $n + 1 \leq j \leq N$. We indicate the division of the degrees of freedom in these two subsystems by the number $n$. In our calculations we set $a = 1$, which is equivalent to measuring time in units of the UV cutoff set by the lattice spacing.

We take advantage of the symmetry property of the entanglement entropy and we always trace out the larger subsystem. This is required in order to apply the method of section \ref{subsec:spectrum_M_method}. Furthermore, this speeds up the numerical calculation, since the matrices involved have the smallest possible dimension. Additionally, the required precision of the numerical calculations is achieved more easily when making this choice. 
The required precision can be high because the local nature of the couplings generates a hierarchy in the eigenvalues of the matrix $\hat{\beta}$ \cite{Katsinis:2017qzh}. As a result, an increase in the dimension of the related matrices, not only increases the volume of the required calculations, but also the required precision of them. Indicatively, our calculation, which includes at most 30$\times$30 matrices, requires about 300 significant digits in order to estimate all eigenvalues accurately.

The overall system has $N$ normal modes. The classical motion of the system when the $i$-th mode is excited is given by
\begin{equation}
\varphi_j^{(i)} \left( t \right) = A^{(i)}_j \sin \omega_i \left( t - t_0 \right) = A^{(i)} \sin \frac{i j \pi}{\left( N + 1 \right)} \sin \omega_i \left( t - t_0 \right) ,
\label{eq:1p1_single_modes}
\end{equation}
where $\omega_i$ is the frequency of the i-th mode. In the following, we squeeze one or more of these modes and study the entanglement entropy.

\subsection{Squeezing a Single Mode}

Following the example of the toy model of the two coupled oscillators, which we presented in section \ref{sec:two_oscillators}, we first study the system lying in a state where only one normal mode is squeezed; the rest are put in their ground states. In this way, the time evolution of the system is periodic, with period equal to half the period of the corresponding mode, and thus its study is more transparent.

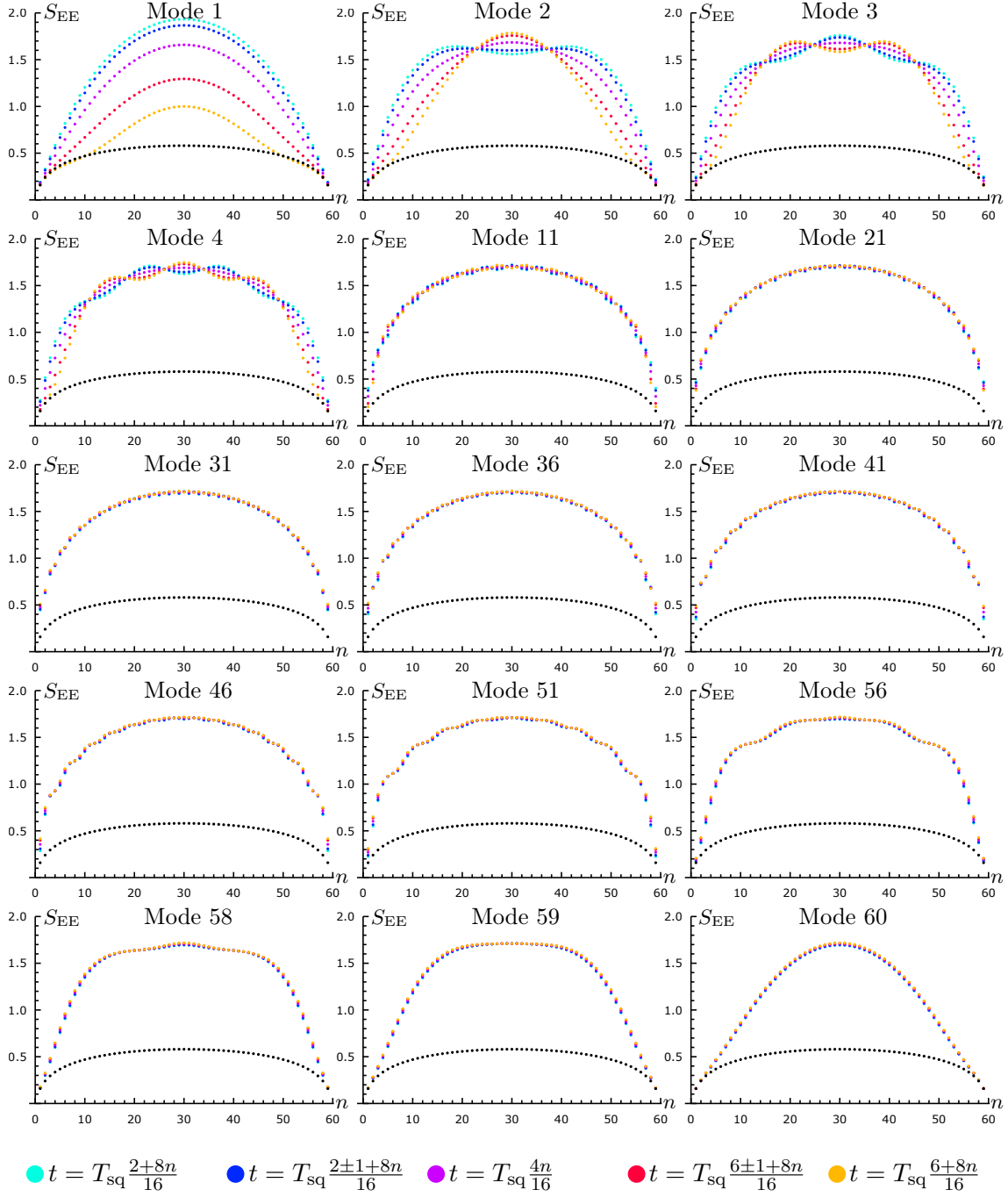
\begin{figure}[pht]
\centering
\begin{picture}(100,125)
\put(0,10){\includegraphics[angle=0,width=\textwidth]{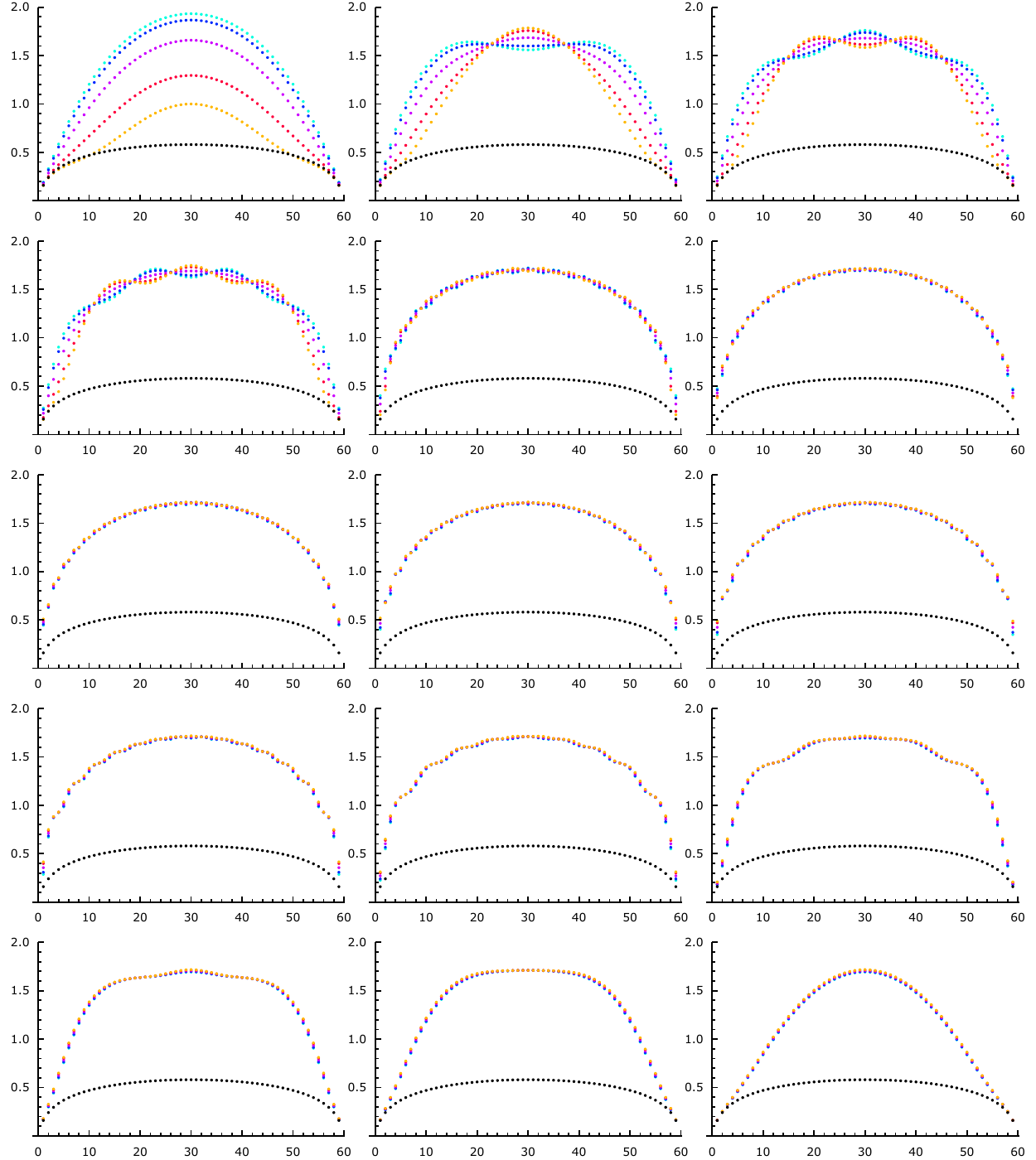}}
\put(0,0){\includegraphics[angle=0,width=\textwidth]{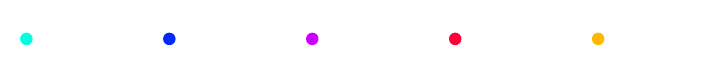}}
\put(33.5,11.75){{\footnotesize $n$}}
\put(66,11.75){{\footnotesize $n$}}
\put(98.5,11.75){{\footnotesize $n$}}
\put(33.5,34.375){{\footnotesize $n$}}
\put(66,34.375){{\footnotesize $n$}}
\put(98.5,34.375){{\footnotesize $n$}}
\put(33.5,57){{\footnotesize $n$}}
\put(66,57){{\footnotesize $n$}}
\put(98.5,57){{\footnotesize $n$}}
\put(33.5,79.675){{\footnotesize $n$}}
\put(66,79.675){{\footnotesize $n$}}
\put(98.5,79.675){{\footnotesize $n$}}
\put(33.5,102.25){{\footnotesize $n$}}
\put(66,102.25){{\footnotesize $n$}}
\put(98.5,102.25){{\footnotesize $n$}}
\put(4.5,30.25){{\footnotesize $S_{\mathrm{EE}}$}}
\put(37,30.25){{\footnotesize $S_{\mathrm{EE}}$}}
\put(69.5,30.25){{\footnotesize $S_{\mathrm{EE}}$}}
\put(4.5,52.875){{\footnotesize $S_{\mathrm{EE}}$}}
\put(37,52.875){{\footnotesize $S_{\mathrm{EE}}$}}
\put(69.5,52.875){{\footnotesize $S_{\mathrm{EE}}$}}
\put(4.5,75.5){{\footnotesize $S_{\mathrm{EE}}$}}
\put(37,75.5){{\footnotesize $S_{\mathrm{EE}}$}}
\put(69.5,75.5){{\footnotesize $S_{\mathrm{EE}}$}}
\put(4.5,98.125){{\footnotesize $S_{\mathrm{EE}}$}}
\put(37,98.125){{\footnotesize $S_{\mathrm{EE}}$}}
\put(69.5,98.125){{\footnotesize $S_{\mathrm{EE}}$}}
\put(4.5,120.75){{\footnotesize $S_{\mathrm{EE}}$}}
\put(37,120.75){{\footnotesize $S_{\mathrm{EE}}$}}
\put(69.5,120.75){{\footnotesize $S_{\mathrm{EE}}$}}
\put(14.5,30.25){{\small Mode 58}}
\put(47,30.25){{\small Mode 59}}
\put(79.5,30.25){{\small Mode 60}}
\put(14.5,52.875){{\small Mode 46}}
\put(47,52.875){{\small Mode 51}}
\put(79.5,52.875){{\small Mode 56}}
\put(14.5,75.5){{\small Mode 31}}
\put(47,75.5){{\small Mode 36}}
\put(79.5,75.5){{\small Mode 41}}
\put(14.75,98.125){{\small Mode 4}}
\put(47,98.125){{\small Mode 11}}
\put(79.5,98.125){{\small Mode 21}}
\put(14.75,120.75){{\small Mode 1}}
\put(47.25,120.75){{\small Mode 2}}
\put(79.75,120.75){{\small Mode 3}}
\put(5,4.5){$t = T_{\mathrm{sq}} \frac{2 + 8n}{16}$}
\put(24.875,4.5){$t = T_{\mathrm{sq}} \frac{2 \pm 1 + 8n}{16}$}
\put(44.75,4.5){$t = T_{\mathrm{sq}} \frac{4n}{16}$}
\put(64.625,4.5){$t = T_{\mathrm{sq}} \frac{6 \pm 1 + 8n}{16}$}
\put(84.5,4.5){$t = T_{\mathrm{sq}} \frac{6 + 8n}{16}$}
\end{picture}
\caption{The entanglement entropy as a function of $n$ for various times when only a single mode has been squeezed with squeezing parameter $z=3$. The black dots correspond to the entanglement entropy at the ground state of the system.}
\label{fig:1p1_single}
\end{figure}

Figure \ref{fig:1p1_single} shows the entanglement entropy as a function of $n$ for various times. Several cases are presented, which differ with respect to the mode that has been squeezed. The squeezing parameter is always the same.
We observe the following:
\begin{itemize}
\item Squeezing generally increases the entanglement entropy in comparison to that in the ground state. However, there are instants when the entanglement entropy is smaller than that of the ground state for specific values of $n$. This is more easily visible in the case that the first mode has been squeezed (top left panel of figure \ref{fig:1p1_single}).
\item The increase of the entanglement entropy by squeezing a single mode does not depend strongly on which mode is squeezed, as long as the squeezing parameter is the same.
\item The entanglement entropy is oscillating in time with a period half that of the corresponding mode, as expected. 
\item The oscillation of entanglement entropy with time is generally more intense when a mode with a smaller index has been squeezed.
\item The pattern of the amplitude of the oscillation of the entanglement entropy as $n$ varies is interesting. It appears that this pattern is strongly related to the form of the squeezed normal mode.
\begin{itemize}
\item The pattern has the form of a stationary wave with several nodes. There are specific values of $n$, where the amplitude of the oscillation of the entanglement entropy vanishes.
\item The nodes of the entanglement entropy oscillation are twice as many as the nodes of the squeezed mode.
\item The nodes appear at $n$ where 
\begin{equation}
\cos \frac{2 i n \pi}{\left( N + 1 \right)} = 0 ,
\end{equation}
where $i$ is the index of the squeezed mode. The nodes appear at locations where the classical amplitude squared of the oscillation due to the squeezed mode is half of the maximum. Nodes appear at all such locations except for the first and last one. In other words, the $k$-th node is located at position
\begin{equation}
n_k = \frac{N + 1}{4 i} \left( 2 k + 1 \right) , \quad k = 1 , 2 , \ldots , 2 \left( i - 1 \right) .
\end{equation}
\item The existence of the nodes justifies why the time dependence of the entanglement entropy is suppressed when the squeezed mode is higher.
\end{itemize}
\end{itemize}

The dependence of the entanglement entropy on the shape of the squeezed mode is visible on the mean entanglement entropy as well. 
\begin{figure}[ht]
\centering
\begin{picture}(88,48)
\put(0,0){\includegraphics[angle=0,width=0.8\textwidth]{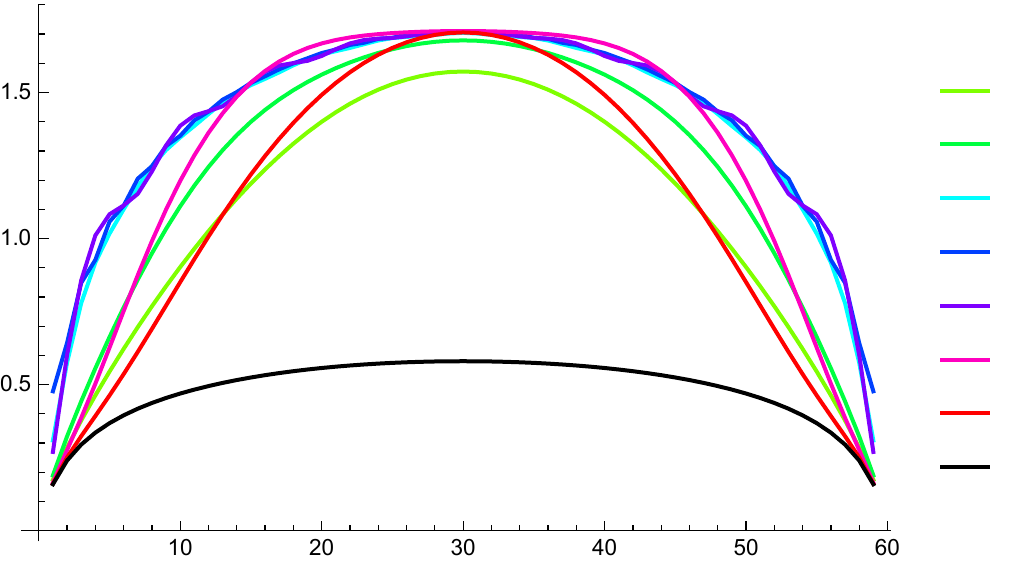}}
\put(2,44.25){$\bar{S}_{\mathrm{EE}}$}
\put(70,2){$n$}
\put(72,5){\line(1,0){15.25}}
\put(72,5){\line(0,1){40}}
\put(72,45){\line(1,0){15.25}}
\put(87.25,5){\line(0,1){40}}
\put(77,6.75){None}
\put(77,10.875){Mode 60}
\put(77,15){Mode 59}
\put(77,19.125){Mode 51}
\put(77,23.25){Mode 31}
\put(77,27.375){Mode 11}
\put(77,31.5){Mode 2}
\put(77,35.625){Mode 1}
\put(76.75,39.75){mode}
\put(74.75,42.125){Squeezed}
\end{picture}
\caption{The mean entanglement entropy as a function of $n$ when squeezing a single mode with squeezing parameter $z=3$.}
\label{fig:1p1_mean_mode}
\end{figure}
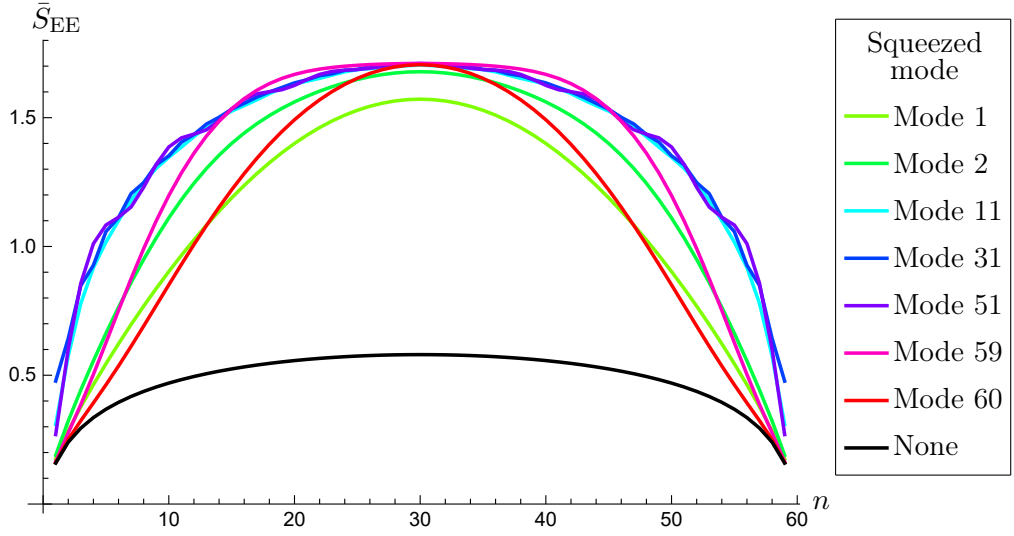
Figure \ref{fig:1p1_mean_mode} depicts the mean entanglement entropy as a function of $n$ for various squeezed modes with the same squeezing parameter. The mean entanglement entropy is always larger than that in the ground state of the system, unlike the entanglement entropy at a given time. We also observe that the mean entanglement entropy is about the same for the vast majority of the modes. Significant differences appear only for the first and last modes. Notice that the curves corresponding to the 11th mode, as well as the 51st mode are almost identical. All intermediate ones are also almost identical, like the one corresponding to the 31st mode, which is also depicted. This can be attributed to the relation between the pattern of entanglement entropy and the shape of the squeezed normal mode that we pointed out above. The classical amplitudes of oscillation of the first and last modes have a strong pattern: there are regions with large and regions with small amplitudes. On the contrary, the classical amplitudes of oscillations for most intermediate modes are more dispersed; thus the similar pattern of the mean entanglement entropy.

\begin{figure}[ht]
\centering
\begin{picture}(87,49)
\put(0,0){\includegraphics[angle=0,width=0.8\textwidth]{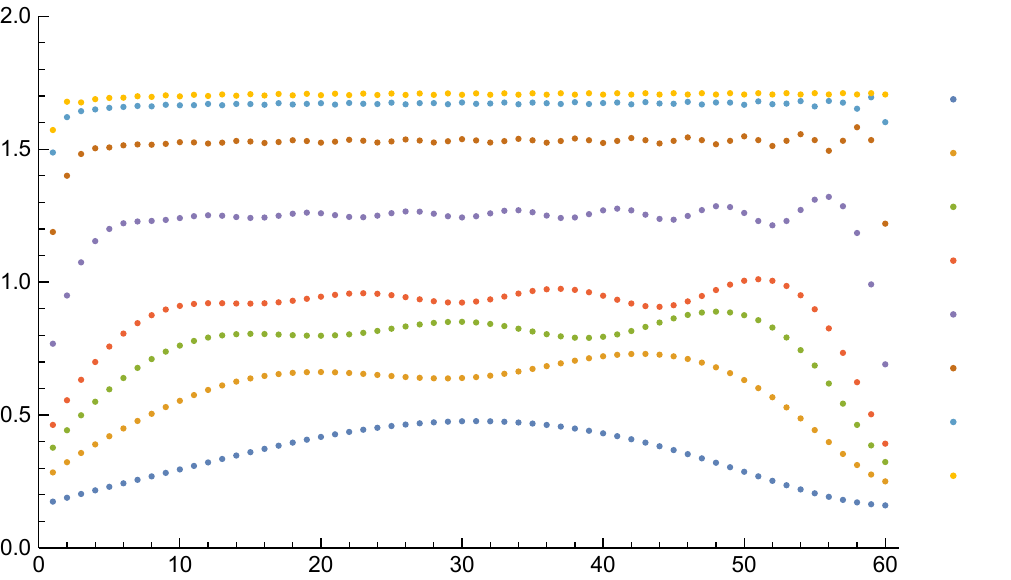}}
\put(2,45.75){$\bar{S}_{\mathrm{EE}}$}
\put(71,4.75){number of}
\put(71.75,2.375){squeezed}
\put(73.75,0){mode}
\put(72.5,6.75){\line(1,0){13.5}}
\put(72.5,6.75){\line(0,1){40}}
\put(72.5,46.75){\line(1,0){13.5}}
\put(86,6.75){\line(0,1){40}}
\put(76.25,7.625){$n = 30$}
\put(76.25,11.875){$n = 23$}
\put(76.25,16.125){$n = 15$}
\put(76.25,20.375){$n = 8$}
\put(76.25,24.625){$n = 4$}
\put(76.25,28.875){$n = 3$}
\put(76.25,33.125){$n = 2$}
\put(76.25,37.375){$n = 1$}
\put(76.5,41.625){node}
\put(73,44){Entangling}
\end{picture}
\caption{The mean entanglement entropy as a function of the index of the squeezed mode for various divisions of the system to two subsystems indicated by the integer $n$. The squeezed mode has always squeezing parameter $z=3$.}
\label{fig:1p1_mean_cite}
\end{figure}
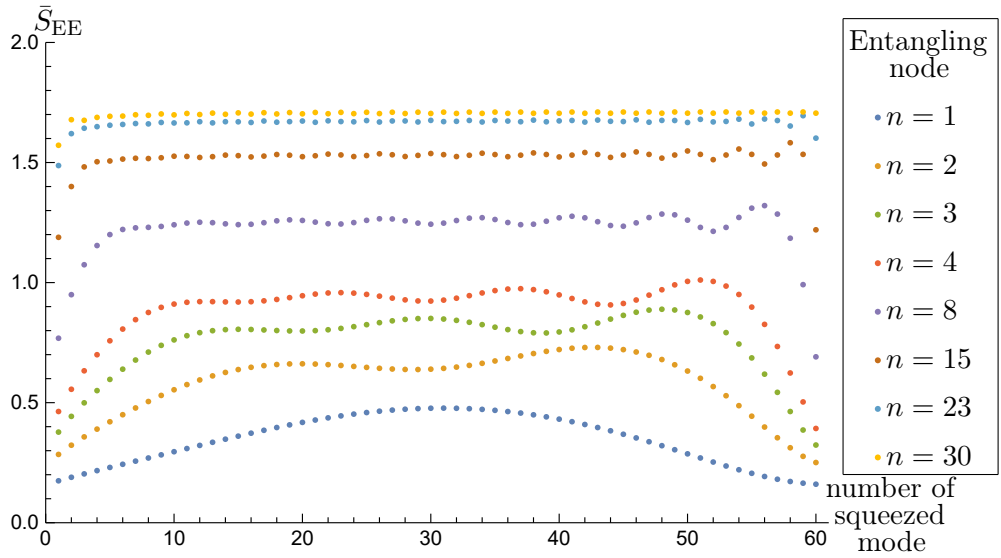

The relation between entanglement entropy and the shape of the squeezed mode is also supported by figure \ref{fig:1p1_mean_cite}, which depicts the mean entanglement entropy as a function of the index of the squeezed mode for several divisions of the system in two subsystems, indicated by the integer $n$.
Indeed the curves vary slowly, especially in the intermediate region. The mean entanglement entropy does not depend strongly on which mode has been squeezed.

The mean entanglement entropy as a function of the order of the squeezed mode $i$ for given $n$ presents as many maxima as $n$. These maxima are almost equidistant. For example, the mean entanglement entropy for $n=1$ has a single maximum around $i = 30$. This may be attributed to the fact that the amplitude of oscillation $A^{(i)}_1$ of the first degree of freedom has this kind of dependence on $i$. Equation \eqref{eq:1p1_single_modes} implies that
\begin{equation}
\left( A^{(i)}_1 \right)^2 \sim \sin^2 \frac{i \pi}{\left( N + 1 \right)} ,
\end{equation}
i.e. indeed $A^{(i)}_1$ has a single maximum around $i = N / 2$.
Similarly we can show that the amplitude of oscillation of the $n$-th degree of freedom has $n$ almost equidistant maxima, since
\begin{equation}
\left( A^{(i)}_n \right)^2 \sim \sin^2 \frac{i n \pi}{\left( N + 1 \right)} ,
\end{equation}
which are as many as the maxima of the mean entanglement entropy.

\begin{figure}[ht]
\centering
\begin{picture}(100,59)
\put(0,0){\includegraphics[angle=0,width=0.89\textwidth]{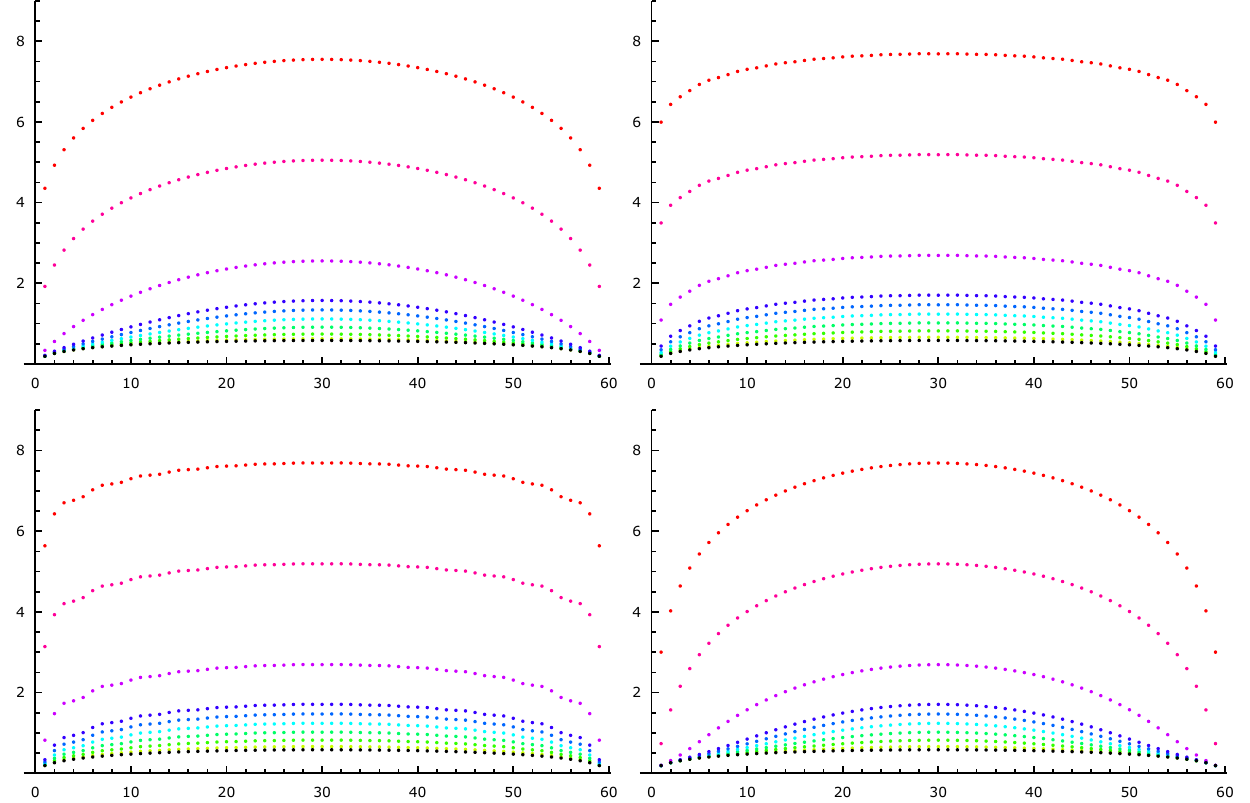}}
\put(89,10){\includegraphics[angle=0,width=0.1\textwidth]{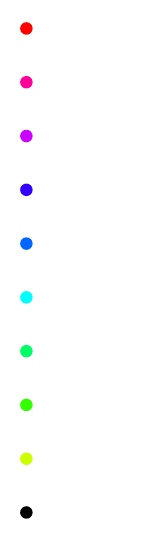}}
\put(3,27.25){{\footnotesize $\bar{S}_{\mathrm{EE}}$}}
\put(46.875,27.25){{\footnotesize $\bar{S}_{\mathrm{EE}}$}}
\put(3,56.375){{\footnotesize $\bar{S}_{\mathrm{EE}}$}}
\put(46.875,56.375){{\footnotesize $\bar{S}_{\mathrm{EE}}$}}
\put(43.75,1.75){{\footnotesize $n$}}
\put(87.625,1.75){{\footnotesize $n$}}
\put(43.75,30.875){{\footnotesize $n$}}
\put(87.625,30.875){{\footnotesize $n$}}
\put(18.5,27.25){{\small Mode 46}}
\put(62.375,27.25){{\small Mode 60}}
\put(19,56.375){{\small Mode 1}}
\put(62.375,56.375){{\small Mode 16}}
\put(90,11){\line(1,0){10}}
\put(90,11){\line(0,1){36.5}}
\put(90,47.5){\line(1,0){10}}
\put(100,11){\line(0,1){36.5}}
\put(91.75,12){$z = 0$}
\put(91.75,15.625){$z = 0.5$}
\put(91.75,19.375){$z = 1$}
\put(91.75,23){$z = 1.5$}
\put(91.75,26.75){$z = 2$}
\put(91.75,30.375){$z = 2.5$}
\put(91.75,34.125){$z = 3$}
\put(91.75,37.75){$z = 5$}
\put(91.75,41.5){$z = 10$}
\put(91.75,45.125){$z = 15$}
\end{picture}
\caption{The mean entanglement entropy as a function of $n$ for various values of the squeezing parameter $z$ when a single mode has been squeezed.}
\label{fig:1p1_mean_z}
\end{figure}
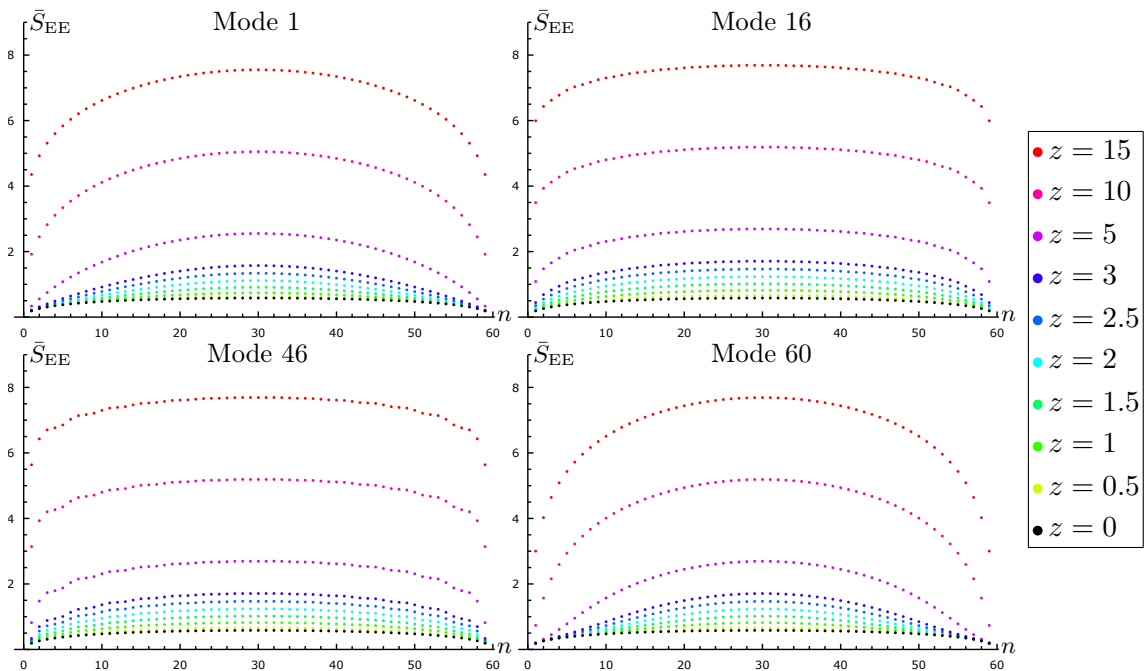

We studied the dependence of the entanglement entropy on time and on the shape of the squeezed mode. It remains to study its dependence on the squeezing parameter $z$. Figure \ref{fig:1p1_mean_z} depicts the mean entanglement entropy as a function of $n$ for various values of $z$ and for a specific choice of the squeezed mode.
We generally observe an increase of the mean entanglement entropy as $z$ increases. For small values of $z$, this increase changes the shape of the curve as a function of $n$. Above some critical value of $z$, a further increase appears to move the curve as a whole. Furthermore, the increase of entanglement entropy appears to be proportional to the increase of $z$.

\begin{figure}[ht]
\centering
\begin{picture}(100,59)
\put(0,0){\includegraphics[angle=0,width=0.89\textwidth]{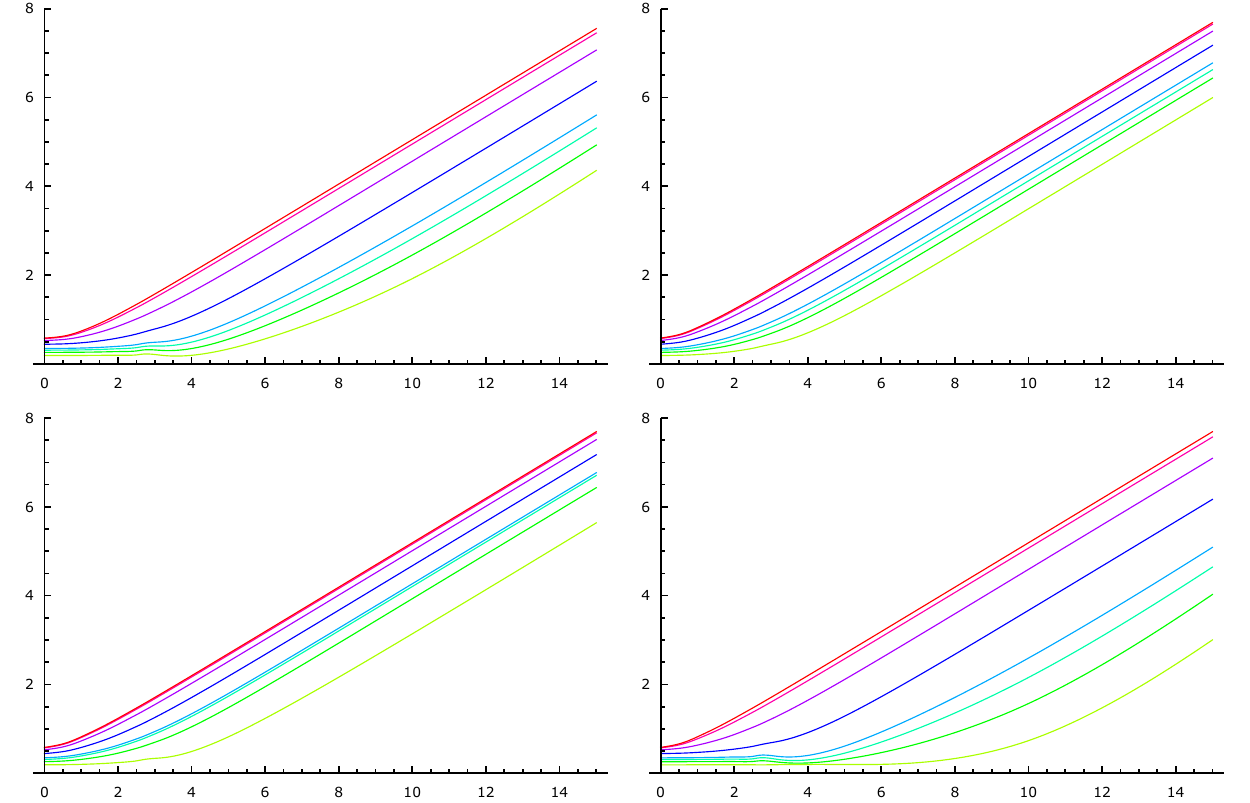}}
\put(89,10){\includegraphics[angle=0,width=0.1\textwidth]{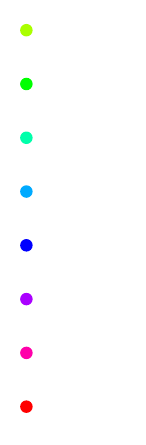}}
\put(3.625,26.75){{\footnotesize $\bar{S}_{\mathrm{EE}}$}}
\put(47.5,26.75){{\footnotesize $\bar{S}_{\mathrm{EE}}$}}
\put(3.625,55.875){{\footnotesize $\bar{S}_{\mathrm{EE}}$}}
\put(47.5,55.875){{\footnotesize $\bar{S}_{\mathrm{EE}}$}}
\put(43.625,1.75){{\footnotesize $z$}}
\put(87.5,1.75){{\footnotesize $z$}}
\put(43.625,30.875){{\footnotesize $z$}}
\put(87.5,30.875){{\footnotesize $z$}}
\put(18.5,27.25){{\small Mode 46}}
\put(62.375,27.25){{\small Mode 60}}
\put(19,56.375){{\small Mode 1}}
\put(62.375,56.375){{\small Mode 16}}
\put(90,11){\line(1,0){10}}
\put(90,11){\line(0,1){29}}
\put(90,40){\line(1,0){10}}
\put(100,11){\line(0,1){29}}
\put(91.75,12){$n = 30$}
\put(91.75,15.625){$n = 23$}
\put(91.75,19.375){$n = 15$}
\put(91.75,23){$n = 8$}
\put(91.75,26.75){$n = 4$}
\put(91.75,30.375){$n = 3$}
\put(91.75,34.125){$n = 2$}
\put(91.75,37.75){$n = 1$}
\end{picture}
\caption{The mean entanglement entropy as a function of the squeezing parameter for various divisions of the system to two subsystems indicated by the integer $n$ when a single mode has been squeezed.}
\label{fig:1p1_mean_z_fixed_node}
\end{figure}
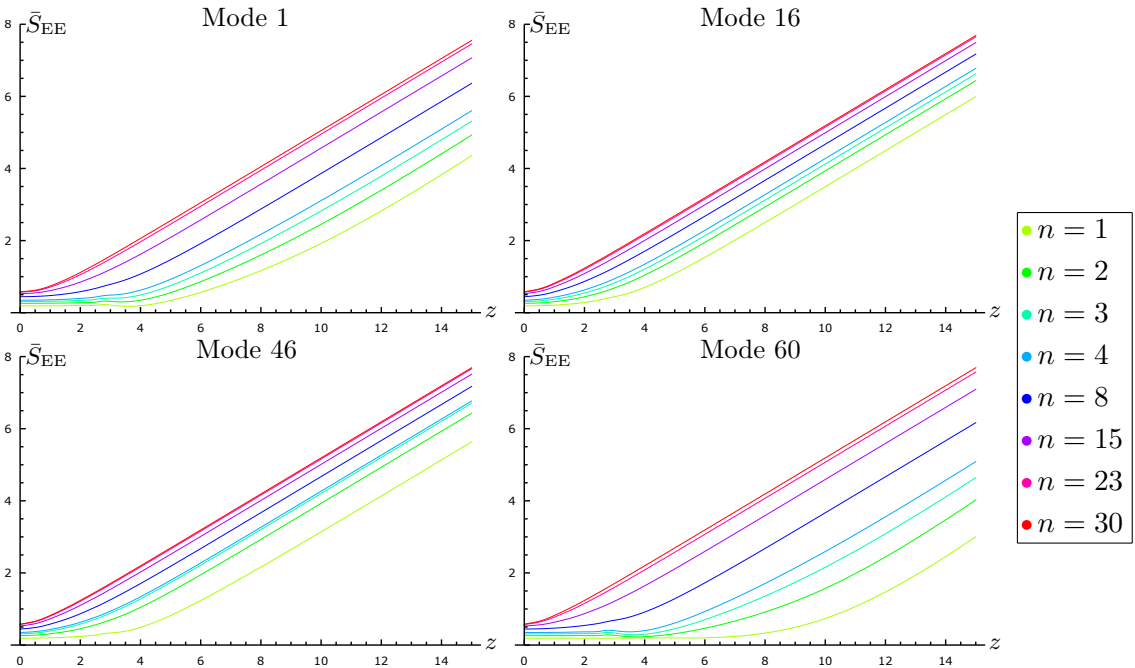

In order to clarify this behaviour, we depict in figure \ref{fig:1p1_mean_z_fixed_node} the mean entanglement entropy as a function of $z$ for fixed values of $n$.
For small $z$ values, the increase of entanglement entropy is quadratic in $z$ and depends on $n$. However, after some critical $z$, this dependence becomes linear and independent of $n$; all curves have the same slope asymptotically for large $z$. Furthermore, this slope does not depend on which normal mode is squeezed. In other words, for large $z$, when only one mode is squeezed
\begin{equation}
S_{\mathrm{EE}} = c z + \mathcal{O} \left( z^0 \right) ,
\end{equation}
where $c$ depends neither on $n$ nor on the order of the squeezed mode.

Our discretized version of 1+1 scalar field theory differs from the continuum field theory in three ways: we have introduced a UV cutoff, an IR cutoff and Dirichlet boundary conditions for the normal modes. As we discussed above, it appears that the dependence of the mean entanglement entropy on which mode is squeezed is due to fact that modes are stationary waves, and thus each mode excites the various degrees of freedom with different amplitudes. We expect that this also happens in the continuum limit if we preserve Dirichlet boundary conditions at some specific point, i.e. if we continue defining the theory in a finite segment or in the infinite half-line. If Dirichlet conditions are abandoned, the normal modes would correspond to travelling waves, which are characterized by identical amplitude of oscillation for all degrees of freedom. Therefore, in free scalar field theory defined on the continuous infinite line we expect that the mean entanglement entropy would not depend at all on which mode is squeezed. This should also be a property of the discretized system if periodic boundary conditions are adopted. In this case a mass should be introduced so that a zero-frequency mode is avoided. This investigation is beyond the scope of this work.

\subsection{Squeezing All Modes}

The large squeezing expansion that we presented in section \ref{sec:expansions} suggests that for large squeezing we should expect that entanglement entropy is dominated by a volume term proportional to the mean squeezing parameter. In the previous subsection, we studied the system of discretized scalar field theory in 1+1 dimensions in a state where only a single mode lies in a squeezed state, whereas all others lie in their ground state. For large squeezing the entanglement entropy is dominated by a term that is proportional to the squeezing parameter. However, this is not a volume term, but rather a constant term. This is not contradictory to the large-squeezing expansion, which requires that the deviation of the squeezing parameter of each mode from the mean is small, while we assumed that only one mode is squeezed. 

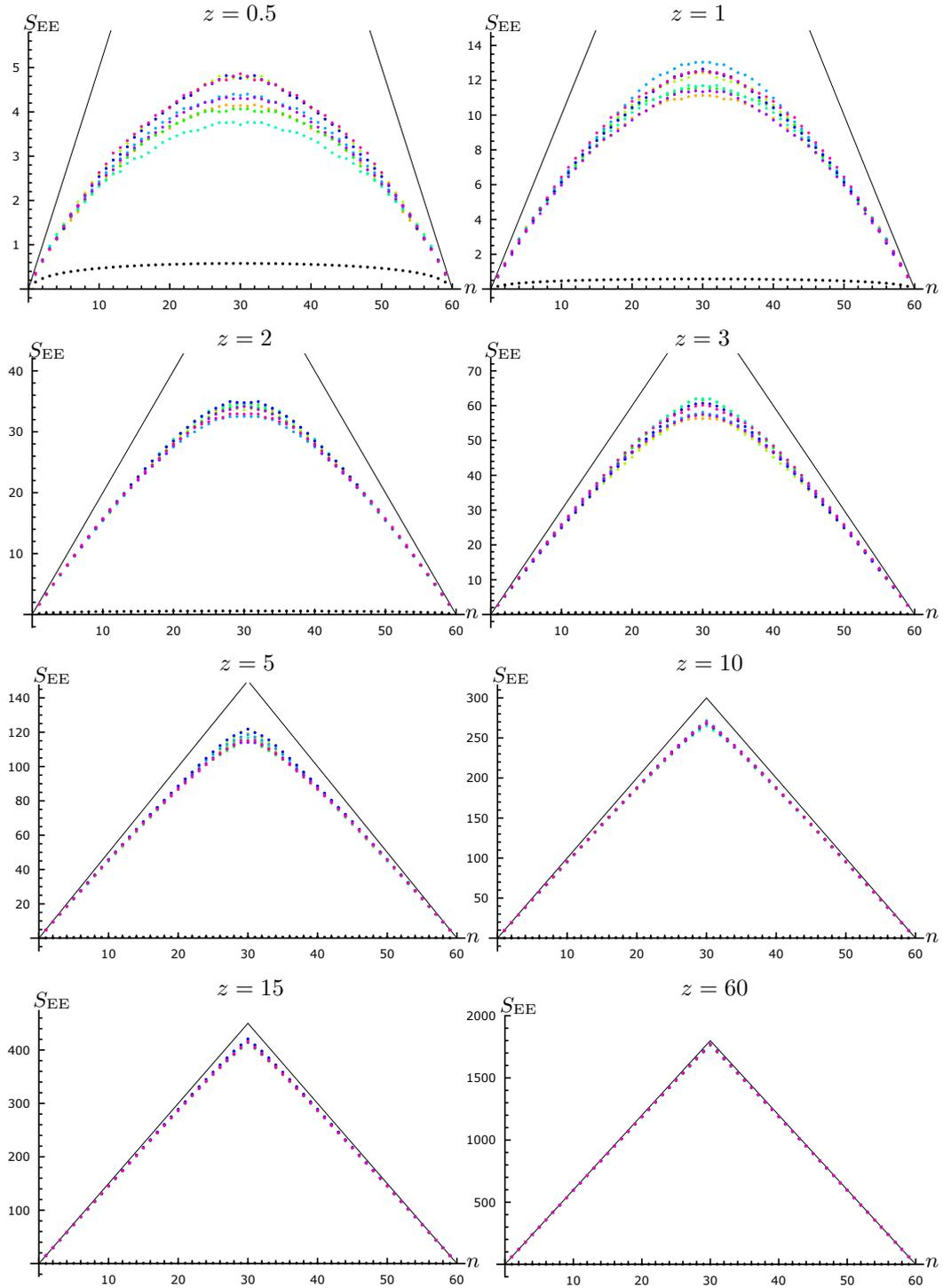
\begin{figure}[pht]
\centering
\begin{picture}(92,128)
\put(0,0){\includegraphics[angle=0,width=0.9\textwidth]{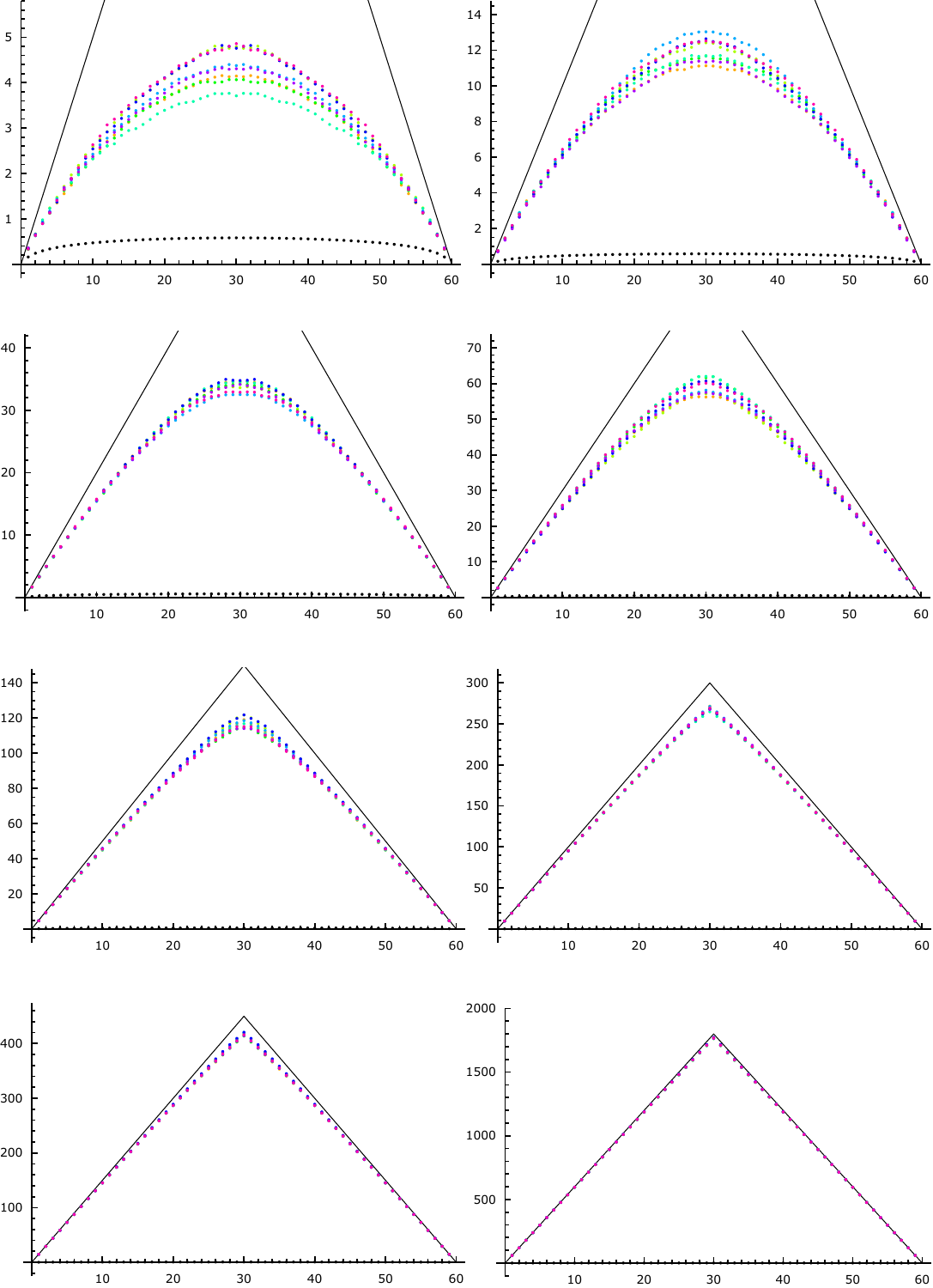}}
\put(45,2.125){{\footnotesize $n$}}
\put(90,2.125){{\footnotesize $n$}}
\put(45,34.25){{\footnotesize $n$}}
\put(90,34.25){{\footnotesize $n$}}
\put(44.875,66.25){{\footnotesize $n$}}
\put(90,66.25){{\footnotesize $n$}}
\put(44.75,98.375){{\footnotesize $n$}}
\put(90,98.375){{\footnotesize $n$}}
\put(2.5,27.875){{\footnotesize $S_{\mathrm{EE}}$}}
\put(48.25,27.375){{\footnotesize $S_{\mathrm{EE}}$}}
\put(2.5,60){{\footnotesize $S_{\mathrm{EE}}$}}
\put(47.5,60){{\footnotesize $S_{\mathrm{EE}}$}}
\put(2,92.125){{\footnotesize $S_{\mathrm{EE}}$}}
\put(46.75,92.125){{\footnotesize $S_{\mathrm{EE}}$}}
\put(1.5,124.5){{\footnotesize $S_{\mathrm{EE}}$}}
\put(46.75,124.5){{\footnotesize $S_{\mathrm{EE}}$}}
\put(20.5,29){{\small $z=15$}}
\put(66,29){{\small $z=60$}}
\put(20.75,61.125){{\small$z=5$}}
\put(65.5,61.125){{\small$z=10$}}
\put(20.5,93.25){{\small$z=2$}}
\put(65.375,93.25){{\small$z=3$}}
\put(19.125,125.375){{\small$z=0.5$}}
\put(65,125.375){{\small$z=1$}}
\end{picture}
\caption{The entanglement entropy as a function of $n$ for various random instants when all modes have been squeezed with the same squeezing parameter. The continuous black line corresponds to the large squeezing approximation given by equation \eqref{sstrong}.}
\label{fig:1p1_all}
\end{figure}

In order to understand the effect of strong squeezing, 
in this subsection we study a system in which all modes lie in a squeezed state with the same squeezing parameter.
The time evolution of the overall system is much more complicated than in the case when only a single mode has been squeezed. In general, the evolution is not periodic. In figure \ref{fig:1p1_all} we show the entanglement entropy at random instants, as a function of the number $n$ that determines the division of the system in two subsystems. We present several cases, which differ in the value of the common squeezing parameter. 
The initial phases for each normal mode are unimportant. As they change at different rates, even if they are selected to be initially equal, in due time they are more or less random. For this reason, the instants displayed in figure \ref{fig:1p1_all} are completely random.

In this figure the black dots depict the entanglement entropy when all modes lie in their ground state. The coloured dots depict the entanglement entropy at the state under study, i.e. 
when all modes lie in a squeezed state with the same squeezing parameter. Different colors correspond to different instants. Finally, the continuous black line depicts the leading term of the large-squeezing approximation for the entanglement entropy, given by equation \eqref{eq:large_squeezing_leading}, i.e. 
\begin{equation}
S_{\mathrm{EE}} \simeq z \min \left( n , N - n \right) .
\label{sstrong}
\end{equation}
Notice that the latter is time-independent and proportional to the volume of the smaller subsystem.

We observe the following:
\begin{itemize}
\item The entanglement entropy generally increases as the squeezing parameter increases.
\item The entanglement entropy approaches the large-squeezing approximation formula as the squeezing parameter increases.
\item The variations of entanglement entropy with time decrease in comparison to the mean entanglement entropy as the squeezing parameter increases. This is in line with the fact that the leading term of the large-squeezing approximation is time-independent.
\end{itemize}

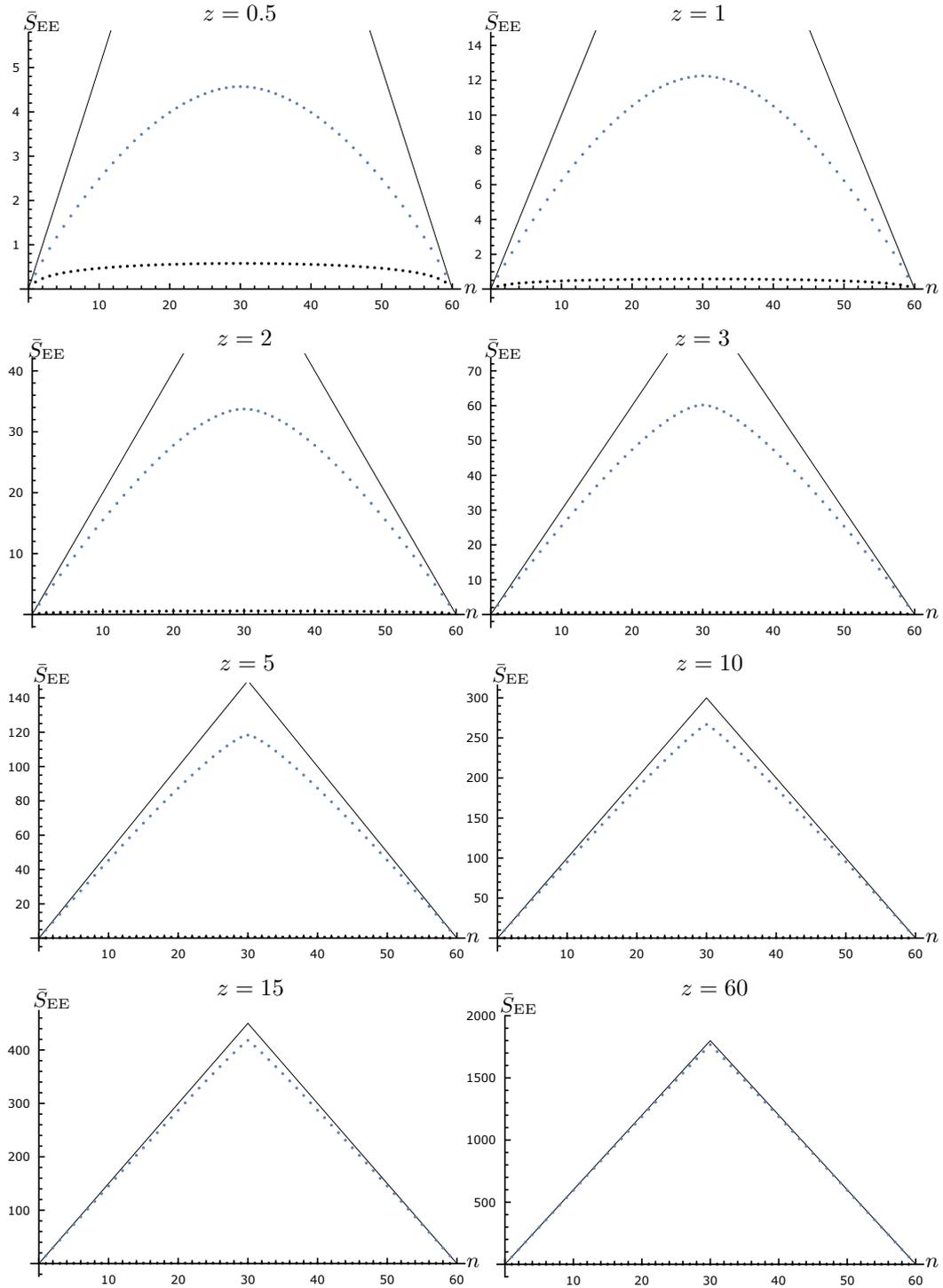
\begin{figure}[pht]
\centering
\begin{picture}(92,128)
\put(0,0){\includegraphics[angle=0,width=0.9\textwidth]{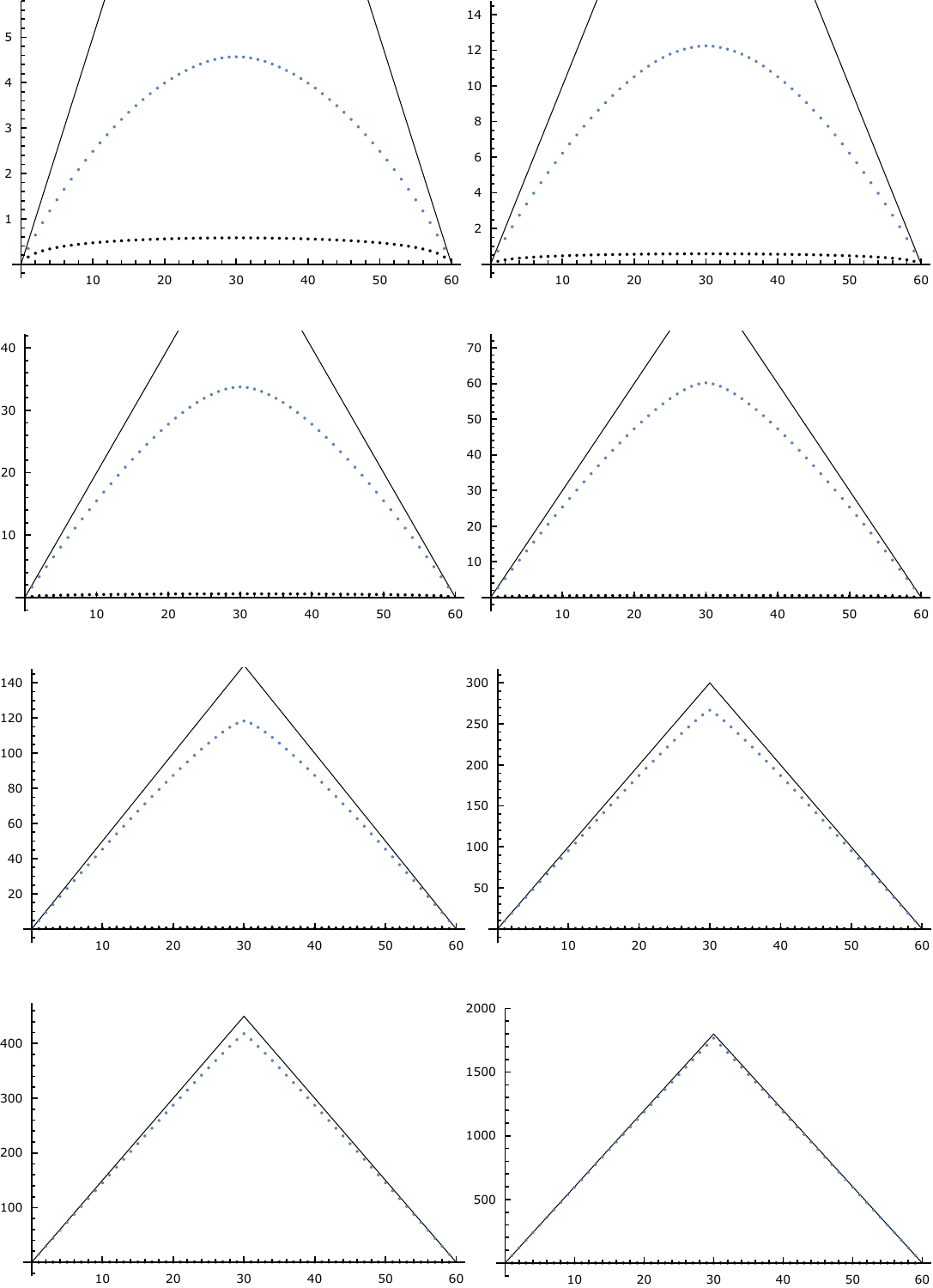}}
\put(45,2.125){{\footnotesize $n$}}
\put(90,2.125){{\footnotesize $n$}}
\put(45,34.25){{\footnotesize $n$}}
\put(90,34.25){{\footnotesize $n$}}
\put(44.875,66.25){{\footnotesize $n$}}
\put(90,66.25){{\footnotesize $n$}}
\put(44.75,98.375){{\footnotesize $n$}}
\put(90,98.375){{\footnotesize $n$}}
\put(2.5,27.875){{\footnotesize $\bar{S}_{\mathrm{EE}}$}}
\put(48.25,27.375){{\footnotesize $\bar{S}_{\mathrm{EE}}$}}
\put(2.5,60){{\footnotesize $\bar{S}_{\mathrm{EE}}$}}
\put(47.5,60){{\footnotesize $\bar{S}_{\mathrm{EE}}$}}
\put(2,92.125){{\footnotesize $\bar{S}_{\mathrm{EE}}$}}
\put(46.75,92.125){{\footnotesize $\bar{S}_{\mathrm{EE}}$}}
\put(1.5,124.5){{\footnotesize $\bar{S}_{\mathrm{EE}}$}}
\put(46.75,124.5){{\footnotesize $\bar{S}_{\mathrm{EE}}$}}
\put(20.5,29){{\small $z=15$}}
\put(66,29){{\small $z=60$}}
\put(20.75,61.125){{\small$z=5$}}
\put(65.5,61.125){{\small$z=10$}}
\put(20.5,93.25){{\small$z=2$}}
\put(65.375,93.25){{\small$z=3$}}
\put(19.125,125.375){{\small$z=0.5$}}
\put(65,125.375){{\small$z=1$}}
\end{picture}
\caption{The mean entanglement entropy as a function of $n$ when all modes have been squeezed with the same squeezing parameter. The continuous black line corresponds to the large squeezing approximation given by equation \eqref{sstrong}.}
\label{fig:1p1_all_mean}
\end{figure}

Figure \ref{fig:1p1_all_mean} depicts the mean entanglement entropy as a function of $n$ for various squeezing parameters. The mean has been calculated as the average of 200 random times.
The blue dots depict the mean entanglement entropy. The black dots and the black continuous line depict the entanglement entropy in the ground state and the leading term of the large-squeezing expansion, as in figure \ref{fig:1p1_all}.
It is evident that the mean entanglement entropy is dominated by a 
\emph{time-independent volume term} when the squeezing parameter $z$ is large. This volume term is proportional to $z$.

In the continuum limit $a \to 0$, $N a \to L$, i.e. the continuum limit with an IR cutoff equal to $1 / L$, the large-squeezing expansion suggests that the dominant term of entanglement entropy is
\begin{equation}
S_{\mathrm{EE}} = \frac{\min \left( r , L - r \right)}{a} z + \mathcal{O} \left( z^0 \right) ,
\end{equation}
where $r$ is the limit of the product $n a$, i.e. the length of the first subsystem. The leading volume term is UV divergent. If we remove the IR cutoff, defining the theory on the infinite half-line, then 
\begin{equation}
S_{\mathrm{EE}} = \frac{r}{a} z + \mathcal{O} \left( z^0 \right) .
\end{equation}
In this scenario the one subsystem has length $r$ and is attached to the end of the infinite half-line.

The dominant term is not only proportional to the volume of the subsystem, but also time-independent, even though the state of the system has non-trivial time dependence. In order to clarify that the variations of entanglement entropy with time are reduced in comparison to the mean entanglement entropy as the squeezing parameter increases, we calculated the standard deviation of the entanglement entropy at $n = 30$ for 200 random times, as a function of $z$. We found that the standard deviation approaches a finite limit as $z$ increases, whereas the mean entanglement entropy increases linearly with $z$.

\section{Discussion}
\label{sec:discussion}

In this work we studied entanglement in coupled harmonic systems lying in squeezed states.
We managed to reduce the problem of the specification of the eigenvalues of the reduced density matrix to a linear eigenproblem, in exactly the same fashion as in the case of a harmonic system lying in its ground state \cite{srednicki} or a coherent state \cite{Katsinis:2022fxu}.

In an interesting way, the spectrum of the reduced density matrix conserves the same structure as for the ground state or a coherent state. Let the number of degrees of freedom of the considered subsystem be $n$. Then, the spectrum of the reduced density matrix describing 
the subsystem is indistinguishable from the spectrum of an effective harmonic system with $n$ degrees of freedom, if each of its normal modes is lying in a thermal state at an appropriate temperature. Notice that this is not a thermal state of the whole effective system; each of 
its normal modes has a different temperature.

However, there are several important differences compared to the ground or coherent state cases. First, the eigenstates of the reduced density matrix have suffered a non-trivial deformation. Although they can be organized in a similar fashion to the Fock space of an effective harmonic system, there is no real linear combination of the physical degrees of freedom of the considered subsystem which can be identified as a normal coordinate. This is due to the fact that the creation operators associated with this Fock space are a linear combination of positions and momenta which are not conjugate to each other (see appendix \ref{subsubsec:spectrum_creation_annihilation}). As a matter of fact it is quite difficult to derive explicit expressions in coordinate representation for the whole set of eigenstates of the reduced density matrix, although it is clear how to construct them iteratively. In other words, in the case of the ground or coherent state of the overall system, the reduced density matrix is separable and it can be written as the tensor product of density matrices that describe a linear combination of the original degrees of freedom each. In the case of the squeezed state, the reduced density matrix is separable, but it is written as a tensor product of matrices that cannot be assigned to any real combination of the original degrees of freedom.

Second, the time evolution of the reduced density matrix, unlike the ground or coherent state case, is non-unitary.

Finally, the spectrum of the reduced density matrix, and, thus, the entanglement entropy is
in general time-dependent. Although at a given instant the entanglement entropy may be smaller than that at the ground state or a coherent state of the system, the examples that we have investigated suggest that the mean entanglement entropy increases with squeezing. 
For states in which all the modes are strongly squeezed, 
the mean entanglement entropy appears to be time-independent and proportional to the 
absolute value of the squeezing parameter and the number of degrees of freedom of the smaller subsystem (see equation (\ref{sstrong})). A large-squeezing expansion supports this conclusion
(see section \ref{sec:expansions}).

Page has shown that the entanglement entropy of an arbitrary quantum state is close to the maximal possible entanglement entropy \cite{Page:1993df}. For systems where the local degrees of freedom have a finite-dimensional Hilbert space, this bound on entanglement entropy as a function of the number of degrees of freedom of the considered subsystem has a characteristic concave form. It vanishes when the subsystem is null or coincides with the whole system, 
while it is maximal when the subsystem contains half of the degrees of freedom of the overall system. In the limit that the total system contains an infinite number of degrees of freedom, this function tends to the union of two linear segments. If $n$ and $N$ denote the number of degrees of freedom of the subsystem and overall system respectively, this curve is approximately
\begin{equation}
S_{\max} \sim \min \left( n , N - n \right) ,
\end{equation}
where the proportionality constant depends on the dimensionality of the Hilbert space of the local degrees of freedom. This curve is of great importance. It has been connected to the Page curve followed by the entropy of black hole radiation \cite{Page:1993wv}, which is a critical piece of the information paradox. This relation has also been established within the framework of holographic duality \cite{Penington:2019npb,Almheiri:2019psf}.

Applying Page's argument to the system of a scalar quantum field theory implies that the entanglement entropy in an arbitrary quantum state should be proportional to the number of degrees of freedom of the smaller subsystem, i.e. proportional to the volume of this subsystem. One has to keep in mind that the Hilbert space of a local degree of freedom in this case is infinite-dimensional, rendering the application of Page's argument in scalar field theory a little hazy. However, assuming that the scaling properties are preserved as we take the limit of the dimension of the Hilbert space to infinity, Page's argument contradicts the seminal results of Bombelli and Srednicki \cite{Bombelli:1986rw,srednicki}, which apply to the ground state, as well as their generalizations to coherent states \cite{Benedict:1995yp,Katsinis:2022fxu}. These studies demonstrate that entanglement entropy scales with the area of the subsystem and not its volume. In this sense, the area-law property of entanglement entropy should be considered as a special property of the most classical states of scalar field theory, i.e. of the coherent states.

In order to understand this issue, we applied our method to the system of free massless scalar field theory in 1+1 dimensions. In agreement with the large-squeezing expansion that we developed for an arbitrary harmonic system, we found that states where all modes have been squeezed with large squeezing parameters give rise to entanglement entropy which is dominated by a term proportional to the volume of the smaller subsystem, in agreement with Page's arguments. Furthermore, this volume term is time-independent, although the state of the system has non-trivial time-dependence. This is consistent with a maximal entanglement entropy bound in line with Page. We expect this behaviour to hold for scalar field theory in higher dimensions, which will be the subject of future work.

We can speculate on the consequences of our results for the interpretation of gravity as an entropic force attributed to quantum entanglement statistics. Such an interpretation is supported by holographic calculations \cite{Lashkari:2013koa,Faulkner:2013ica}. However, there are more general arguments that suggest how such an entropic force operates. In 1995 Jacobson argued that the dynamic metric of a theory with two specific properties is subject to Einstein dynamics \cite{Jacobson:1995ab}. The first property is the validity of a first law of thermodynamics. The second one is that the entropies of the horizons are proportional to their area. In other words, the scaling properties of entropy determine the gravitational dynamics. Entanglement is fertile ground in which to realize such a mechanism. The entanglement entropy scales with area (at least when the overall system lies in a coherent state) and also obeys a first law of entanglement thermodynamics with the expectation value of the modular Hamiltonian.

Our investigation suggests that the only Gaussian states that give rise to entanglement entropy dominated by an area law are the minimal uncertainty states, i.e. the coherent states. There is no indication that this property extends beyond the Gaussian states. In this spirit, our results imply that Einstein gravity emerges as a quantum entropic force only when the overall system lies in a closest-to-classical state, i.e. in the ground or a coherent state. When a more ``arbitrary'' quantum state is considered, the emergent dynamics will be more complicated than Einstein gravity.

As a final comment we point out that the squeezing of quantum states plays an important 
role in early cosmology. During inflation, a momentum mode of a massless field
gets stretched by the rapid expansion.
When the mode wavelength becomes larger than the horizon, the scalar fluctuation 
loses its oscillatory form (it freezes) \cite{physrep}. 
After horizon crossing, the field can be viewed as a classical stochastic 
field, and its quantum expectation value can be replaced by the classical 
stochastic average.
The quantum properties of the field are considered invisible in late-time 
observations, which focus on classical local quantities \cite{albrecht,classical1}.
However, from a quantum mechanical point of view, the modes of the scalar field 
evolve from a simple oscillator ground state to an increasingly squeezed state
\cite{Grishchuk}. 
Quantum entanglement is a purely quantum non-local phenomenon
that cannot be encoded in the classical probability distributions. 
The squeezing of canonical modes increases the entanglement between 
local degrees of freedom and is expected to increase the entanglement entropy \cite{squeeze1,squeeze2,squeeze3,squeeze4,squeeze5,squeeze6,squeeze7,squeeze8}.
The techniques we presented in this work were employed in \cite{Boutivas:2023ksg} 
in order to compute 
the entanglement entropy resulting from tracing out local degrees of freedom of a quantum scalar field in an expanding universe. It was shown that the entanglement entropy grows continuously 
during inflation, as successive modes cross the horizon. 
The resulting entropy is proportional to the total duration of inflation, and 
is preserved during a subsequent era of radiation or matter domination.
The emergence of a volume term in the entanglement entropy as a result of squeezing was observed in \cite{Boutivas:2023ksg} 
in the context of a toy model in 1+1 dimensions, in agreement
with our findings here.

\acknowledgments

The research of D.Katsinis was supported by FAPESP Grant No. 2021/01819-0. The research of N. Tetradis was supported by the Hellenic Foundation for
Research and Innovation (H.F.R.I.) under the “First Call for H.F.R.I.
Research Projects to support Faculty members and Researchers and
the procurement of high-cost research equipment grant” (Project
Number: 824).

\appendix

\section{The Dependence of Entanglement on Squeezing in the Case of Two Oscillators}
\label{sec:app_squeezing_entanglement}

In the case of two oscillators, we have found simple formulae describing the entanglement entropy as a function of the squeezing parameters and time. We can study them in order to gain intuition on how squeezing affects entanglement. In this appendix we present all the details of this analysis, whose summary was presented in subsection \ref{subsec:squeezing_entanglement}.

Instead of searching for extrema of the entanglement entropy, it is easier to search for extrema of the ratio $r$, which is defined as
\begin{equation}\label{eq:r_definition}
r:=\frac{\re \left(\gamma\right)+\beta }{\re \left(\gamma\right)-\beta} ,
\end{equation}
where $\gamma$ and $\beta$ are given by equations \eqref{eq:two_osc_gamma} and \eqref{eq:two_osc_beta}. The entanglement entropy is a strictly increasing function of $r$. It follows that extrema of $r$ correspond to extrema of the entanglement entropy.

The ratio $r$ is by definition always larger than 1. Using its definition, we can express the entanglement entropy in a symmetric form, namely
\begin{equation}\label{eq:entanglment_formula_2}
S =\frac{\sqrt{r}+1}{2}\ln\left(\frac{\sqrt{r}+1}{2}\right)-\frac{\sqrt{r}-1}{2}\ln\left(\frac{\sqrt{r}-1}{2}\right).
\end{equation}
This expression is related to the calculation of entanglement entropy in terms of correlation functions (see e.g. \cite{Ppeschel}). 

Before presenting our analysis, let us first briefly review the case of the ground/coherent state of the overall system, so we can use it as a basis for comparison. In this case the ratio $r$ is given by
\begin{equation}
r_0 = \frac{1}{4} \left( \sqrt{\frac{\omega_+}{\omega_-}} + \sqrt{\frac{\omega_-}{\omega_+}} \right)^2 .
\label{eq:two_r0}
\end{equation}
It is evident that $r_0$ and consequently the entanglement entropy depend only on the ratio of the two eigenfrequencies $\omega_+$ and $\omega_-$. Furthermore, it is invariant under the interchange $\omega_+\leftrightarrow \omega_-$. Let
\begin{equation}
\rho_0 := \frac{\omega_-}{\omega_+} .
\end{equation}
Considering that $\rho_0 > 1$, the ratio $r$ and thus the entanglement entropy are strictly increasing functions of $\rho_0$. The symmetry $\omega_+\leftrightarrow \omega_-$ obviously implies that, when $\rho_0 < 1$, $\xi_0$ is a strictly decreasing function of $\rho_0$.

\subsection{Squeezing a Single Mode}
\label{subsubsec:squeeze_one}
For simplicity let us first consider the case $z_- = 0$, i.e. we ``squeeze'' only the symmetric mode. This is a more transparent case, since the time evolution of the parameter $\xi$, and thus of the entanglement entropy, is periodic with period $T_+ / 2 = \pi / \omega_+$.

In this case, the ratio $r$ assumes the simple form
\begin{equation}
r = \frac{1}{2} + \frac{\omega_-^2 + \omega_+^2}{4 \omega_+ \omega_-} \cosh z_+ + \frac{\omega_-^2 - \omega_+^2}{4 \omega_+ \omega_-} \sinh z_+ \sin 2 \omega_+ t ,
\label{eq:ratio}
\end{equation}
which is manifestly positive and larger than 1, as required. This equation directly implies that $r$ is bounded between two values, $r_\pm$,
\begin{equation}
r_\pm = \frac{1}{4} \left( \sqrt{\frac{\omega_+}{\omega_-}} e^{\pm \frac{z_+}{2}} + \sqrt{\frac{\omega_-}{\omega_+}} e^{\mp \frac{z_+}{2}} \right)^2 .
\label{eq:two_rpm}
\end{equation}
It directly follows that the entanglement entropy is bounded between two values $S_\pm$ that 
are determined by $r_\pm$ and equation \eqref{eq:entanglment_formula_2}. The value $r_+$ is obtained at the instants when $\sin 2\omega_+ t = 1$, whereas the value $r_-$ is obtained at the instants when $\sin 2\omega_+ t = - 1$. At these instants the squeezed state is a minimal uncertainty state, i.e. $\Delta x_+\Delta p_+ = \hbar / 2$ (see equation \eqref{eq:squeezed_uncertainty_product}). It follows that, at these instants, the symmetric mode is described by a wavefunction that is indistinguishable from an appropriate coherent state of an effective oscillator with eigenfrequency $\omega_+ e^{\pm z_+}$. Recall that the spectrum of the reduced density matrix when the system lies in a coherent state is identical to that when it lies in the ground state. Therefore, at these instants, the symmetric mode is effectively described by the ground state wavefunction of this effective oscillator, at least as long as entanglement is concerned.

As is evident from the above discussion, as well as the comparison of equations \eqref{eq:two_r0} and \eqref{eq:two_rpm}, the two bounds on the entanglement entropy in the case of a squeezed symmetric mode are identical to the entanglement entropies of two equivalent effective systems of coupled oscillators at their ground state. The ratios of eigenfrequencies of each of these two equivalent systems are
\begin{equation}
\rho_\pm = \rho_0 e^{\pm z_+} .
\end{equation}
We assume that $\rho_0 > 1$ and $z_+ > 0$. It directly follows from equation \eqref{eq:ratio} that $r_+ > r_-$. Therefore, the value of $r_+$ determines the maximal value of entanglement entropy, whereas the value of $r_-$ determines the minimal value of entanglement entropy.

Recalling the monotonicity of the relation between $r$ and the ratio of eigenfrequencies $\rho$ for a system at its ground state that we analysed above, since $\rho_0 > 1$ and $\rho_+ > \rho_0$, the maximal entanglement entropy is larger than that at the ground state of the system, $S_0$. As long as the minimal entanglement entropy is concerned, there is a change of the qualitative behaviour as the squeezing parameter $z_+$ increases. As $z_+$ increases from zero to positive values, the ratio $\rho_-$ gets smaller, but 
it remains larger than 1 until the critical squeezing parameter
\begin{equation}
z_0 = \ln\frac{\omega_-}{\omega_+} .
\end{equation}
 As a result, the minimal entanglement entropy is smaller than $S_0$ and a decreasing function of the squeezing parameter. For $z_+ = z_0$ the minimum entanglement entropy vanishes, i.e. there are instants when the wavefunction of the system is separable. As the squeezing parameter further increases, the ratio $\rho_-$ further decreases and it is smaller than 1. As a result, the minimal entanglement entropy becomes an increasing function of the squeezing parameter $z_+$. There is a critical value of the squeezing parameter $z_+$, which results in $\rho_- = 1 / \rho_0$, namely
\begin{equation}
z_{\mathrm{vac}} = 2 \ln\frac{\omega_-}{\omega_+} = 2 z_0 ,
\end{equation}
where the minimal entanglement entropy coincides with $S_0$.

For small values of the squeezing parameter, it is not difficult to show that the ratio $r$ and the entanglement entropy perform a sinusoidal oscillation around a mean value, and also calculate this mean value. In particular, one can show that for $z_+ \ll 1$ the ratio $r$ assumes the form
\begin{equation}
r = r_0 \left( 1 + z_+ \, \frac{\omega_- - \omega_+}{\omega_- + \omega_+} \sin 2 \omega_+ t + \frac{z^2_+}{2} \frac{\omega_-^2 + \omega_+^2}{\omega_- + \omega_+} \right) + \mathcal{O} \left( z^3 \right) .
\end{equation}
For $z_+ \gg 1$ we obtain
\begin{equation}
r = \frac{\omega_-^2 + \omega_+^2 + \left( \omega_-^2 - \omega_+^2 \right) \sin 2 \omega_+ t}{8 \omega_+ \omega_-} e^{z_+} + \frac{1}{2} + \mathcal{O} \left( e^{- z_+} \right) .
\end{equation}
Substituting these expressions in \eqref{eq:entanglment_formula} one can obtain the following expressions for the mean entanglement entropy\footnote{In order to perform this calculation the formula
\begin{equation}\label{eq:ln_mean_value}
\frac{1}{2\pi}\int_0^{2\pi}dt\ln\left(1+a\sin t\right)=\ln\frac{\sqrt{1-a^2}+1}{2},\qquad \vert a\vert<1
\end{equation}
is required.}$^{\textrm{,}}$,
\footnote{Notice that the coefficient diverges in the $\omega_+=\omega_-$ limit. This is an artefact of the order of the limits, i.e. the small $ z_+$ limit and the degenerate limit do not commute. To see why this is the case, we turn to the degenerate limit. For $\omega_+=\omega_-$ the parameter $\xi$ and the entanglement entropy simplify a lot. It is straightforward to show that they read
\begin{equation*}
\xi_{\textrm{degenerate}}=\tanh^2\frac{ z_+}{4}
\end{equation*}
and
\begin{equation*}
S_{\textrm{degenerate}}=\cosh^2\frac{ z_+}{4}\ln\cosh^2\frac{ z_+}{4}-\sinh^2\frac{ z_+}{4}\ln\sinh^2\frac{ z_+}{4}.
\end{equation*}
So, the entanglement spectrum of the reduced density matrix and the entanglement entropy change at will.}
\begin{equation}
\bar{S}=\begin{cases}
S_0 - \frac{z_+^2}{16} \left( 1 + \frac{1 + 4 \xi_0 + \xi_0^2}{1 - \xi_0^2} \ln \xi_0 \right) , \quad & z_+ \ll 1 , \\
\frac{z_+}{2}+ \frac{1}{2} \ln r_0 + 1 - 2 \ln2 , \quad & z_+ \gg 1 .
\end{cases}
\label{eq:app_two_mean_SEE_asymtotics}
\end{equation}

Using the inequality $x \geq \ln \left( 1 + x \right)$ one can show that the coefficient of $z_+^2$ in the small $z_+$ expansion of the mean entanglement entropy is manifestly positive. Its minimal value is equal to $1/8$, and it is obtained in the limit $\xi_0 \to 1$. This 
implies that the mean entanglement entropy is increasing with squeezing for small squeezing parameters.

\subsubsection{The Mean Entanglement Entropy}
\label{sec:app_mean}
In the special case we are studying, i.e. only the symmetric mode has been squeezed, it is possible to derive an analytic formula for the mean entanglement entropy for an arbitrary value of the squeezing parameter. 

We begin by rewriting the entanglement entropy, which is given by \eqref{eq:entanglment_formula_2}, as
\begin{equation}
S = \frac{1}{2} \ln \left( \frac{r - 1}{4} \right) + \sqrt{r} \tanh^{-1} \left( \frac{1}{\sqrt{r}} \right) .
\end{equation}
We use the relatively simple expression \eqref{eq:ratio} for $r$ and we split the entanglement entropy into three terms as
\begin{align}
S_1 &= - \ln2 + \frac{1}{2} \ln \left( - \frac{1}{2} + \frac{\omega_+^2 + \omega_-^2}{4 \omega_+ \omega_-} \cosh z_+ \right) \label{eq:bar_s1} , \\
S_2 &= \frac{1}{2} \ln \left( 1 + \frac{\frac{\omega_-^2 - \omega_+^2}{4 \omega_+ \omega_-} \sinh z_+}{- \frac{1}{2} + \frac{\omega_-^2 + \omega_+^2}{4 \omega_+ \omega_-} \cosh z_+}\sin 2 \omega_+t \right) \label{eq:s_2} , \\
S_3 &= \sqrt{r} \tanh^{-1} \left( \frac{1}{\sqrt{r}} \right) \label{eq:s_3} .
\end{align}

The first term is constant, so $\bar{S}_1 = S_1$. We are left with the calculation of $\bar{S}_2$ and $\bar{S}_3$. To proceed, notice that
\begin{equation}
-\frac{1}{2}+\frac{\omega_-^2+\omega_+^2}{4\omega_+\omega_-}\cosh z_+\pm \frac{\omega_-^2-\omega_+^2}{4\omega_+\omega_-}\sinh z_+=\left(\frac{\omega_-e^{\pm z/2}-\omega_+e^{\mp z/2}}{2\sqrt{\omega_+\omega_-}}\right)^2.
\end{equation}
Thus the absolute value of the coefficient of $\sin2\omega_+t$ in \eqref{eq:s_2} in smaller than one. Therefore, $\bar{S}_2$ can be calculated directly by equation \eqref{eq:ln_mean_value}: 
\begin{equation}\label{eq:bar_s2}
\bar{S}_2=\frac{1}{2}\ln\left[\frac{1}{2}\frac{\left\vert \left(\omega_++\omega_-\right)^2\sinh^2\frac{ z_+}{2}-\left(\omega_+-\omega_-\right)^2\cosh^2\frac{ z_+}{2}\right\vert}{\left(\omega_++\omega_-\right)^2\sinh^2\frac{ z_+}{2}+\left(\omega_+-\omega_-\right)^2\cosh^2\frac{ z_+}{2}}+\frac{1}{2}\right] .
\end{equation}

We are left with the calculation of $\bar{S}_3$. We expand $\sqrt{r}\tanh^{-1}(1/\sqrt{r})$ and get
\begin{equation}
S_3=\sum_{k=0}^\infty\frac{1}{2k+1}\frac{1}{b^k}\frac{1}{\left(1+c\sin2\omega_+t\right)^k} ,
\end{equation}
where
\begin{equation}
b=\frac{1}{2}+\frac{\omega_-^2+\omega_+^2}{4\omega_+\omega_-}\cosh z_+ \quad \textrm{and} \quad c=\frac{\frac{\omega_-^2-\omega_+^2}{4\omega_+\omega_-}\sinh z_+}{\frac{1}{2}+\frac{\omega_-^2+\omega_+^2}{4\omega_+\omega_-}\cosh z_+}.
\end{equation}
The integration can be performed using the formula
\begin{equation}\label{eq:integral_Legendre}
\frac{1}{2\pi}\int_0^{2\pi}dt\frac{1}{\left(1+c \sin t\right)^k}=\frac{1}{\left(1-c^2\right)^{k/2}}P_{k-1}\left(1/\sqrt{1-c^2}\right),
\end{equation}
where $P_k$ is the Legendre polynomial of order $k$, to obtain
\begin{equation}
\bar{S}_3=\sum_{k=0}^\infty\frac{1}{2k+1}y^k P_{k-1}\left(x\right),
\end{equation}
where $x$ and $y$ are given by
\begin{align}
x&=\frac{1}{\sqrt{1-c^2}}=\frac{2\omega_+\omega_-+\left(\omega_+^2+\omega_-^2\right)\cosh z_+}{\omega_+^2+\omega_-^2+2\omega_+\omega_-\cosh z_+},\\
y&=\frac{1}{b\sqrt{1-c^2}}=\frac{4\omega_+\omega_-}{2\omega_+\omega_-+\left(\omega_+^2+\omega_-^2\right)\cosh z_+}.
\end{align}
We separate the $k=0$ term, which is constant, to write
\begin{equation}
\bar{S}_3=1+\sum_{k=0}^\infty\frac{1}{2k+3}y^{k+1}P_{k}\left(x\right).
\end{equation}
We are unable to perform this summation directly. So we introduce a Schwinger parameter and interchange the summation and integration, to obtain
\begin{equation}
\bar{S}_3=1+ y\int_0^1 dw \sum_{k=0}^\infty w^{2}\left(w^2y\right)^k P_{k}\left(x\right).
\end{equation}
The summation can be performed using the generating function of Legendre polynomials to arrive at
\begin{equation}
\bar{S}_3=1+y\int_0^1 dw\frac{w^2}{\sqrt{w^4y^2-2 y w^2 x+1}},
\end{equation}
or with a trivial change of variable
\begin{equation}
\bar{S}_3=1+\frac{1}{\sqrt{y}}\int_0^{\sqrt{y}} dw\frac{w^2}{\sqrt{w^4-2w^2 x+1}}.
\end{equation}
Since we are calculating a physical quantity, the result should be real for all values of the parameters. To verify this fact, recall that $x>1$ and $1>y>0$. The quantity under the square root gets its minimum value for $w=\sqrt{y}$, while we have
\begin{equation}
y^2-2 x y+1=\left[\frac{\omega_+^2+\omega_-^2-2\omega_+\omega_-\cosh z_+}{\omega_+^2+\omega_-^2-2\omega_+\omega_-\cosh z_+}\right]^2.
\end{equation}
Thus, the result is manifestly real for any value of the parameters. Setting $x=\cosh u$ we obtain
\begin{equation}
\bar{S}_3=1+\frac{1}{\sqrt{y}}\int_0^{\sqrt{y}} dw\frac{w^2}{\sqrt{w^2-e^{-u}}\sqrt{w^2-e^{u}}}.
\end{equation}
We perform another change of the integration variable to obtain
\begin{equation}
\begin{split}
\bar{S}_3&=1+\frac{e^{u/2}}{\sqrt{y}}\int_0^{\sqrt{y}e^{u/2}} dw\frac{e^{-2u}w^2}{\sqrt{1-w^2}\sqrt{1-e^{-2u} w^2}}\\
&=1+\frac{e^{u/2}}{\sqrt{y}}\int_0^{\sqrt{y}e^{u/2}} dw\left[\frac{1}{\sqrt{1-w^2}\sqrt{1-e^{-2u} w^2}}-\frac{\sqrt{1-e^{-2u} w^2}}{\sqrt{1-w^2}}\right]\\
&=1+\frac{e^{u/2}}{\sqrt{y}}\left[F\left(\sin^{-1}\sqrt{y}e^{u/2};e^{-2u}\right)-E\left(\sin^{-1}\sqrt{y}e^{u/2};e^{-2u}\right)\right]
\end{split}
\end{equation}
where $F$ and $E$ are the incomplete elliptic integrals of first and second kind respectively, and $e^{-2u}$ is their elliptic modulus. The careful reader would have noticed that initially the sign of $u$ was irrelevant, but we have treated $e^u$ and $e^{-u}$ differently. Had we made the opposite choice, we would have ended up with a symmetric formula. Of course, this result can also be obtained using the transformations of the elliptic integrals under the inversion of the elliptic modulus, which is required for consistency. In order to substitute the original parameters we use
\begin{equation}
u = \ln \frac{e^{z_+ / 2} \omega_- + e^{- z_+ / 2} \omega_+}{e^{- z_+ / 2} \omega_-+ e^{z_+ / 2} \omega_+}.
\end{equation}

Gathering all the terms, i.e. the above result, along with \eqref{eq:bar_s1} and \eqref{eq:bar_s2}, we obtain
\begin{equation}
\begin{split}
\bar{S}=-2\ln2+\frac{1}{2}\ln\left[\frac{\left\vert \omega_1^2\sinh^2\frac{ z_+}{2}-\omega_2^2\cosh^2\frac{ z_+}{2}\right\vert+\omega_1^2\sinh^2\frac{ z_+}{2}+\omega_2^2\cosh^2\frac{ z_+}{2}}{2\omega_+\omega_-}\right]\\+1+\frac{e^{ z_+/2}\omega_-+e^{- z_+/2}\omega_+}{2\sqrt{\omega_+\omega_-}}\left[F\left(\phi;k^2\right)-E\left(\phi;k^2\right)\right], 
\end{split}
\label{eq:average_S}
\end{equation}
where $\omega_1^2=\left(\omega_++\omega_-\right)^2$, $\omega_2^2=\left(\omega_+-\omega_-\right)^2$ and
\begin{equation}
\phi=\sin^{-1}\frac{2\sqrt{\omega_+\omega_-}}{e^{- z_+/2}\omega_-+e^{ z_+/2}\omega_+},\qquad k^2=\frac{e^{- z_+/2}\omega_-+e^{ z_+/2}\omega_+}{e^{ z_+/2}\omega_-+e^{- z_+/2}\omega_+}.
\end{equation}
This expression for the mean entanglement entropy is an increasing function of the squeezing parameter.

Notice that \eqref{eq:average_S} is manifestly invariant under the interchange $\omega_+\leftrightarrow \omega_-$ along with $ z_+\rightarrow - z_+$. Had we performed only one of these transformations, it would require the transformation of the elliptic integrals under the inversion of the elliptic modulus to show that \eqref{eq:average_S} is indeed invariant. Equation \eqref{eq:average_S} also explains why the small $ z_+$ and the degenerate $\omega_+=\omega_-$ limits do not commute. For $\omega_+\neq\omega_-$ in the small $ z_+$ limit the quantity inside the absolute value is negative, whereas for finite $ z_+$ this quantity is positive in the $\omega_+-\omega_-\rightarrow 0$ limit. On the contrary, in the large $\vert  z_+\vert$ limit this quantity is positive, just like in the $\omega_+-\omega_-\rightarrow 0$ limit, thus these limits commute. Finally, the elliptic modulus takes any positive value, but it is equal to 1 either for $ z_+=0$ or $\omega_+=\omega_-$.

\subsection{Squeezing Both Modes}
When both the symmetric and antisymmetric modes are squeezed, the parameter $r$ equals
\begin{equation}
\begin{split}
r=\frac{1}{2}-&\frac{1}{2}\cos\Phi_-\cos\Phi_+\sinh z_-\sinh z_+ \\
+&\left(\cosh z_++\sin\Phi_+\sinh z_+\right)\left(\cosh z_--\sin\Phi_-\sinh z_-\right)\frac{\omega_-}{4\omega_+}\\
+&\left(\cosh z_-+\sin\Phi_-\sinh z_-\right)\left(\cosh z_+-\sin\Phi_+\sinh z_+\right)\frac{\omega_+}{4\omega_-},
\end{split}
\end{equation}
where $\Phi_\pm$ are the phases of the two modes, namely, $\Phi_\pm=2\omega_\pm\left(t-t_{0\pm}\right)$. Without loss of generality, in the following we consider that $z_+$ and $z_-$ are both positive. The introduction of a negative squeezing parameter is equivalent to a shift of the corresponding phase $\Phi_\pm$ by $\pi$. Furthermore, we assume that $\omega_- > \omega_+$.

It is a matter of algebra to show that
\begin{multline}
r=1+\frac{1}{4}\left[\frac{\sqrt{\cosh z_++\sin\Phi_+\sinh z_+}}{\sqrt{\cosh z_-+\sin\Phi_-\sinh z_-}}\frac{\sqrt{\omega_-}}{\sqrt{\omega_+}}-\left(+\leftrightarrow-\right)\right]^2\\
+\frac{1}{4}\left[\frac{\sqrt{\cosh z_++\sin\Phi_+\sinh z_+}}{\sqrt{\cosh z_-+\sin\Phi_-\sinh z_-}}\frac{\sqrt{\omega_-}}{\sqrt{\omega_+}}\cos\Phi_-\sinh z_--\left(+\leftrightarrow-\right)\right]^2 ,
\label{eq:2_modes_ratio_r}
\end{multline}
which implies that $r$ is manifestly greater or equal to 1, as required.

Unlike the case where we had squeezed only the symmetric mode, the ratio $r$ and thus the entanglement entropy is not necessarily a periodic function of time. Actually it is periodic if and only if the ratio of the eigenfrequencies of the two normal modes is rational. In such a case, the phases $\Phi_\pm$ follow a one-dimensional closed path in the $\Phi_+ \Phi_-$ plane. Otherwise, they follow an open trajectory, which in infinite time will cover the whole region of possible $\left( \Phi_+, \Phi_- \right)$ pairs, i.e. the two phases get arbitrarily close to any given pair of admissible values at some instant. In this spirit, we search for the extrema of the ratio $r$, and thus of entanglement entropy, treating the two phases as independent, although actually they are not; they are both given functions of time.

There are stationary points of the ratio $r$ when the phases $\Phi_\pm$ satisfy the equations\footnote{Notice that there is another mathematical solution, namely 
\begin{align*}
\sin\Phi_\pm \sinh z_\pm&=\frac{\left(\omega_+^2+\omega_-^2\right)\cosh z_\pm+2\omega_+\omega_- \cosh z_\mp}{\omega_\pm^2-\omega_\mp^2},\\
\cos \Phi_-\sinh z_-&=-\cos\Phi_+\sinh z_+.
\end{align*}
However, it is unphysical, since for any values of the parameters it does not correspond to real $\Phi_+$ and $\Phi_-$.}
\begin{align}
\sin\Phi_\pm \sinh z_\pm&=\frac{\left(\omega_+^2+\omega_-^2\right)\cosh z_\pm-2\omega_+\omega_- \cosh z_\mp}{\omega_\pm^2-\omega_\mp^2}, \label{eq:2_modes_vacuum_sin}\\
\cos \Phi_-\sinh z_-&=\cos\Phi_+\sinh z_+,\label{eq:2_modes_vacuum_cos}
\end{align}
or
\begin{align}
\cos\Phi_+=\cos\Phi_-=0.
\label{eq:2_modes_vacuum_trivial}
\end{align}
Notice that the square of equation \eqref{eq:2_modes_vacuum_cos} is automatically satisfied if the two equations \eqref{eq:2_modes_vacuum_sin} are satisfied. Equation \eqref{eq:2_modes_vacuum_cos} only specifies the relative sign of $\cos\Phi_\pm$. 

Equations \eqref{eq:2_modes_vacuum_sin} and \eqref{eq:2_modes_vacuum_cos} do not always have a real solution. This depends on the values of the squeezing parameters. Demanding that $\left| \sin \Phi_\pm \right| \leq 1$ leads to
\begin{equation}
\cosh\left(z_\pm+\ln\frac{\omega_-}{\omega_+}\right) \geq \cosh z_\mp \geq \cosh\left(z_\pm-\ln\frac{\omega_-}{\omega_+}\right) ,
\end{equation}
i.e.
\begin{align}
z_\pm+\ln\frac{\omega_-}{\omega_+}\geq z_\mp \geq \left\vert z_\pm-\ln\frac{\omega_-}{\omega_+}\right\vert
\end{align}
or
\begin{equation}
z_++z_-\geq \ln\frac{\omega_-}{\omega_+}\geq \left\vert z_+-z_- \right\vert .
\label{eq:2_modes_vacuum_condition}
\end{equation}
Whenever the above condition holds, equations \eqref{eq:2_modes_vacuum_sin} and \eqref{eq:2_modes_vacuum_cos} do have two solutions, both corresponding to the ratio $r$ being equal to 1. Actually, it is evident from equation \eqref{eq:2_modes_ratio_r} that these are the only values of the phases $\Phi_\pm$ where the ratio $r$ can be equal to 1. Since this is the minimal possible value of $r$, whenever these solutions exist, they provide the global minimum of the ratio $r$.

Equations \eqref{eq:2_modes_vacuum_trivial} have always 4 solutions in a trivial manner, namely
\begin{equation}
\Phi_+ = \pm \frac{\pi}{2} , \quad \Phi_- = \pm \frac{\pi}{2} .
\end{equation}
They correspond to the following values of the ratio $r$
\begin{equation}
r \left(\Phi_+ = s_+ \frac{\pi}{2} , \Phi_- = s_- \frac{\pi}{2} \right) \equiv r_{s_+ s_-} = \cosh^2 \left[ \frac{1}{2} \left( s_+ z_+ - s_- z_- + \ln \frac{\omega_-}{\omega_+} \right) \right] ,
\label{eq:2_modes_r_extrema}
\end{equation}
where the symbols $s_\pm$ take the values $\pm 1$.

In an obvious manner $r_{+-}$ is larger than $r_{++}$, $r_{-+}$ and $r_{--}$. Therefore, $\Phi_+ = \frac{\pi}{2}$ and $\Phi_- = - \frac{\pi}{2}$ is the position of the global maximum of the ratio $r$,
\begin{equation}
r_{\max} = r_{+-} .
\end{equation}
 When $z_+ < \ln \frac{\omega_-}{\omega_+}$ and $z_- < \ln \frac{\omega_-}{\omega_+}$ the smallest of the four $r_{s_+ s_-}$ is $r_{-+}$. When the above does not hold and $z_+ > z_-$ the smallest is $r_{--}$, whereas when $z_+ < z_-$ the smallest is $r_{++}$. The smallest of the four $r_{s_+ s_-}$ is the global minimum of the ratio $r$, whenever condition \eqref{eq:2_modes_vacuum_condition} does not hold, i.e.
\begin{equation}
r_{\min} = \begin{cases} 1 , & z_++z_-\geq \ln\frac{\omega_-}{\omega_+}\geq \left\vert z_+-z_- \right\vert , \\
r_{-+} , & z_+ + z_- < \ln\frac{\omega_-}{\omega_+} , \\
r_{--} , & z_+ - z_- > \ln\frac{\omega_-}{\omega_+} , \\
r_{++} , & z_- - z_+ > \ln\frac{\omega_-}{\omega_+} .
\end{cases}
\end{equation}

Since the value of the ratio $r$ for the vacuum state of the two oscillators can be written as $r_0 = \cosh^2 \left( \frac{1}{2} \ln \frac{\omega_-}{\omega_+} \right) $, it turns out that the globally minimal value of the ratio $r$ and thus of the entanglement entropy coincides with that of the vacuum state if
\begin{equation}
\left\vert z_+-z_- \right\vert = 2 \ln\frac{\omega_-}{\omega_+} .
\end{equation}

The values of the phases $\Phi_\pm$ that correspond to the extrema $r_{\pm\pm}$, namely $\Phi_\pm = \pm \pi / 2$, are not arbitrary. When a phase takes one of these two values, the wavefunction that describes the corresponding mode is a minimal uncertainty state with maximal or minimal position uncertainty, respectively.

The fact that the ratio $r$ is stationary when the two modes are both minimal uncertainty states suggests that these are the times that contributions of squeezed modes to entanglement are maximal. However, these contributions add when the two modes are in opposite phases. When the two phases are equal, the contributions cancel each other, resulting in weak entanglement. The quantity that receives negative or positive contributions directly equal to the squeezing parameter of each mode at these instants is $\arccosh\sqrt{r}$.

In the above we dealt with the two phases as independent variables. Actually, they are not; they are both functions of time. As time flows, the system follows a specific one dimensional trajectory within the two-dimensional space of the phases of the two modes. The trajectory depends on the ratio of the frequencies of the two modes. Because of this fact, the ratio $r$ may present local minima or maxima with time, which do not coincide with the theoretical minima and maxima $r_{\min}$ and $r_{\max}$ that we specified above. However, the ratio $r$ is always bound by these values.

\section{Algebraic Construction of the Reduced Density Matrix Eigenstates}
\label{subsubsec:spectrum_creation_annihilation}

In section \ref{subsec:many_eigenstates}, we showed that the eigenfunctions of the matrix $\tilde{\rho}_2$, and thus those of the reduced density matrix, form a tower of states, in many aspects similar to the tower of eigenstates of a coupled harmonic system. Actually, we know that when the matrix $\beta$ is real, the above statement is exact \cite{srednicki,Katsinis:2022fxu}. It would be nice to construct creation and annihilation operators which would relate the eigenstates of $\tilde{\rho}_2$ in the same sense that they relate the eigenstates of a coupled harmonic system.

However, we know that this cannot be that simple. For example, we know that the ``second excited'' eigenstate of the reduced density matrix, which corresponds to different eigenvectors of the matrix $\Xi$, cannot be produced by the action of two ``ordinary'' creation operators on the ground state, as we showed in section \ref{subsubsec:spectrum_tower}. In general, the required operators have to be linear combinations of positions and momenta. They differ though from the ``ordinary'' ones, as the combination of momenta that appears in one of them cannot be the conjugate momentum of the combination of positions that appears.

Therefore, we search for annihilation operators of the form
\begin{equation}
A_i= C_{ik}\left(\partial_k+ \mathcal{A}_{kj}x_j\right) ,
\end{equation}
so that they annihilate the ``ground'' eigenstate $\Psi_0$ of $\tilde{\rho}_2$ \eqref{eq:spectrum_ground_state}.
These operators should act on the ``first excited'' eigenstates $\Psi_{1 \ell}$, which are given by equation \eqref{eq:spectrum_first excited_eigenstate}, as
\begin{equation}
A_i \Psi_{1 \ell} = \delta_{i \ell} \Psi_0 .
\label{eq:spectrum_A_on_first}
\end{equation}
Introducing the notation $\left(\mathbf{v}_\ell\right)_k = v^\ell_k$, equation \eqref{eq:spectrum_A_on_first} yields
\begin{equation}
C_{ik}v^\ell_k=\frac{1}{\sqrt{2}}\delta_{i\ell}.
\end{equation}
Bearing in mind that the eigenvectors of the matrix $\Xi$ are normalized so that
\begin{equation}
\mathbf{v}^\dagger_{i}\re\left(\mathcal{A}\right)^{-1}\mathbf{v}_{j}=v^{i*}_k \left( \re \left( \mathcal{A} \right)^{-1} \right)_{k\ell} v^j_\ell = \delta_{ij} ,
\end{equation}
the above equation implies that
\begin{equation}
C_{ik}=\frac{1}{\sqrt{2}}v^{i*}_{n} \left(\re\left(\mathcal{A}\right)^{-1}\right)_{n k} .
\end{equation}
Therefore, the annihilation operators $A_i$ and their Hermitian conjugates, the creation operators $A_i^\dagger$, read
\begin{align}
A_i &= \frac{1}{\sqrt{2}}v^{i*}_n \left(\re\left(\mathcal{A}\right)^{-1}\right)_{n k}\left(\partial_k+ \mathcal{A}_{km}x_m\right) , \\
A_i^\dagger &= \frac{1}{\sqrt{2}}v^i_n \left(\re\left(\mathcal{A}\right)^{-1}\right)_{n k}\left(-\partial_k+ \mathcal{A}_{k m}^{*}x_m\right) .
\end{align}
One can trivially show that $A_i^\dagger \Psi_0 = \Psi_{1 i}$.

Let us study the commutation relations of the creation and annihilation operators. It is a matter of algebra to show that
\begin{align}
\left[A_i,A_j^\dagger\right] &= v^{i*}_n \left(\re\left(\mathcal{A}\right)^{-1}\right)_{n k}v^j_k = \delta_{ij} , \\
\left[A_i,A_j\right] &= \frac{1}{2}v^{i*}_n \left( \re\left(\mathcal{A}\right)^{-1}\right)_{n k}v^{j*}_\ell \left( \re\left(\mathcal{A}\right)^{-1}\right)_{\ell r}\left[ \mathcal{A}_{r k}- \mathcal{A}_{kr}\right]=0 .
\end{align}

Let us assume that $\Psi \left( \mathbf{x} \right)$ is an eigenfunction of the matrix $\tilde{\rho}_2$ with eigenvalue $\lambda$, i.e.
\begin{equation}
\begin{split}
\tilde{\rho}_2\Psi \left( \mathbf{x} \right) &= \int d^n \mathbf{x^\prime} \tilde{\rho}_2 \left( \mathbf{x} ; \mathbf{x}^\prime \right) \Psi \left( \mathbf{x^\prime} \right)\\
&= c \int d^n \mathbf{x^\prime} \exp \left[ - \frac{1}{2} \left( \mathbf{x}^T \mathbf{x} + \mathbf{x}^{\prime T}\mathbf{x}^\prime \right) + \mathbf{x}^{\prime T} \hat{\beta} \mathbf{x} \right]\Psi \left( \mathbf{x^\prime} \right) =\lambda  \Psi \left( \mathbf{x} \right).
\end{split}
\end{equation}
Differentiating this relation with respect to $x_k$ yields
\begin{multline}
\int d^n \mathbf{x^\prime} \exp \left[ - \frac{1}{2} \left( \mathbf{x}^T \mathbf{x} + \mathbf{x}^{\prime T}\mathbf{x}^\prime \right) + \mathbf{x}^{\prime T} \hat{\beta} \mathbf{x} \right]x^\prime_m \Psi \left( \mathbf{x^\prime} \right)\\
=\frac{\lambda}{c} \left(\hat{\beta}^{-1}\right)_{km}\left[\partial_k \Psi \left( \mathbf{x} \right)+x_k\Psi \left( \mathbf{x} \right)\right] .
\label{eq:spectrum_A_der_of_eigen}
\end{multline}

The question that we would like to answer is whether the state $A_i^\dagger\Psi \left( \mathbf{x} \right)$ is an eigenfunction of the matrix $\tilde{\rho}_2$. If the answer is yes, then what is the corresponding eigenvalue? It is a matter of algebra to show that
\begin{multline}
\tilde{\rho}_2  A_i^\dagger\Psi \left( \mathbf{x} \right) = \frac{c}{\sqrt{2}}v^i_n \left(\re\left(\mathcal{A}\right)^{-1}\right)_{n k}\\
\times\int d^n \mathbf{x^\prime} \exp \left[ - \frac{1}{2} \left( \mathbf{x}^T \mathbf{x} + \mathbf{x}^{\prime T}\mathbf{x}^\prime \right) + \mathbf{x}^{\prime T} \hat{\beta} \mathbf{x} \right]\left(-\partial_k^\prime+ \mathcal{A}_{km}^*x_m^\prime\right)\Psi \left( \mathbf{x^\prime} \right) .
\end{multline}
The right hand side contains two terms. We can perform by parts integration to the term containing the derivative of $\Psi \left( \mathbf{x} \right)$,
\begin{multline}
\int d^n \mathbf{x^\prime} \exp \left[ - \frac{1}{2} \left( \mathbf{x}^T \mathbf{x} + \mathbf{x}^{\prime T}\mathbf{x}^\prime \right) + \mathbf{x}^{\prime T} \hat{\beta} \mathbf{x} \right]\partial_k^\prime\Psi \left( \mathbf{x^\prime} \right)\\
=\int d^n \mathbf{x^\prime} \Psi \left( \mathbf{x^\prime} \right)\left(x_k^\prime-\hat{\beta}_{k\ell} x_\ell\right)\exp \left[ - \frac{1}{2} \left( \mathbf{x}^T \mathbf{x} + \mathbf{x}^{\prime T}\mathbf{x}^\prime \right) + \mathbf{x}^{\prime T} \hat{\beta} \mathbf{x} \right]\\
=-\frac{\lambda}{c}\hat{\beta}_{k\ell} x_\ell \Psi \left( \mathbf{x} \right)+\int d^n \mathbf{x^\prime} \exp \left[ - \frac{1}{2} \left( \mathbf{x}^T \mathbf{x} + \mathbf{x}^{\prime T}\mathbf{x}^\prime \right) + \mathbf{x}^{\prime T} \hat{\beta} \mathbf{x} \right]x_k^\prime\Psi \left( \mathbf{x^\prime} \right) .
\end{multline}
This implies that
\begin{multline}
\tilde{\rho}_2  A_i^\dagger\Psi \left( \mathbf{x} \right) = \frac{c}{\sqrt{2}}v^i_n \left( \re\left(\mathcal{A}\right)^{-1}\right)_{n k} \Bigg[ \hat{\beta}_{k\ell} x_\ell\lambda \Psi \left( \mathbf{x} \right) \\
- \left(\delta_{km}-\mathcal{A}_{km}^*\right) \int d^n \mathbf{x^\prime} \exp \left[ - \frac{1}{2} \left( \mathbf{x}^T \mathbf{x} + \mathbf{x}^{\prime T}\mathbf{x}^\prime \right) + \mathbf{x}^{\prime T} \hat{\beta} \mathbf{x} \right]x_m^\prime\Psi \left( \mathbf{x^\prime} \right) \Bigg].
\end{multline}
Finally, using equation \eqref{eq:spectrum_A_der_of_eigen} we obtain
\begin{multline}
\tilde{\rho}_2  A_i^\dagger\Psi \left( \mathbf{x} \right) = \frac{\lambda}{\sqrt{2}}v^i_n \left( \re\left(\mathcal{A}\right)^{-1} \right)_{n k} \\
\times \left[ \hat{\beta}_{k\ell} x_\ell\Psi \left( \mathbf{x} \right)-\left(\delta_{km}-\mathcal{A}_{km}^*\right) \left( \hat{\beta}^{-1} \right)_{\ell m} \left[\partial_\ell \Psi \left( \mathbf{x} \right)+x_\ell\Psi \left( \mathbf{x} \right)\right] \right] .
\label{eq:spectrum_rho2_Adagger_Psi}
\end{multline}

The defining property of the matrix $\mathcal{A}$ \eqref{eq:spectrum_omega_def} and the definition of the matrix $\Xi$ \eqref{eq:spectrum_Xi_def} imply that $\left(I-\mathcal{A}^{*}\right)\left(\hat{\beta}^T\right)^{-1}=\hat{\beta}\left(I+\mathcal{A}^{*}\right)^{-1}=\Xi^*$. Furthermore, we have shown that the matrix $\Xi^\prime$, defined in \eqref{eq:spectrum_Xi_prime} is Hermitian. This implies that $\re\left(\mathcal{A}\right)^{-1}\Xi^{*} = \Xi^T\re\left(\mathcal{A}\right)^{-1}$. As a direct consequence, it follows that
\begin{equation}
\left( \re\left(\mathcal{A}\right)^{-1}\right)_{n k} \left(\delta_{km}-\mathcal{A}_{km}^*\right) \left( \hat{\beta}^{-1} \right)_{\ell m} = \left(\Xi^T\re\left(\mathcal{A}\right)^{-1}\right)_{n\ell} .
\end{equation}
Recalling that the vector $\mathbf{v}_i$ is an eigenvector of the matrix $\Xi$ with eigenvalue $\xi_i$, the above relation allows the re-writing of equation \eqref{eq:spectrum_rho2_Adagger_Psi} as
\begin{multline}
\tilde{\rho}_2  A_i^\dagger\Psi \left( \mathbf{x} \right) = \frac{\lambda}{\sqrt{2}}v^i_n \left( \re\left(\mathcal{A}\right)^{-1} \right)_{n k}\left[\left(\hat{\beta}_{k\ell} - \xi_i\delta_{k\ell}\right)x_\ell\Psi \left( \mathbf{x} \right)- \xi_i\partial_k \Psi \left( \mathbf{x} \right)\right]\\
=\lambda \xi_i A_i^\dagger \Psi \left( \mathbf{x} \right) + \frac{\lambda}{\sqrt{2}} v^i_n \left( \re\left(\mathcal{A}\right)^{-1} \right)_{n k}\left[\delta_{km} - \xi_i\left(\delta_{kr}+\mathcal{A}_{k r}^{*}\right)\left( \hat{\beta}^{-1} \right)_{rm}\right] \hat{\beta}_{m\ell}x_\ell\Psi \left( \mathbf{x} \right) .
\label{eq:spectrum_rho2_Adagger_Psi_2}
\end{multline}
Similarly to the algebra that we used in the previous step, the definition of the matrix $\Xi$ \eqref{eq:spectrum_Xi_def} implies that $\left( I+\mathcal{A}^* \right) \hat{\beta}^{-1} = \left( \Xi^* \right)^{-1}$. Furthermore, the fact that the matrix $\Xi^\prime$, defined in \eqref{eq:spectrum_Xi_prime}, is Hermitian implies that $\re\left(\mathcal{A}\right)^{-1}\left(\Xi^*\right)^{-1} = \left(\Xi^T\right)^{-1} \re\left(\mathcal{A}\right)^{-1}$. As a direct consequence, it follows that
\begin{equation}
\left( \re\left(\mathcal{A}\right)^{-1} \right)_{n k}\left(\delta_{kr}+\mathcal{A}_{k r}^{*}\right) \left( \hat{\beta}^{-1}\right)_{rm} 
=\left(\left(\Xi^{T}\right)^{-1}\re\left(\mathcal{A}\right)^{-1}\right)_{nm} .
\end{equation}
Once again, recalling that the vector $\mathbf{v}_i$ is an eigenvector of the matrix $\Xi$ with eigenvalue $\xi_i$, we get
\begin{multline}
v^i_n \left( \re\left(\mathcal{A}\right)^{-1} \right)_{n k} \left[\delta_{km} - \xi_i\left(\delta_{kr}+\mathcal{A}_{k r}^{*}\right) \left( \hat{\beta}^{-1} \right)_{rm} \right] \\
= v^i_n \left( \left(\re\left(\mathcal{A}\right)^{-1} \right)_{n m} - \xi_i \left( \left( \Xi^{T} \right)^{-1} \right)_{nk} \left( \re\left(\mathcal{A}\right)^{-1} \right)_{km} \right] = 0 .
\end{multline}
This implies that equation \eqref{eq:spectrum_rho2_Adagger_Psi_2} assumes the form
\begin{equation}
\tilde{\rho}_2 A_i^\dagger\Psi \left( \mathbf{x} \right) = \lambda \xi_i A_i^\dagger \Psi \left( \mathbf{x} \right) .
\end{equation}

We proved that if the state $\Psi \left( \mathbf{x} \right)$ is an eigenstate of the matrix $\tilde{\rho}_2$ with eigenvalue $\lambda$, then the state $A_i^\dagger \Psi \left( \mathbf{x} \right)$ is also an eigenstate with eigenvalue $\lambda \xi_i$. Inductively, this means that the states
\begin{equation}
\Psi_{\left\{ m_1 , m_2 , \ldots , m_n \right\}} \left( \mathbf{x} \right) = \frac{\left( A_1^\dagger \right)^{m_1}}{\sqrt{m_1 !}} \frac{\left( A_2^\dagger \right)^{m_2}}{\sqrt{m_2 !}} \ldots \frac{\left( A_{N-n}^\dagger \right)^{m_{N-n}}}{\sqrt{m_{N-n} !}} \Psi_0 \left( \mathbf{x} \right)
\end{equation}
are normalized eigenstates of the matrix $\tilde{\rho}_2$ with eigenvalues given by equation \eqref{eq:spectrum_final_spectrum}. The eigenstates of the reduced density matrix $\rho_2$ can be trivially found, recalling their relation to the eigenstates of $\tilde{\rho}_2$, which is given by equation \eqref{eq:eigenrho_eigenrhotilde}.

\section{A Solvable Example}
\label{sec:app_solvable}
In this appendix we analyse a solvable example, in order to clarify the properties of the spectrum of the matrix $M$ that we introduced in section \ref{subsec:spectrum_M_method}, and furthermore to verify that the asymptotic form of the eigenvalues of the matrix $\Xi$ for large squeezing parameters are indeed of the form that is predicted by the large squeezing expansion developed in section \ref{sec:expansions}. Let us consider the case that all modes lie in a squeezed state with the same squeezing parameter $z$, and further assume that we study the system at a instant when the phase of the oscillation of all modes is the same and equal to 0. In this case, we have
\begin{equation}
W = \left(\frac{1}{\cosh z} - i \tanh z\right)\Omega \, .
\end{equation}
Naturally, when $z = 0$, $W = \Omega$, as in the usual ground state calculation.

It follows that the matrices $\gamma$ and $\beta$ assume the form
\begin{align}
\gamma &= \left(\frac{1}{\cosh z} - i \tanh z\right)\left(\gamma_0+i \sinh z \beta_0\right), \\
\beta &= \cosh z\beta_0 ,
\end{align}
where $\gamma_0$ and $\beta_0$ are the matrices $\gamma$ and $\beta$ in the case of the ground state. The first of the two equations implies that
\begin{equation}
\re\left(\gamma\right)=\frac{\gamma_0 + \beta_0 \sinh^2z}{\cosh z} .
\end{equation}

Finally, the above imply that the matrix $M^\prime$, which defined in equation \eqref{eq:large_M_prime_def} and is similar to the matrix $M$, assumes the form
\begin{equation}
M^\prime = \begin{pmatrix} 2 \beta^{-1} \re\left(\gamma\right)& -\beta^{-1}\beta^T\\
I & 0
\end{pmatrix}=\begin{pmatrix} \frac{2}{\cosh^2 z} \left(\beta_0^{-1}\gamma_0 +\tanh^2 z I\right)& -I\\
I & 0
\end{pmatrix} .
\end{equation}
The eigenvalues of this matrix are
\begin{equation}
\lambda_{i\pm}=\frac{\hat{\beta}_i\cosh^2z}{1+\hat{\beta}_i\sinh^2z\pm\sqrt{(1-\hat{\beta}_i)(1+\hat{\beta}_i\cosh 2 z)}} ,
\end{equation}
where $\hat{\beta}_i$ are the eigenvalues of the matrix $\gamma_0^{-1}\beta_0$. These eigenvalues come in pairs of the form $\left( \lambda , 1 / \lambda \right)$, since $\lambda_{i+} \lambda_{i-} =1$. The eigenvalues which are smaller than 1 are the $\lambda_{i+}$, and, thus,
\begin{equation}
\xi_i=\frac{\hat{\beta}_i\cosh^2z}{1+\hat{\beta}_i\sinh^2z+\sqrt{(1-\hat{\beta}_i)(1+\hat{\beta}_i\cosh 2 z)}}=\frac{\sqrt{\frac{1+\hat{\beta}_i\cosh 2 z}{1-\hat{\beta}_i}}-1}{\sqrt{\frac{1+\hat{\beta}_i\cosh 2 z}{1-\tilde{\beta}_i}}+1}.
\end{equation}

For $\vert z\vert \gg1$ we have
\begin{equation}
\xi_i=1-\frac{2(1-\xi^{0}_i)}{\sqrt{\xi^{0}_i}}e^{-\vert z\vert}+\mathcal{O}\left(e^{-2\vert z\vert}\right) .
\end{equation}
This formula is in agreement with the large-squeezing expansion developed in section \ref{sec:expansions}. Similarly, for $\vert z\vert \ll 1$ we obtain
\begin{equation}
\xi_i=\xi^{0}_i\left(1+\frac{1-\xi^{0}_i}{1+\xi^{0}_i}z^2\right)+\mathcal{O}\left(z^4\right).
\end{equation}
The correction to $\xi^{0}_i$ is always non-negative.

\section{Entanglement Entropy through \Renyi Entropies - a Toy Case}
\label{sec:app_renyi}

In section \ref{sec:many_oscillators} we developed a method to calculate the entanglement entropy in harmonic systems with an arbitrary number of degrees of freedom that lie in a squeezed state. It would be nice if we could verify this method via an independent calculation in a non-trivial case.

There is an alternative method to calculate entanglement entropy via the so called entanglement \Renyi entropies. In this appendix we use this alternative method as a verifying example in a special case that the reduced system contains two degrees of freedom and the reduced density matrix is complex, but has a specific form.

\Renyi entropies constitute a family of entropies which extend the notion of Shannon's entropy. For a probability distribution $p_i$, the \Renyi entropy of order $a$ is defined as
\begin{equation}
S_a := \frac{1}{1 - a} \ln \left( \sum_i p_i^a \right) .
\end{equation}
Shannon's entropy is the limit $a \to 1$ of \Renyi entropies, i.e.
\begin{equation}
S = \lim_{a \to 1} S_a .
\end{equation}

As a direct generalization, we may define the entanglement \Renyi entropies of order $a$ as
\begin{equation}
S_a^{\mathrm{EE}} := \frac{1}{1 - a} \ln \Tr \rho_2^a
\label{eq:renyi_def_renyi}
\end{equation}
and recover the entanglement entropy as the limit
\begin{equation}
S_{\mathrm{EE}} = \lim_{a \to 1} S_a^{\mathrm{EE}} .
\end{equation}

In the case we study, namely when the overall oscillatory system lies in a squeezed state, we know that the reduced density matrix is of the form
\begin{equation}
\rho_2 \left( \mathbf{x} ; \mathbf{x}^\prime \right) = c \exp \left[ - \frac{1}{2} \left( \mathbf{x}^T \gamma \mathbf{x} + \mathbf{x}^{\prime T} \gamma^* \mathbf{x}^\prime \right) + \mathbf{x}^T \beta \mathbf{x}^\prime \right] ,
\end{equation}
where $\gamma$ is a complex symmetric matrix, $\beta$ is a Hermitian matrix. The normalization constant is given by $c = \left( {\det \re \left( \gamma - \beta \right)}/{\pi^d} \right)^{\frac{1}{2}}$, where $d$ is the number of degrees of freedom of the reduced system.

It is not difficult to show that the powers of the reduced density matrix are of the same form
\begin{equation}
\rho_2^n \left( \mathbf{x} ; \mathbf{x}^\prime \right) = c_n \exp \left[ - \frac{1}{2} \left( \mathbf{x}^T \gamma_n \mathbf{x} + \mathbf{x}^{\prime T} \gamma_n^* \mathbf{x}^\prime \right) + \mathbf{x}^T \beta_n \mathbf{x}^\prime \right] ,
\end{equation}
where the matrices $\gamma_n$ are complex symmetric and the matrices $\beta_n$ are Hermitian. Obviously, $\gamma_1 = \gamma$, $\beta_1 = \beta$ and $c_1 = c$. 

The matrices $\gamma_n$, $\beta_n$ and the coefficients $c_n$ obey some recursion relations. It holds that
\begin{equation}
\rho_2^{n + 1} \left( \mathbf{x} ; \mathbf{x}^{\prime\prime} \right) = \int d^d \mathbf{x}^\prime \rho_2^n \left( \mathbf{x} ; \mathbf{x}^\prime \right) \rho_2 \left( \mathbf{x}^\prime ; \mathbf{x}^{\prime\prime} \right) .
\end{equation}
But,
\begin{multline}
\rho_2^n \left( \mathbf{x} ; \mathbf{x}^\prime \right) \rho_2 \left( \mathbf{x}^\prime ; \mathbf{x}^{\prime\prime} \right)
= c_n c \exp \left[ - \frac{1}{2} \left( \mathbf{x}^T \gamma_n \mathbf{x} + \mathbf{x}^{\prime\prime T} \gamma^* \mathbf{x}^{\prime\prime} \right) + \frac{1}{2} \mathbf{v}^T \left( \gamma_n^* + \gamma \right) \mathbf{v} \right] \\
\times \exp \left[ - \frac{1}{2} \left( \mathbf{x}^\prime + \mathbf{v} \right)^T \left( \gamma_n^* + \gamma \right) \left( \mathbf{x}^\prime + \mathbf{v} \right) \right] ,
\end{multline}
where
\begin{equation}
\mathbf{v} = \left( \gamma_n^* + \gamma \right)^{- 1} \left( \beta_n^* \mathbf{x} + \beta \mathbf{x}^{\prime\prime} \right) .
\end{equation}
It directly follows that
\begin{multline}
\rho_2^{n + 1} \left( \mathbf{x} ; \mathbf{x}^{\prime\prime} \right) = c_n c \left( \frac{2^d \pi^d}{\det \left( \gamma_n^* + \gamma \right)} \right)^{\frac{1}{2}} \exp \bigg[ - \frac{1}{2} \bigg( \mathbf{x}^T \left( \gamma_n - \beta_n \left( \gamma_n^* + \gamma \right)^{- 1} \beta_n^* \right) \mathbf{x} \\
+ \mathbf{x}^{\prime\prime T} \left( \gamma^* - \beta^* \left( \gamma_n^* + \gamma \right)^{- 1} \beta \right) \mathbf{x}^{\prime\prime} \bigg) + \mathbf{x}^T \beta_n \left( \gamma_n^* + \gamma \right)^{- 1} \beta \mathbf{x}^{\prime\prime} \bigg] .
\label{eq:renyi_rhon1}
\end{multline}

The fact that $\rho_2^{n + 1}$ is Hermitian implies that
\begin{align}
\gamma_n - \beta_n \left( \gamma_n^* + \gamma \right)^{- 1} \beta_n^* &= \gamma - \beta \left( \gamma_n + \gamma^* \right)^{- 1} \beta^* , \\
\beta_n \left( \gamma_n^* + \gamma \right)^{- 1} \beta &= \beta \left( \gamma_n + \gamma^* \right)^{- 1} \beta_n .
\end{align}
The recursive relations for $\gamma_n$, $\beta_n$ and $c_n$ can be directly read from the expression of the matrix $\rho_2^{n + 1}$ \eqref{eq:renyi_rhon1}. They read
\begin{align}
\gamma_{n + 1} &= \gamma - \beta \left( \gamma_n + \gamma^* \right)^{- 1} \beta^* , \\
\beta_{n + 1} &= \beta \left( \gamma_n + \gamma^* \right)^{- 1} \beta_n , \\
c_{n + 1} &= c_n c \left( \frac{2^d \pi^d}{\det \left( \gamma_n^* + \gamma \right)} \right)^{\frac{1}{2}} .
\end{align}

We have shown that the imaginary part of the matrix $\gamma$ does not affect the eigenvalues of the reduced density matrix. It only alters its eigenstates in a trivial way. Therefore, without loss of generality we may assume that the matrix $\gamma$ is a real symmetric matrix. Of course, the matrices $\gamma_n$, $n > 1$ may still be complex symmetric matrices. This assumption simplifies the recursive formulae,
\begin{align}
\gamma_{n + 1} &= \gamma - \beta \left( \gamma_n + \gamma \right)^{- 1} \beta^* , \label{eq:renyi_recursive_gamma} \\
\beta_{n + 1} &= \beta \left( \gamma_n + \gamma \right)^{- 1} \beta_n , \label{eq:renyi_recursive_beta} \\
c_{n + 1} &= c_n c \left( \frac{2^d \pi^d}{\det \left( \gamma_n^* + \gamma \right)} \right)^{\frac{1}{2}} . \label{eq:renyi_recursive_c}
\end{align}

We define
\begin{align}
\hat{\beta} &\equiv \gamma^{- \frac{1}{2}} \beta \gamma^{- \frac{1}{2}} , \label{eq:renyi_def_beta_tilde} \\
\hat{\gamma}_n &\equiv \gamma^{- \frac{1}{2}} \gamma_n \gamma^{- \frac{1}{2}} , \label{eq:renyi_def_gamman_tilde} \\
\hat{\beta}_n &\equiv \gamma^{- \frac{1}{2}} \beta_n \gamma^{- \frac{1}{2}} . \label{eq:renyi_def_betan_tilde}
\end{align}
Using these definitions, the recursive formula \eqref{eq:renyi_recursive_gamma} assumes the form
\begin{equation}
\hat{\gamma}_{n + 1} = I - \hat{\beta} \left( I + \hat{\gamma}_n \right)^{- 1} \hat{\beta}^* .
\label{eq:renyi_recursive_gamma_tilde}
\end{equation}
The initial condition for this recursive relation is obviously $\gamma_1 = \gamma$, which implies $\hat{\gamma}_1 = I$. Similarly, the recursive formula \eqref{eq:renyi_recursive_beta} assumes the form
\begin{equation}
\hat{\beta}_{n + 1} = \hat{\beta} \left( I + \hat{\gamma}_n \right)^{- 1} \hat{\beta}_n .
\label{eq:renyi_recursive_beta_tilde}
\end{equation}
The initial condition for this is simply $\beta_1 = \beta$, which implies that $\hat{\beta}_1 = \hat{\beta}$. Finally, the recursive relation \eqref{eq:renyi_recursive_c} can be written as
\begin{equation}
c_{n + 1} = c_n c \left( \frac{2^d \pi^d}{\det \gamma \det \left( I + \hat{\gamma}_n^* \right)} \right)^{\frac{1}{2}} = c_n \left( \frac{2^d \det \left( 1 - \re \hat{\beta} \right)}{\det \left( I + \hat{\gamma}_n^* \right)} \right)^{\frac{1}{2}} .
\end{equation}

In the case that the matrix $\hat{\beta}$ is real, e.g. when the overall system lies in the ground state, all matrices $\hat{\gamma}_n$ and $\hat{\beta}_n$ are functions of a single matrix, namely of $\hat{\beta}$, and these recursion relations can be solved as recursion relations for numbers. Their solution leads to simple explicit formulas for the entanglement \Renyi entropies and the exact same formula for entanglement entropy found in \cite{srednicki}. However, in our case of study the recursion relation for $\hat{\gamma}_n$ contains the two matrices $\hat{\beta}$ and $\hat{\beta}^*$, which in general do not commute.

However, the above formulae can be solved in a special case, where the matrix $\hat{\beta}$ is complex. Assume the case where the reduced system contains two degrees of freedom and
\begin{equation}
\hat{\beta} = \beta_0 I + \beta_2 \sigma_2 .
\label{eq:renyi_beta_tilde_specific}
\end{equation}
Let us define the eigenvalues of the matrix $\hat{\beta}$ as
\begin{equation}
\lambda_1 = \beta_0 + \beta_2 , \quad \lambda_2 = \beta_0 - \beta_2 .
\end{equation}
In an obvious manner,
\begin{equation}
\hat{\beta} \hat{\beta}^* = \left( \beta_0^2 - \beta_2^2 \right) I = \lambda_1 \lambda_2 I = \det \hat{\beta} I .
\end{equation}

Since the initial condition for the recursion relation for $\hat{\gamma}_n$ \eqref{eq:renyi_recursive_gamma_tilde} is $\hat{\gamma}_1 = I$, it follows that this recursion relation has the trivial solution
\begin{equation}
\hat{\gamma}_n = a_n I ,
\end{equation}
where the coefficients $a_n$ obey
\begin{equation}
a_{n + 1} = a_n + \frac{\det \hat{\beta}}{1 + a_n} .
\end{equation}
This recursion relation has the solution
\begin{equation}
a_n = \sqrt{1 - \det \hat{\beta}} \frac{1 + \xi^n}{1 - \xi^n} ,
\end{equation}
which implies that
\begin{equation}
\hat{\gamma}_n = \sqrt{1 - \det \hat{\beta}} \frac{1 + \xi^n}{1 - \xi^n} I ,
\label{eq:renyi_gamman_tilde_solution}
\end{equation}
where
\begin{equation}
\xi = \frac{1 - \sqrt{1 - \det \hat{\beta}}}{1 + \sqrt{1 - \det \hat{\beta}}} .
\end{equation}

The above form of the matrices $\hat{\gamma}_n$ implies that
\begin{equation}
\left( I + \hat{\gamma}_n \right)^{- 1} = \frac{1}{1 + \sqrt{1 - \det \hat{\beta}}} \frac{1 - \xi^n}{1 - \xi^{n + 1}} I .
\label{eq:renyi_ipgamma_inverse}
\end{equation}

The recursion relation for $\hat{\beta}_n$ \eqref{eq:renyi_def_betan_tilde}, combined with equation \eqref{eq:renyi_ipgamma_inverse}, implies that
\begin{equation}
\hat{\beta}_n = \frac{1}{\left( 1 + \sqrt{1 - \det \hat{\beta}} \right)^{n - 1}} \frac{1 - \xi}{1 - \xi^n} \hat{\beta}^n .
\label{eq:renyi_betan_tilde_solution}
\end{equation}
It is not difficult to show that
\begin{equation}
\hat{\beta}^n = \frac{1}{2} \left( \lambda_1^n + \lambda_2^n \right) I + \frac{1}{2} \left( \lambda_1^n - \lambda_2^n \right) \sigma_2 .
\end{equation}

We may define a sequence of matrices $\hat{\Gamma}_n$, so that
\begin{equation}
c_n = c \left( \det \hat{\Gamma}_n \right)^{\frac{1}{2}} .
\label{eq:renyi_def_Gamman}
\end{equation}
In an obvious manner, the determinants of these matrices should obey the recursion relation
\begin{equation}
\det \hat{\Gamma}_{n + 1} = \det \hat{\Gamma}_n \left( \frac{2^d \det \left( 1 - \re \hat{\beta} \right)}{\det \left( I + \hat{\gamma}_n \right)} \right)^{\frac{1}{2}}
\end{equation}
and the initial condition $\det \hat{\Gamma}_1 = 1$. A simple way to satisfy this is to find the sequence of matrices that obey the recursion relation
\begin{equation}
\hat{\Gamma}_{n + 1} = 2 \left( I - \re \hat{\beta} \right) \left( I + \hat{\gamma}_n \right)^{- 1} \hat{\Gamma}_n
\end{equation}
and the initial condition $\hat{\Gamma}_1 = I$. This equation combined with \eqref{eq:renyi_ipgamma_inverse} directly implies that
\begin{equation}
\hat{\Gamma}_n = \frac{1 - \beta_0}{\pi} \frac{2 \left( 1 - \beta_0 \right)}{\left( 1 + \sqrt{1 - \det \hat{\beta}} \right)^{n - 1}} \frac{1 - \xi}{1 - \xi^n} I
\label{eq:renyi_Gamman_solution}
\end{equation}

The trace of $\rho_2^n$ is simply
\begin{equation}
\Tr \rho_2^n = \int d^d \mathbf{x} \rho_2^n \left( \mathbf{x} ; \mathbf{x} \right) 
= c_n \left( \frac{\pi^d}{\det \left( \gamma_n - \re \beta_n \right)} \right)^{\frac{1}{2}} .
\end{equation}
Using the definitions \eqref{eq:renyi_def_beta_tilde}, \eqref{eq:renyi_def_gamman_tilde} and \eqref{eq:renyi_def_betan_tilde}, as well as the definition \eqref{eq:renyi_def_Gamman}, we find
\begin{equation}
\Tr \rho_2^n 
= \left( \frac{\det \left( I - \re \hat{\beta} \right) \det \hat{\Gamma}_n}{\det \left( \hat{\gamma}_n - \re \hat{\beta}_n \right)} \right)^{\frac{1}{2}} .
\end{equation}

Using the form of the matrices $\hat{\gamma}_n$, $\hat{\beta}_n$ and $\hat{\Gamma}_n$ from equations \eqref{eq:renyi_gamman_tilde_solution}, \eqref{eq:renyi_betan_tilde_solution} and \eqref{eq:renyi_Gamman_solution} and putting everything together yields
\begin{equation}
\Tr \rho_2^n = \frac{\left( 2 - \lambda_1 - \lambda_2 \right)^n}{\left( 1 + \sqrt{1 - \lambda_1 \lambda_2} \right)^n + \left( 1 - \sqrt{1 - \lambda_1 \lambda_2} \right)^n - \lambda_1^n - \lambda_2^n}
\label{eq:renyi_trrhon}
\end{equation}

The Renyi entanglement entropy is defined by equation \eqref{eq:renyi_def_renyi}. It reads
\begin{equation}
S_a^{\mathrm{EE}} = \frac{1}{1 - a} \ln \frac{\left( 2 - \lambda_1 - \lambda_2 \right)^a}{\left( 1 + \sqrt{1 - \lambda_1 \lambda_2} \right)^a + \left( 1 - \sqrt{1 - \lambda_1 \lambda_2} \right)^a - \lambda_1^a - \lambda_2^a} .
\end{equation}
The form of the trace $\Tr \rho_2^a$ clearly implies that $\lim_{n \to 1} \Tr \rho_2^n = 1$, as expected. It follows that
\begin{equation}
S_{\mathrm{EE}} = - \lim_{n \to 1} \Tr \frac{\partial \Tr \rho_2^n}{\partial n}
\end{equation}
It is a matter of algebra to show that
\begin{multline}
S_{\mathrm{EE}} = \ln \left( 2 - \lambda_1 - \lambda_2 \right) - \frac{1}{2 - \lambda_1 - \lambda_2} \bigg[ - \lambda_1 \ln \lambda_1 - \lambda_2 \ln \lambda_2 \\
+ \left( 1 - \sqrt{1 - \lambda_1 \lambda_2} \right) \ln \left( 1 - \sqrt{1 - \lambda_1 \lambda_2} \right) \\
+ \left( 1 + \sqrt{1 - \lambda_1 \lambda_2} \right) \ln \left( 1 + \sqrt{1 - \lambda_1 \lambda_2} \right) \bigg]
\end{multline}
or
\begin{multline}
S_{\mathrm{EE}} = \ln \left( 2 - \lambda_1 - \lambda_2 \right) - \frac{1}{2 - \lambda_1 - \lambda_2} \Bigg[ - \lambda_1 \ln \lambda_1 - \lambda_2 \ln \lambda_2 \\
+ \ln \left( \lambda_1 \lambda_2 \right) + \sqrt{1 - \lambda_1 \lambda_2} \ln \frac{1 + \sqrt{1 - \lambda_1 \lambda_2}}{1 - \sqrt{1 - \lambda_1 \lambda_2}} \Bigg].
\label{eq:renyi_See_renyi}
\end{multline}

In order to verify that this result is consistent with the general method that we developed in section \ref{sec:many_oscillators}, we need to solve the non-linear eigenvalue equation of the eigenvalues of the matrix $M$, namely
\begin{equation}
\det \left( \xi^2 \hat{\beta}^T - 2 \xi I + \hat{\beta} \right) = 0 .
\end{equation}
In our example, the matrix $\hat{\beta}$ is given by equation \eqref{eq:renyi_beta_tilde_specific}. The above equation gives
\begin{equation}
\det \left( \left( \xi^2 \beta_0 - 2 \xi + \beta_0 \right) I + \beta_2 \left( - \xi^2 + 1 \right) \sigma_2 \right) = 0 .
\end{equation}
This reads
\begin{equation}
\left( \xi^2 \beta_0 - 2 \xi + \beta_0 \right)^2 - \beta_2^2 \left( - \xi^2 + 1 \right)^2 = 0
\end{equation}
or
\begin{equation}
\left( \xi^2 \lambda_1 - 2 \xi + \lambda_2 \right) \left( \xi^2 \lambda_2 - 2 \xi + \lambda_1 \right) = 0 .
\end{equation}
The last equation has four solutions,
\begin{equation}
\xi = \frac{1 \pm \sqrt{1 - \lambda_1 \lambda_2}}{\lambda_1} \equiv \xi_{1 \pm} , \quad \mathrm{or} \quad \xi = \frac{1 \pm \sqrt{1 - \lambda_1 \lambda_2}}{\lambda_2} \equiv \xi_{2 \pm} .
\end{equation}
These indeed form two pairs of solutions that are inverse to each other. Namely $\xi_{1 +} = 1 / \xi_{2 -}$ and $\xi_{2 +} = 1 / \xi_{1 -}$.
The solutions that are smaller than $1$ are the solutions $\xi_{1 -}$ and $\xi_{2 -}$.
It follows that the entanglement entropy reads
\begin{equation}
S_{\mathrm{EE}} = - \ln \left( 1 - \xi_{1 -} \right) - \frac{\xi_{1 -}}{1 - \xi_{1 -}} \ln \xi_{1 -} - \ln \left( 1 - \xi_{2 -} \right) - \frac{\xi_{2 -}}{1 - \xi_{2 -}} \ln \xi_{2 -} .
\end{equation}
It is a matter of tedious algebra to show that the above expression is identical to equation \eqref{eq:renyi_See_renyi}.

\section{Entanglement in terms of Correlation Functions}
\label{subsec:correlation}
For Gaussian states there exist an alternative method for the calculation of entanglement entropy based on correlation functions \cite{Ppeschel}, see also \cite{Casini:2009sr}. This method is based on the fact that for Gaussian states the correlation functions are expressed as products of 2-point functions. Therefore, specifying a modular Hamiltonian that reproduces the correct 2-point functions guaranties that this is indeed the modular Hamiltonian corresponding to the particular density matrix. Via this process the spectrum of the modular Hamiltonian is related to the eigenvalues of the correlation functions. In the case of the vacuum, the matrix $W$ appearing in \eqref{eq:many_overall_state} (in which case $W=\Omega$, where $\Omega$ is the frequency matrix) is real. The matrices $\gamma$ and $\beta$ are real as well. One can show that\footnote{In this section we use a slightly different notation for the blocks of a matrix $Q$, namely
\begin{equation*}
Q=\begin{pmatrix}
Q_A & Q_B\\
Q_B^T & Q_C
\end{pmatrix}.
\end{equation*}
}
\begin{equation}\label{eq:gamma_beta_cor}
\gamma^{-1}\beta=\frac{\left(\Omega^{-1}\right)_C \Omega_C-I}{\left(\Omega^{-1}\right)_C \Omega_C+I}.
\end{equation}
To derive this relation one has to observe that
\begin{equation}
\gamma+\beta=\Omega_C,\qquad \gamma-\beta=\Omega_C -\Omega_B^T\Omega_A^{-1}\Omega_B=\left( \left(\Omega^{-1}\right)_C\right)^{-1},
\end{equation}
where the last equation is a property of the Schur complement $\Omega_C -\Omega_B^T\Omega_A^{-1}\Omega_B$. This property enables us to use directly blocks of the matrix $\Omega^{-1}$, such as $\left( \Omega^{-1} \right)_C$, rather than the inverses of blocks of $\Omega$, such as $\Omega_A^{-1}$. Using the fact that $\Omega$ and $\Omega^{-1}$ are the momentum and position 2-point functions respectively, more specifically
\begin{align}
X_{ij}=\left\langle x_ix_j\right\rangle&=\Tr\left[x_ix_j\rho\right]=\frac{1}{2}\left(\Omega^{-1}\right)_{ij},\\
\Pi_{ij}=\left\langle \pi_i\pi_j\right\rangle&=-\Tr\left[\partial_i\partial_j\rho\right]=\frac{1}{2}\Omega_{ij},\\
\left\langle x_i\pi_j\right\rangle&=-i\Tr\left[x_i\partial_j\rho\right]=\frac{i}{2}\delta_{ij},
\end{align}
we relate the spectrum of the reduced density matrix to the spectrum of the matrix $X_C \Pi_C$. In particular, the eigenvalues of the matrix $\Xi$ are given by
\begin{equation}\label{eq:xi_cor}
\xi_i=\frac{\Lambda_i-\frac{1}{2}}{\Lambda_i+\frac{1}{2}},
\end{equation}
where $\Lambda_i$ are the eigenvalues of $\sqrt{X_C \Pi_C}$. As a final remark, in order to be on the same page, we remind the reader that one may calculate the correlation functions for the overall system and then restrict the indices to the subsystem under consideration. This is denoted by the index $C$.

After this short introduction let us turn to the case of interest. The vacuum state is characterized by the fact that $\left\langle x_i\pi_j+\pi_j x_i\right\rangle=0$, which of course is equivalent to $\re\left\langle x_i\pi_j\right\rangle=0$. The method based on the correlation functions can be generalized appropriately for $\re\left\langle x_i\pi_j\right\rangle\neq0$. One considers the matrices
\begin{equation}
\mathcal{M}=\begin{pmatrix}
\left\langle x_ix_j\right\rangle & \left\langle x_i\pi_j\right\rangle\\
\left\langle x_i\pi_j\right\rangle^T & \left\langle \pi_i\pi_j\right\rangle
\end{pmatrix},\qquad J=\begin{pmatrix}
0 & I\\
-I & 0
\end{pmatrix}
\end{equation}
and calculates the eigenvalues of $i J\re(\mathcal{M})$, see for instance \cite{Coser:2014gsa,Cotler:2016acd,Bianchi:2015fra,Sorkin:2012sn}.

It is easy to show that
\begin{equation}
\left\langle x_ix_j\right\rangle=\Tr\left[x_ix_j\rho\right]=\frac{1}{2}\left(\re\left(W\right)^{-1}\right)_{ij} .
\end{equation}
In a similar manner, one can show that
\begin{equation}
\left\langle \pi_i\pi_j\right\rangle=-\Tr\left[\partial_i\partial_j\rho\right]=\frac{1}{2}\left[\re\left(W\right)+\im\left(W\right)\re\left(W\right)^{-1}\im\left(W\right)\right]_{ij}
\end{equation}
and
\begin{align}
\left\langle x_i\pi_j\right\rangle&=-i \Tr\left[x_i\partial_j\rho\right]=\frac{1}{2}\left[iI-\re\left(W\right)^{-1}\im\left(W\right)\right]_{ij},\\
\left\langle \pi_jx_i\right\rangle&=-i \Tr\left[\partial_jx_i\rho\right]=-\frac{1}{2}\left[iI+\re\left(W\right)^{-1}\im\left(W\right)\right]_{ij}.
\end{align}
Using these correlation functions the matrix $i J\re(\mathcal{M})$ reads
\begin{equation}
i J\,\re(\mathcal{M})=\frac{i}{2}\begin{pmatrix}
-\im\left(W\right)\re\left(W\right)^{-1} & \re\left(W\right)+\im\left(W\right)\re\left(W\right)^{-1}\im\left(W\right)\\
-\re\left(W\right)^{-1} & \re\left(W\right)^{-1}\im\left(W\right)
\end{pmatrix}.
\end{equation}
Unfortunately, we are not able to find a direct relation between the matrix $M$, defined in \eqref{eq:many_M_def}, and the matrix $i J\,\re(\mathcal{M})$ or its blocks. So, the best we can do is to show that we obtain the same spectrum, by relating the characteristic polynomials of the matrices.
The characteristic polynomial of the matrix $i J\re(\mathcal{M})$ is
\begin{multline}
\det\left(i J\,\re(\mathcal{M})-\Lambda I\right)=\det\left(\left(\re\left(W\right)^{-1}\right)_C\right)\\
\det\left[\Lambda^2\left(\left(\re\left(W\right)^{-1}\right)_C\right)^{-1}-\frac{1}{4}\left(\re\left(W\right)_C+\im\left(W\right)_B^T\left(\re\left(W\right)_A\right)^{-1}\im\left(W\right)_B\right)\right.\\
-\frac{i}{2}\Lambda\left(\left(\left(\re\left(W\right)^{-1}\right)_C\right)^{-1}\left(\re\left(W\right)^{-1}\right)_B^T\im\left(W\right)_B\right.\\-\left.\left.\im\left(W\right)_B^T\left(\re\left(W\right)^{-1}\right)_B\left(\left(\re\left(W\right)^{-1}\right)_C\right)^{-1}\right)\right],
\end{multline}
where we used the fact that
\begin{equation}
\left(\re\left(W\right)_A\right)^{-1}=\left(\re\left(W\right)^{-1}\right)_A-\left(\re\left(W\right)^{-1}\right)_B^T\left(\left(\re\left(W\right)^{-1}\right)_C\right)^{-1}\left(\re\left(W\right)^{-1}\right)_B.
\end{equation}
Similarly, it also holds true that
\begin{equation}
\left(\left(\re\left(W\right)^{-1}\right)_C\right)^{-1}=\re\left(W\right)_C-\re\left(W\right)_B^T\left(\re\left(W\right)_A\right)^{-1}\re\left(W\right)_B.
\end{equation}
Defining $\Lambda=\frac{1}{2}\frac{1+\lambda}{1-\lambda}$ we obtain
\begin{equation}
\det\left(i J\,\re(\mathcal{M})-\Lambda I\right)=0\Rightarrow \det\left(\mathcal{M}_2-\frac{\lambda}{4} \mathcal{M}_0-\frac{1}{4\lambda}\mathcal{M}_0^T\right)=0,
\end{equation}
where
\begin{multline}
\mathcal{M}_2=\re\left(W\right)_C-\frac{1}{2}\re\left(W\right)_B^T \left(\re\left(W\right)_A\right)^{-1}\re\left(W\right)_B \\ + \frac{1}{2}\im\left(W\right)_B^T \left(\re\left(W\right)_A\right)^{-1}\im\left(W\right)_B=\re\left(\gamma\right)
\end{multline}
and
\begin{equation}
\mathcal{M}_0=\Omega_B^\dagger \left(\re\left(W\right)_A\right)^{-1}\Omega_B+i \left(\mathcal{M}_0^\prime-\mathcal{M}_0^{\prime T}\right)= 2\beta+i \left(\mathcal{M}_0^\prime-\mathcal{M}_0^{\prime T}\right),
\end{equation}
with
\begin{multline}
\mathcal{M}_0^\prime=\im\left(\Omega\right)_B^T\left[\left(I-\left(\re\left(\Omega\right)^{-1}\right)_B\re\left(\Omega\right)_B^T\right)\left(\re\left(\Omega\right)_A\right)^{-1}\re\left(\Omega\right)_B^T\right.\\ 
+\left.\left(\re\left(\Omega\right)^{-1}\right)_B \re\left(\Omega\right)_C\right].
\end{multline}
The trivial relation $\re\left(\Omega\right)^{-1}\re\left(\Omega\right)=I$ implies that
\begin{align}
&\left(\re\left(\Omega\right)^{-1}\right)_A\re\left(\Omega\right)_A+\left(\re\left(\Omega\right)^{-1}\right)_B\re\left(\Omega\right)_B^T=I,\\
&\left(\re\left(\Omega\right)^{-1}\right)_A\re\left(\Omega\right)_B^T+\left(\re\left(\Omega\right)^{-1}\right)_B \re\left(\Omega\right)_C=0.
\end{align}
Thus, $\mathcal{M}_0^\prime$ vanishes and we arrive at
\begin{equation}
\det\left(i J\,\re(\mathcal{M})-\Lambda I\right)=0\Rightarrow \det\left(2\re\left(\gamma\right)-\lambda\beta-\frac{1}{\lambda}\beta^T\right)=0,
\end{equation}
where $\lambda=\frac{\Lambda-\frac{1}{2}}{\Lambda+\frac{1}{2}}$. As a result, we have shown that the method based on the correlation functions and the direct calculation, see \eqref{eq:spectrum_eigenvalues_Xi_equation}, result in the same spectrum. Also, notice that the eigenvalues of the matrices used in these methods are related in the same way as in the case of the vacuum, see \eqref{eq:xi_cor}. 

Interestingly enough, we can relate the admissible eigenvalues of $M$ to the eigenvalues of another matrix. It can be shown that the matrix $M$ is similar to another matrix with the same structure, namely
\begin{equation}
M^\prime=\begin{pmatrix}
2\beta^{-1}\re\left(\gamma\right)& -\beta^{-1} \beta^T \\
I & 0
\end{pmatrix}.
\end{equation}
Since $M$ and $M^\prime$ are related by a similarity transformation, they share the same spectrum.

The eigenvalues we are interested in are the solutions of the equation
\begin{equation}
\det\left(M^\prime-\lambda I\right)=\frac{1}{\det\left(-\beta/2\right)}\det\left(\lambda\re\left(\gamma\right)-\frac{\lambda^2}{2} \beta-\frac{1}{2}\beta^T\right)=0.
\end{equation}
It is a matter of algebra to show that
\begin{multline}
\det\left(\lambda\re\left(\gamma\right)-\frac{\lambda^2}{2} \beta-\frac{1}{2}\beta^T\right)=\det\Bigg[\lambda\re\left(C\right)\\
-\left(\frac{1+\lambda}{2}\re\left(B\right)+i\frac{1-\lambda}{2}\im\left(B\right)\right)^T\re\left(A\right)^{-1}\left(\frac{1+\lambda}{2}\re\left(B\right)-i\frac{1-\lambda}{2}\im\left(B\right)\right)\Bigg].
\end{multline}
As a result, we obtain
\begin{equation}
\det\left(M^\prime-\lambda I\right)\propto\det\begin{pmatrix}
\re\left(A\right) & \frac{1+\lambda}{2}\re\left(B\right)-i\frac{1-\lambda}{2}\im\left(B\right)\\
\frac{1+\lambda}{2}\re\left(B\right)^T+i\frac{1-\lambda}{2}\im\left(B\right)^T & \lambda \re\left(C\right)
\end{pmatrix}
\end{equation}
or
\begin{equation}
\det\left(M^\prime-\lambda I\right)\propto\det \left[\frac{1+\lambda}{2}\re\left(W\right) -\frac{1-\lambda}{2}\begin{pmatrix}
-\re\left(A\right) & i\im\left(B\right)\\
-i\im\left(B\right)^T & \re\left(C\right)
\end{pmatrix}\right].
\end{equation}
Thus, the eigenvalues of the matrix $M$, denoted by $\lambda$, are related to the eigenvalues $\tilde{\lambda}$ of the matrix $\tilde{M}$, where
\begin{equation}
\tilde{M}=\re\left(W\right)^{-1}\begin{pmatrix}
-\re\left(A\right) & i\im\left(B\right)\\
-i\im\left(B\right)^T & \re\left(C\right)
\end{pmatrix},
\end{equation}
via the equation
\begin{equation}\label{eq:mtilde}
\lambda=\frac{\tilde{\lambda}-1}{\tilde{\lambda}+1}.
\end{equation}
Notice that the matrix $\tilde{M}$ is $N\times N$, thus we have introduced spurious eigenvalues. However, there is also an advantage. Recall that the approach based on the matrix $M$ works only when we trace out the larger subsystem and we have to rely on the fact that entanglement entropy satisfies $S_A=S_{A^C}$ for pure states. The calculation based on $\tilde{M}$ works in both cases: either when we trace out the larger subsystem or the smaller one. One has to pick out the admissible eigenvalues, i.e. the ones that are larger than $1$.

In the vacuum case, in which case $\im\left(B\right)=0$, the structure of the eigenvalues is as follows: For $n<N/2$ the eigenvalues have the structure $\tilde{\lambda}_i=\pm\frac{1}{2}\Lambda_i$, which gives in total $2n$ eigenvalues, along with $N-2n$ eigenvalues which are equal to $1$. When $n>N/2$ the eigenvalues have the structure $\tilde{\lambda}_i=\pm\frac{1}{2}\Lambda_i$, which gives in total $2(N-n)$ eigenvalues, along with $2n-N$ eigenvalues which are equal to $-1$. This structure implies that the full spectrum of $\tilde{M}$ contains the eigenvalues of both $-\sqrt{\Omega_A^{-1}\Omega_A}$ and $\sqrt{\Omega_C^{-1}\Omega_C}$. Of course this is expected by comparing \eqref{eq:mtilde} and \eqref{eq:xi_cor}.

\section{Small-Squeezing Expansion}
\label{subsec:small_expansion}

Similarly to the large squeezing expansion that we presented in section \ref{sec:expansions}, we expand the parameter $w$ as a series in the squeezing parameter. This reads
\begin{equation}
w = \omega - i \omega z e^{- 2 i \omega t} + \mathcal{O} \left( z^2 \right) .
\label{eq:small_z_w_exp}
\end{equation}
The zeroth order term is real, time-independent and equal to the eigenfrequency of the mode. Unlike the case of the large squeezing expansion, all terms in the expansion of $w$ (apart the zeroth order one) contain both a real and an imaginary part. We will use a similar notation to that we used in section \ref{sec:expansions}.

Equation \eqref{eq:small_z_w_exp} implies that the matrix $W$ has an expansion of the form
\begin{equation}
W = \sum_{i = 0}^\infty z^{i} W^{(i)} ,
\end{equation}
where $W^{(i)}$ are in general complex. It follows that its blocks have a similar expansion and the same holds for the matrices $\gamma$ and $\beta$,
\begin{align}
\gamma &= \sum_{i = 0}^\infty z^{i} \gamma^{(i)} , \\
\beta &= \sum_{i = 0}^\infty z^{i} \beta^{(i)} .
\end{align}
In all these expansions, the small squeezing parameter $z$ may be the squeezing parameter of a single mode or even a small parameter in terms of which the small squeezing parameters of all modes can be expressed.

We would like to perform textbook first order perturbation theory to the spectrum of the matrix $M$. This would be simpler if the matrix $M$ were Hermitian, at least at zeroth order, so that its eigenvectors are orthogonal. Actually, we can find a matrix $\hat{M}$, which is similar to $M$ and Hermitian. This reads
\begin{equation}
\hat{M}=\begin{pmatrix}
I & \hat{\Xi}_+ \\
\hat{\Xi}_+ & I
\end{pmatrix}^{-1}\begin{pmatrix}
\left(\gamma^{(0)}\right)^{1/2} & 0\\
0 & \left(\gamma^{(0)}\right)^{1/2}
\end{pmatrix} M^\prime \begin{pmatrix}
\left(\gamma^{(0)}\right)^{1/2} & 0\\
0 & \left(\gamma^{(0)}\right)^{1/2}
\end{pmatrix}^{-1}\begin{pmatrix}
I & \hat{\Xi}_+ \\
\hat{\Xi}_+ & I
\end{pmatrix},
\end{equation}
where $\hat{\beta}=\left(\gamma^{(0)}\right)^{-1/2} \left(\beta^{(0)}\right) \left(\gamma^{(0)}\right)^{-1/2}$ and $\hat{\Xi}_\pm=\frac{\hat{\beta}}{I\pm\sqrt{I-\hat{\beta}^2}}$. Trivially $\hat{M}$ shares the same eigenvalues with $M^\prime$, which is similar to $M$ and is defined in \eqref{eq:large_M_prime_def}. Given that
\begin{equation}
\begin{pmatrix}
I & \hat{\Xi}_+ \\
\hat{\Xi}_+ & I
\end{pmatrix}^{-1}=\frac{\hat{\beta}}{2\sqrt{I-\hat{\beta}^2}}\begin{pmatrix}
\hat{\Xi}_- & -I\\
-I & \hat{\Xi}_-
\end{pmatrix},
\end{equation}
we obtain
\begin{equation}
\hat{M}^{(0)}=\begin{pmatrix}
\hat{\Xi}_- & 0\\
0 & \hat{\Xi}_+.
\end{pmatrix}
\end{equation}

The matrix $\hat{M}^{(0)}$ is not only Hermitian but also block-diagonal and its eigenvectors are trivially constructed from the eigenvectors of $\hat{\beta}$. Let $x_i$ be the eigenvectors of the matrix $\hat{\beta}$ with corresponding eigenvalues equal to $\hat{\beta}_i$, i.e.
\begin{equation}
\hat{\beta} x_i = \hat{\beta}_i x_i .
\end{equation}
Then, there are two kinds of eigenvalues and eigenvectors of the matrix $\hat{M}^{(0)}$, namely
\begin{align}
&\hat{v}_i=\begin{pmatrix}
0 \\ x_i
\end{pmatrix},\quad \textrm{with eigenvalues} \quad \lambda_i=\frac{\hat{\beta}_i}{1+\sqrt{1-\hat{\beta}_i^2}},\\
&\hat{v}_i^\prime=\begin{pmatrix}
x_i \\ 0
\end{pmatrix},\quad \textrm{with eigenvalues} \quad \lambda_i^\prime=\frac{\hat{\beta}_i}{1-\sqrt{1-\hat{\beta}_i^2}}.
\end{align}
As expected the eigenvalues come in pairs of the form $\left( \lambda , 1 / \lambda\right)$. Indeed, $\lambda_i \lambda_i^\prime = 1$. The eigenvalues that are smaller than 1 are the ones corresponding to eigenvectors of the first kind, namely the $\lambda_i$. Indeed, they coincide with the values of the parameters $\xi$ in the original calculation by Srednicki \cite{srednicki}. Notice also that the matrix $\hat{\beta}$ is real and symmetric, and thus its eigenvectors $x_i$ are real.

It is a matter of algebra to show that
\begin{equation}
\hat{M}^{(1)}=\frac{\hat{\beta}}{2\sqrt{I-\hat{\beta}^2}}\begin{pmatrix}
\hat{\Xi}_- \left(\hat{M}_{11}^{(1)}+\hat{M}_{12}^{(1)}\hat{\Xi}_+\right) & \hat{\Xi}_- \left(\hat{M}_{11}^{(1)}\hat{\Xi}_++\hat{M}_{12}^{(1)}\right), \\
-\left(\hat{M}_{11}^{(1)}+\hat{M}_{12}^{(1)}\hat{\Xi}_+\right) & -\left(\hat{M}_{11}^{(1)}\hat{\Xi}_++\hat{M}_{12}^{(1)}\right)
\end{pmatrix},
\end{equation}
where 
\begin{align}
\hat{M}_{11}^{(1)}&=2\hat{\beta}^{-1}\left(\re\left(\hat{\gamma}^{(1)}\right)-\hat{\beta}^{(1)}\hat{\beta}^{-1}\right),\\
\hat{M}_{12}^{(1)}&=2i\hat{\beta}^{-1}\im\left(\hat{\beta}^{(1)}\right)
\end{align}
and
\begin{align}
\hat{\gamma}^{(1)}&=\left(\gamma^{(0)}\right)^{-1/2}\gamma^{(1)}\left(\gamma^{(0)}\right)^{-1/2},\\
\hat{\beta}^{(1)}&=\left(\gamma^{(0)}\right)^{-1/2}\beta^{(1)}\left(\gamma^{(0)}\right)^{-1/2}.
\end{align}

Now we can apply perturbation theory to find the eigenvalues of the matrix $M$ at first order. Considering that $\xi_i=\xi_i^{(0)}+z\xi_i^{(1)} + \mathcal{O} \left( z^2 \right)$, and assuming that the zeroth order eigenvectors of the matrix $\hat{M}$ have been defined so that they are normalized, the $\xi_i^{(1)}$ are given by the usual first order perturbation theory formula $\xi_i^{(1)} = \hat{v}_i^T \hat{M}^{(1)} \hat{v}_i$, which yields
\begin{equation}
\xi_i^{(1)}=\frac{1}{\sqrt{I-\hat{\beta}_i^2}} x_i^T\left(\re\left(\hat{\gamma}^{(1)}\right)-\hat{\beta}^{(1)}\hat{\beta}_i^{-1}+i \, \im\left(\hat{\beta}^{(1)}\right)\right)x_i.
\end{equation}
Since $\im\left(\hat{\beta}^{(1)}\right)$ is antisymmetric, the above expression simplifies to
\begin{equation}
\xi_i^{(1)}=\frac{1}{\sqrt{I-\hat{\beta}_i^2}} x_i^T\left(\re\left(\hat{\gamma}^{(1)}\right)-\re\left(\hat{\beta}^{(1)}\right)\hat{\beta}_i^{-1}\right)x_i.
\end{equation}
This implies that the entanglement entropy at a given time contains corrections which are first order in $z$, namely
\begin{equation}
S_{\mathrm{EE}} = S_{\mathrm{EE}}^{(0)} + \sum_i \left( \left. \frac{\partial S_{\mathrm{EE}}}{\partial \xi_i} \right|_{\xi_i = \xi_i^{(0)}} \xi_i^{(1)} \right) z + \mathcal{O} \left( z^2 \right) .
\end{equation}
However, this is not the case for the mean entanglement entropy. The quantities  $\xi_i^{(1)}$ depend linearly on $i \omega e^{- 2 i \omega t}$, as it results from equation \eqref{eq:small_z_w_exp}. However, they are real, therefore they depend linearly on a combination of $\cos \left( 2 \omega t \right)$ and $\sin \left( 2 \omega t \right)$. As a result, the mean value of $\xi_i^{(1)}$ vanishes and so does the correction of the mean entanglement entropy at first order. It follows that the ground state is a stationary point for the mean entanglement entropy within the space of squeezed states.



\begin{thebibliography}{99}

\bibitem{Jacobson:1995ab}
T.~Jacobson,
``Thermodynamics of space-time: The Einstein equation of state'',
Phys. Rev. Lett. \textbf{75}, 1260-1263 (1995)
[arXiv:gr-qc/9504004 [gr-qc]].

\bibitem{VanRaamsdonk:2010pw}
M.~Van Raamsdonk,
``Building up spacetime with quantum entanglement'',
Gen. Rel. Grav. \textbf{42}, 2323-2329 (2010)
[arXiv:1005.3035 [hep-th]].

\bibitem{Jacobson:2015hqa}
T.~Jacobson,
``Entanglement Equilibrium and the Einstein Equation'',
Phys. Rev. Lett. \textbf{116}, no.20, 201101 (2016)
[arXiv:1505.04753 [gr-qc]].

\bibitem{Lashkari:2013koa}
N.~Lashkari, M.~B.~McDermott and M.~Van Raamsdonk,
``Gravitational dynamics from entanglement {``thermodynamics''}'',
JHEP \textbf{04}, 195 (2014)
[arXiv:1308.3716 [hep-th]].

\bibitem{Faulkner:2013ica}
T.~Faulkner, M.~Guica, T.~Hartman, R.~C.~Myers and M.~Van Raamsdonk,
``Gravitation from Entanglement in Holographic CFTs''.
JHEP \textbf{03}, 051 (2014)
[arXiv:1312.7856 [hep-th]].

\bibitem{Bombelli:1986rw}
L.~Bombelli, R.~K.~Koul, J.~Lee and R.~D.~Sorkin,
``A Quantum Source of Entropy for Black Holes'',
Phys. Rev. D \textbf{34}, 373-383 (1986)

\bibitem{srednicki} 
M.~Srednicki,
``Entropy and area'',
Phys.\ Rev.\ Lett.\  {\bf 71}, 666 (1993)
[hep-th/9303048].

\bibitem{Calabrese:2004eu}
P.~Calabrese and J.~L.~Cardy,
``Entanglement entropy and quantum field theory'',
J. Stat. Mech. \textbf{0406}, P06002 (2004)
[arXiv:hep-th/0405152 [hep-th]].

\bibitem{Casini:2009sr}
H.~Casini and M.~Huerta,
``Entanglement entropy in free quantum field theory'',
J. Phys. A \textbf{42}, 504007 (2009)
[arXiv:0905.2562 [hep-th]].

\bibitem{Calabrese:2009qy}
P.~Calabrese and J.~Cardy,
``Entanglement entropy and conformal field theory'',
J. Phys. A \textbf{42}, 504005 (2009)
[arXiv:0905.4013 [cond-mat.stat-mech]].



\bibitem{Katsinis:2017qzh}
D.~Katsinis and G.~Pastras,
``An Inverse Mass Expansion for Entanglement Entropy in Free Massive Scalar Field Theory'',
Eur. Phys. J. C \textbf{78}, no.4, 282 (2018)
[arXiv:1711.02618 [hep-th]].
  
\bibitem{Katsinis:2019vhk}
D.~Katsinis and G.~Pastras,
``Area Law Behaviour of Mutual Information at Finite Temperature'',
[arXiv:1907.04817 [hep-th]].

\bibitem{Katsinis:2019lis}
D.~Katsinis and G.~Pastras,
``An Inverse Mass Expansion for the Mutual Information in Free Scalar QFT at Finite Temperature'',
JHEP \textbf{02}, 091 (2020)
[arXiv:1907.08508 [hep-th]].

\bibitem{Benedict:1995yp}
E.~Benedict and S.~Y.~Pi,
``Entanglement entropy of nontrivial states'',
Annals Phys. \textbf{245}, 209-224 (1996)
[arXiv:hep-th/9505121 [hep-th]].
  
\bibitem{Katsinis:2022fxu}
D.~Katsinis and G.~Pastras,
``Entanglement in harmonic systems at coherent states'',
[arXiv:2206.05781 [hep-th]].


\bibitem{Page:1993df}
D.~N.~Page,
``Average entropy of a subsystem'',
Phys. Rev. Lett. \textbf{71}, 1291-1294 (1993)
[arXiv:gr-qc/9305007 [gr-qc]].

\bibitem{Page:1993wv}
D.~N.~Page,
``Information in black hole radiation'',
Phys. Rev. Lett. \textbf{71}, 3743-3746 (1993)
[arXiv:hep-th/9306083 [hep-th]].

\bibitem{Eisert:2008ur}
J.~Eisert, M.~Cramer and M.~B.~Plenio,
``Area laws for the entanglement entropy - a review'',
Rev. Mod. Phys. \textbf{82}, 277-306 (2010)
[arXiv:0808.3773 [quant-ph]].

\bibitem{Bianchi:2021lnp}
E.~Bianchi, L.~Hackl and M.~Kieburg,
``Page curve for fermionic Gaussian states'',
Phys. Rev. B \textbf{103}, no.24, L241118 (2021)
[arXiv:2103.05416 [quant-ph]].


\bibitem{Bianchi:2015fra}
E.~Bianchi, L.~Hackl and N.~Yokomizo,
``Entanglement entropy of squeezed vacua on a lattice'',
Phys. Rev. D \textbf{92}, no.8, 085045 (2015)
[arXiv:1507.01567 [hep-th]].

\bibitem{Adesso:2014}
G.~Adesso, S.~Ragy and A. R. Lee,
``Continuous Variable Quantum Information: Gaussian States and Beyond'',
Open Systems and Information Dynamics 21 01n02 1440001 (2014)
[arXiv:1401.4679 [quant-ph]]


\bibitem{Penington:2019npb}
G.~Penington,
``Entanglement Wedge Reconstruction and the Information Paradox'',
JHEP \textbf{09}, 002 (2020)
[arXiv:1905.08255 [hep-th]].

\bibitem{Almheiri:2019psf}
A.~Almheiri, N.~Engelhardt, D.~Marolf and H.~Maxfield,
``The entropy of bulk quantum fields and the entanglement wedge of an evaporating black hole'',
JHEP \textbf{12}, 063 (2019)
[arXiv:1905.08762 [hep-th]].

\bibitem{physrep}
V.~F.~Mukhanov, H.~Feldman and R.~H.~Brandenberger,
``Theory of cosmological perturbations. Part 1. Classical perturbations. Part 2. Quantum theory of perturbations. Part 3. Extensions'',
Phys. Rept. \textbf{215} (1992), 203-333.


\bibitem{albrecht}
A.~Albrecht, P.~Ferreira, M.~Joyce and T.~Prokopec,
``Inflation and squeezed quantum states'',
Phys. Rev. D \textbf{50} (1994), 4807-4820
[arXiv:astro-ph/9303001 [astro-ph]].

\bibitem{classical1}
D.~Polarski and A.~A.~Starobinsky,
``Semiclassicality and decoherence of cosmological perturbations'',
Class. Quant. Grav. \textbf{13} (1996), 377-392
[arXiv:gr-qc/9504030 [gr-qc]].


\bibitem{Grishchuk}
L.~P.~Grishchuk and Y.~V.~Sidorov,
``Squeezed quantum states of relic gravitons and primordial density fluctuations'',
Phys. Rev. D \textbf{42} (1990), 3413-3421.

\bibitem{squeeze1}
R.~H.~Brandenberger, T.~Prokopec and V.~F.~Mukhanov,
``The Entropy of the gravitational field'',
Phys. Rev. D \textbf{48} (1993), 2443-2455
[arXiv:gr-qc/9208009 [gr-qc]].

\bibitem{squeeze2}
R.~H.~Brandenberger, V.~F.~Mukhanov and T.~Prokopec,
``Entropy of a classical stochastic field and cosmological perturbations'',
Phys. Rev. Lett. \textbf{69} (1992), 3606-3609
[arXiv:astro-ph/9206005 [astro-ph]].

\bibitem{squeeze3}
T.~Prokopec,
``Entropy of the squeezed vacuum'',
Class. Quant. Grav. \textbf{10} (1993), 2295-2306.

\bibitem{squeeze4}
A.~L.~Matacz,
``The Coherent state representation of quantum fluctuations in the early universe'',
Phys. Rev. D \textbf{49} (1994), 788-798
[arXiv:gr-qc/9212008 [gr-qc]].

\bibitem{squeeze5}
M.~Gasperini and M.~Giovannini,
``Entropy production in the cosmological amplification of the vacuum fluctuations'',
Phys. Lett. B \textbf{301} (1993), 334-338
[arXiv:gr-qc/9301010 [gr-qc]].

\bibitem{squeeze6}
M.~Gasperini and M.~Giovannini,
``Quantum squeezing and cosmological entropy production'',
Class. Quant. Grav. \textbf{10} (1993), L133-L136
[arXiv:gr-qc/9307024 [gr-qc]].

\bibitem{squeeze7}
C.~Kiefer, D.~Polarski and A.~A.~Starobinsky,
``Entropy of gravitons produced in the early universe'',
Phys. Rev. D \textbf{62} (2000), 043518
[arXiv:gr-qc/9910065 [gr-qc]].

\bibitem{squeeze8}
D.~Campo and R.~Parentani,
``Decoherence and entropy of primordial fluctuations. I: Formalism and interpretation'',
Phys. Rev. D \textbf{78} (2008), 065044
[arXiv:0805.0548 [hep-th]].


\bibitem{Boutivas:2023ksg}
K.~Boutivas, G.~Pastras and N.~Tetradis,
``Entanglement and expansion'',
[arXiv:2302.14666 [hep-th]].


\bibitem{Lindblad}
G.~Lindblad, 
``On the generators of quantum dynamical semigroups'',
Comm.Math.Phys, 48, 119(1976)

\bibitem{Ppeschel}
I.~Peschel,
``Calculation of reduced density matrices from
correlation functions'',
J. Phys. A 36 (2003) L205
[arXiv:cond-mat/0212631 [hep-th]].

\bibitem{Coser:2014gsa}
A.~Coser, E.~Tonni and P.~Calabrese,
``Entanglement negativity after a global quantum quench,''
J. Stat. Mech. \textbf{1412} (2014) no.12, P12017
[arXiv:1410.0900 [cond-mat.stat-mech]].

\bibitem{Cotler:2016acd}
J.~S.~Cotler, M.~P.~Hertzberg, M.~Mezei and M.~T.~Mueller,
``Entanglement Growth after a Global Quench in Free Scalar Field Theory,''
JHEP \textbf{11} (2016), 166
[arXiv:1609.00872 [hep-th]].

\bibitem{Sorkin:2012sn}
R.~D.~Sorkin,
``Expressing entropy globally in terms of (4D) field-correlations'',
J. Phys. Conf. Ser. \textbf{484} (2014), 012004
[arXiv:1205.2953 [hep-th]].




\end{thebibliography}
\end{document}